\documentclass{CSML}

\def\dOi{12(4:2)2016}
\lmcsheading%
{\dOi}
{1--51}
{}
{}
{Feb.~\phantom08, 2016}
{Oct.~11, 2016}
{}

\ACMCCS{[{\bf Theory of computation}]: Logic---Logic and
  verification\,/\,Modal and temporal logics\,/\,Verification by model
  checking; Semantics and reasoning; [{\bf Software and its engineering}]: Software organization
  and properties---Software functional properties---Formal
  methods---Software verification;
  }

\usepackage{tabto}
\usepackage{amsfonts}
\usepackage{amsmath}
\usepackage{mathrsfs}
\usepackage{supertabular}
\usepackage{xspace}
\usepackage{bookmark}
\usepackage[utf8]{inputenc}
\usepackage{url}
\usepackage{amssymb}
\usepackage{amsmath}
\usepackage{tikz}
\usepackage{stmaryrd}
\usepackage{graphicx}
\usepackage[noline,noend,norelsize]{algorithm2e}
\newcommand{\clsexa}{}
\newcommand{\pow}{\wp}
\newcommand{\open}{O}

\newcommand{\eval}{\mathcal{V}}
\newcommand{\props}{AP}

\newcommand{\model}{\mathcal{M}}
\newcommand{\nmodels}{\nvDash}

\newcommand{\reals}{\mathbb{R}}

\newcommand{\logicSfourU}{\mathcal{S}4_u}

\newcommand{\boundary}{\mathcal{B}}
\newcommand{\iboundary}{\boundary^-}
\newcommand{\cboundary}{\boundary^+}
\newcommand{\closure}{\mathcal{C}}
\newcommand{\interior}{\mathcal{I}}

\newcommand{\nats}{\mathbb{N}}

\newcommand{\lboundary}{\mathcal{\delta}}
\newcommand{\liboundary}{\lboundary^-}
\newcommand{\lcboundary}{\lboundary^+}
\newcommand{\rats}{\mathbb{Q}}

\renewcommand{\succ}{\mathit{Succ}}
\newcommand{\pto}[2]{\overset{#1}{\underset{#2}{\rightsquigarrow}}\infty}
\DeclareMathOperator{\ldualuntil}{\mathcal{R}} 
\DeclareMathOperator{\ldualdiff}{\mathcal{A}} 
\DeclareMathOperator{\leverywhere}{\mathcal{E}}
\newcommand{\lsomewhere}{\mathcal{F}}

\newcommand{\mneigh}{N}
\newcommand{\dertab}{.9cm}
\newcommand{\dervskip}{.2cm}

\newcommand{\der}{$\vskip \dervskip\noindent\tabto{\dertab}$}

\newcommand{\stp}[1]{\\*[\dervskip]#1$\tabto{\dertab}$}
\newcommand{\jstp}[2]{\\*[\dervskip]#1$\tabto{\dertab}$[\,#2\,]\\*[\dervskip]$\tabto{\dertab}$}
\newcommand{\crstp}{\stp{}}
\newcommand{\slcs}{SLCS}
\newcommand{\cslcs}{CSLCS}
\newcommand{\lShare}{\ -\hskip-6pt<}

\DeclareMathOperator{\lSSurr}{\mathcal{CS}}
\DeclareMathOperator{\lPartitioned}{\mathcal{CP}}

\newcommand{\lGroup}{\mathcal{G}}
\newcommand{\lnear}{\mathcal{N}}
\DeclareMathOperator{\lsurr}{\mathcal{S}}
\DeclareMathOperator{\ldiff}{\mathcal{P}}

\newcommand{\closenv}{\hfill$\bullet$}

\newcommand{\lForall}{\forall}
\newcommand{\lExists}{\exists}

\newcommand{\linterior}{\mathcal{I}}

\DontPrintSemicolon
\SetKwFunction{check}{Sat}
\SetKwFunction{checkUntil}{CheckSurr}
\SetKwFunction{checkDiff}{CheckProp}
\SetKwFunction{checkGroup}{CheckGroup}
\SetKwFunction{visit}{Visit}
\SetKwProg{myfun}{Function}{}{}
\SetKwSwitch{Match}{Case}{Other}{Match}{}{case}{otherwise}{~ }{}
\SetKwIF{If}{ElseIf}{Else}{if}{}{else if}{else}{endif}
\SetKw{Let}{let}
\SetKw{In}{}
\SetKw{Var}{var}
\SetKw{Conj}{and}
\SetKw{Neg}{not}


\usepackage{cleveref}
\theoremstyle{plain}
\newtheorem{theorem}[thm]{Theorem}\crefname{theorem}{Theorem}{Theorems}
\newtheorem{proposition}[thm]{Proposition}\crefname{proposition}{Proposition}{Propositions}
\newtheorem{lemma}[thm]{Lemma}\crefname{lemma}{Lemma}{Lemmas}
\crefname{corollary}{Corollary}{Corollaries}

\theoremstyle{definition}
\newtheorem{remark}[thm]{Remark}\crefname{remark}{Remark}{Remarks}
\newtheorem{definition}[thm]{Definition}\crefname{definition}{Definition}{Definitions}
\newtheorem{example}[thm]{Example}\crefname{example}{Example}{Examples}

\crefname{section}{Section}{Sections}
\crefname{figure}{Figure}{Figures}
\crefname{algorithm}{Algorithm}{Algorithms}
\crefname{equation}{Equation}{Equations}

\renewcommand{\autoref}[1]{\cref{#1}}

\usepackage{hyperref}
\hypersetup{pdfborder={0 0 0}}

\begin{document}

\title[Model Checking Spatial Logics for Closure Spaces]
      {Model Checking Spatial Logics for Closure Spaces\rsuper*}
\titlecomment{{\lsuper*}Research partially funded by EU project QUANTICOL (nr. 600708)}

\author[V.~Ciancia]{Vincenzo Ciancia\rsuper a}
\address{{\lsuper{a,b,d}}Istituto di Scienza e Tecnologie dell'Informazione ``A. Faedo'' - CNR, Pisa}	
\email{\{vincenzo.ciancia, diego.latella, mieke.massink\}@isti.cnr.it}

\author[D.~Latella]{Diego Latella\rsuper b}	
\address{\vspace{-18 pt}}

\author[M.~Loreti]{Michele Loreti\rsuper c}	
\address{{\lsuper{c}}Università degli Studi di Firenze and IMT Alti
  Studi, Lucca}	%
\email{michele.loreti@unifi.it}

\author[M.~Massink]{Mieke Massink\rsuper d}
\address{\vspace{-18 pt}}

\begin{abstract}
Spatial aspects of computation are becoming increasingly relevant in Computer Science, especially in the field of \emph{collective adaptive systems} and when dealing with systems distributed in physical space. Traditional formal verification techniques are well suited to analyse the temporal evolution of programs; however, properties of space are typically not taken into account explicitly. We present a topology-based approach to formal verification of spatial properties depending upon physical space. We define an appropriate logic, stemming from the tradition of topological interpretations of modal logics, dating back to earlier logicians such as Tarski, where modalities describe neighbourhood. We lift the topological definitions to the more general setting of \emph{closure spaces}, also encompassing discrete, graph-based structures. We extend the framework with a spatial \emph{surrounded} operator, a \emph{propagation} operator and with some \emph{collective} operators. The latter are interpreted over arbitrary \emph{sets} of points instead of individual points in space. We define efficient model checking procedures, both for the individual and the collective spatial fragments of the logic and provide a proof-of-concept tool.
\end{abstract}

\maketitle




\section{Introduction}

Much attention has been devoted in Computer Science to formal verification of process behaviour. Several techniques have been studied and developed that are based on a formal understanding of system requirements through \emph{modal} logics. Such logics typically  have a \emph{temporal} flavour, describing the flow of events, and are interpreted in various kinds of transition structures. Among those techniques {\em model checking} is one of the most successful (for an extensive overview see e.g.~\cite{BK08} and references therein).

In recent times, aspects of computation related to the distribution of systems in physical space have become increasingly relevant. An example is provided by so called \emph{collective adaptive systems}\footnote{See e.g. the web site of the QUANTICOL project: \url{http://www.quanticol.eu}, and that of the FOCAS Coordination Action: \url{http://www.focas.eu}.}. Such systems are typically composed of a large number of interacting objects located in space. Their global behaviour critically depends on interactions which are often local in nature. The aspect of locality immediately poses issues of spatial distribution of objects. Abstraction from spatial distribution may sometimes provide insights in the system behaviour, but this is not always the case. For example, consider a bike (or car) sharing system having several parking stations, and featuring twice as many parking slots as there are vehicles in the system. Ignoring the spatial dimension, on average, the probability to find completely full or empty parking stations at an arbitrary station is very low; however, this kind of analysis may be misleading, as in practice some stations are much more popular than others, often depending on nearby points of interest. This leads to quite different probabilities to find stations completely full or empty, depending on the examined location. In other cases, it may be important to be able to specify spatial properties concerning {\em groups} of points in space rather than of individual points. For example, the property that agents associated to points in space are able to connect to one another and act as a group, or that they are located all together in a protected environment, or that they can share part of the same route to reach a common exit or goal. In all such situations, it is important to be able to predicate over spatial aspects, and eventually find methods to certify that a given collective adaptive system satisfies specific requirements in this respect. 

In Logics, there is a considerable amount of literature focused on so called \emph{spatial} logics, that is, a spatial interpretation of modal logics~\cite{HBSL}. Dating back to early logicians such as Tarski, modalities may be interpreted using the concept of \emph{neighbourhood} in a topological space. The field of spatial logics is well developed in terms of descriptive languages and decidability or complexity aspects. However, in this field, scant attention has been devoted to date to the development of formal and automatic verification methods, e.g. model checking. Furthermore, the formal treatment of \emph{discrete} models of space is still a relatively unexplored field, with notable exceptions such as the work by Rosenfeld~\cite{KR89,Ros79}, Galton (e.g.\cite{Gal14,Gal03,Gal99}) and by Smyth and Webster~\cite{SW07}. Kovalevsky~\cite{Kov08} studied alternative axioms for topological spaces in order to recover well-behaved notions of neighbourhood. The outcome is that one may impose closure operators on top of a topology, that do not coincide with topological closure.

In~\cite{Ci+14b} we proposed the logic \slcs\ (Spatial Logic for Closure Spaces), extending the topological semantics of modal logics to \emph{closure spaces}. The work follows up on the research line of Galton and Smyth and Webster, enhancing it with a modal logic perspective. Closure spaces (also called \emph{\v Cech closure spaces} or \emph{preclosure spaces} in the literature) are based on a 
single operator on sets of points, namely the {\em closure} operator, and are a generalisation of standard topological spaces. 
In addition, finite spaces and graphs are subclasses of closure spaces and the graph-theoretical notion of neighbourhood coincides with the notion of neighbourhood defined in the context of closure spaces. 
Thus, closure spaces provide a uniform framework for the treatment of all major models of space. 

We provided a logical operator corresponding to the closure operator on sets of points in space, and  a
spatial interpretation of the temporal \emph{until} operator, fundamental in the classical temporal setting, arriving at the definition of a logic which is able to describe unbounded areas of space. Intuitively, the spatial until operator, which in the present paper we call \emph{surrounded}, describes a situation in which it is not possible to ``escape'' an area of points satisfying a certain property, unless by passing through at least one point that satisfies another given formula. This operator is similar in spirit to the spatial until operator for topological spaces discussed 
by Aiello and van Benthem in~\cite{A02,vBB07}. In~\cite{Ci+14b} we also presented a model-checking algorithm for \slcs\ when interpreted on finite models. The combination of \slcs\ with temporal operators from the well-known branching time logic CTL (Computation Tree Logic)~\cite{ClE82}, has been explored in~\cite{CGLLM15,CLMP15} and provides spatio-temporal reasoning and model checking. 

In the present paper we extend \slcs\ with a further operator, $ \ldiff$, capturing the notion of spatial propagation;
intuitively the formula $\phi \, \ldiff \, \psi$ describes a situation in which the points satisfying $\psi$ can be reached 
by paths rooted in points satisfying $\phi$ and, for the rest, composed only of points satisfying $\psi$.
We furthermore extend the logic with operators for \emph{collective properties}, namely properties which are satisfied by \emph{connected sets} of points, rather than points in isolation. The formal semantics of the extended logic---\cslcs{,} Collective \slcs\---are provided in the form of a satisfiability relation defined using the notion of infinite path in closure spaces.
We finally extend the model-checking algorithm in order to treat the newly introduced operators, 
and we present several examples of use of \slcs\ and \cslcs\ from the domain of collective adaptive systems using a prototype implementation of the spatial model-checker. 


\subsection*{Related work.} 
Variants of spatial logics have also been proposed for the symbolic representation of the contents of images, and, combined with temporal logics, for sequences of images~\cite{DVD95}. The latter approach is based on a discretisation of the space of the images in rectangular regions and the orthogonal projection of objects and regions onto Cartesian coordinate axes such that their possible intersections can be analysed from different perspectives. It involves two spatial until operators defined on such projections considering spatial shifts of regions along the positive, respectively negative, direction of the coordinate axes
and it is very different from the topological spatial logic approach.

In~\cite{grosu_learning_2008,GSCWEB09,bartocci2014} another variant of spatial logic is proposed in which spatial properties are expressed using ideas from image processing, namely quad trees. This variant is equipped with practical model checking algorithms and with machine learning procedures and allows one to capture very complex spatial structures. However, this comes at the price of a complex formulation of spatial properties, which need to be learned from some template image. The combination of this spatial logic with linear time signal temporal logic, defined with respect to continuous-valued signals, has recently led to the spatio-temporal logic SpaTeL~\cite{bartocci2015}. 

In the specific setting of complex and collective adaptive systems, techniques for efficient approximation have been developed in the form of mean-field or fluid-flow analysis (see \cite{BHLM13} for a tutorial introduction). Recently (see for example \cite{CLR09}), the importance of spatial aspects has been recognised and studied in this context. In~\cite{NB14} a first step towards the combination of signal temporal logic with spatial operators such as `somewhere' and `everywhere' has been performed. These two operators were also proposed in work by Reif and Sistla \cite{RS85}. In further joint work along
these 
lines~\cite{NBCLM15} some of the spatial operators based on closure spaces from \slcs{,} such as the `surrounded' operator, have been added to the signal temporal logic fraction. Both boolean semantics and quantitative semantics of the spatio-temporal logic have been provided. The quantitative semantics provide a measure of the robustness with which a spatio-temporal property holds in a given point in space at a particular time. The approach has been applied to investigate the emergence and persistence of Turing patterns in animal fur based on reaction diffusion models.

In~\cite{CG02} a geometric process algebra based on affine geometry has been proposed for describing the concurrent evolution of geometric structures in 3D space.
Spatial dynamics of systems have also been studied in the context of Systems Biology applying suitable modelling and simulation approaches. In~\cite{JEU08} a spatial (and temporal) extension of the $\pi$-Calculus is proposed. The notion of space is expressed by associating each process with its current position in $\mathbb{R}^d$. The formal semantics of the language is given, based on which simulation tools have been developed. In~\cite{BHMU11} an attributed, multi-level, rule-based language, ML-Space, is presented that allows one to integrate different types of spatial dynamics within one model. The associated simulator combines several stochastic simulation methods. This allows for the simulation of reaction diffusion systems as well as taking excluded volume effects into account. Formal verification and analysis, e.g. model checking, is not addressed.

In the Computer Science literature, some spatial logics have been proposed, that typically describe situations in which modal operators are interpreted \emph{syntactically} against the structure of agents in a process calculus. We refer to~\cite{CG00,CC03} for some classical examples. In the same line, a recent example is given by~\cite{TPGN15}, concerning model checking of security aspects in cyber-physical systems, in a spatial context based on the idea of bigraphical reactive systems introduced by Milner~\cite{Mil09}. The objects of discussion in the latter research lines are operators that for example quantify over the parallel sub-components of a system, the containment relation between places, or the hidden resources of an agent. The meaning of the terminology ``spatial logics'' in that case is different from that used in the present paper, where the ``topological'' interpretation of \cite{vBB07} is intended. The influence of space on agents interaction is also considered in the literature on process calculi using \emph{named locations}~\cite{DFP98},
where every process interaction primitive is enriched with the indication of (the name of) the location where the action operates. In that paper, space is modelled as a discrete, finite set of points.

Logics for graphs have been studied in the context of databases and process calculi (see \cite{CGG02,GL07}, and the references therein), even though the relationship with physical space is  often only implicit, if considered at all. 

Graph-based spatial logics for collective adaptive systems are also proposed in~\cite{AnS15a}. In that approach the logic extends a chemical-based coordination model based on logic inference. Properties are expressed in the form of combinations of logic programs. The spatial operators distribute such programs over the nodes of a graph to infer information local to each node. The locally inferred  data is 
logically aggregated at local spatial locations. Evaluated properties involve collective aspects either with a local scope (neighbourhood) or with a global scope. The approach relies on \emph{a priori} defined spatial patterns.

A successful attempt to bring topology and digital imaging together is represented by the field of \emph{digital topology} \cite{Ros79,KR89}. In spite of its name, this area studies digital images using models inspired by topological spaces, but neither generalising nor specialising these structures. Rather recently, closure spaces have been proposed as an alternative foundation of digital imaging by various authors, especially Smyth and Webster \cite{SW07} and Galton \cite{Gal03}; we continue that research line in the present paper, enhancing it with a (modal) logic perspective. 

In~\cite{Gal14}, a sub-class of closure spaces, namely {\em adjacency spaces}, 
is presented. An adjacency space is characterized by a set of entities together with a reflexive and symmetric relation. In the above
mentioned paper, adjacency spaces are used as the basis for the definition of {\em regions}, i.e. {\em sets of} entities, and the construction of a discrete interpretation of  logical operators typical of  region calculi, based on the  notion of region {\em connectedness}
derived from the notion of entity adjacency. Region calculi operators predicate on regions (see \cite{KKWZ07} for a comprehensive overview), using boolean connectives like  ``part of'', ``boundary'', ``overlap" and so on. 
An important aspect of adjacency spaces is that they can be easily turned into topological spaces, without loosing any information on their internal structure, which makes them rather attractive. Verification issues, e.g. model-checking, are not addressed in~\cite{Gal14}.

The structure of the present paper is as follows. \autoref{sec:ClosureSpaces} recalls basic concepts and definitions related to closure spaces, their sub-classes of topological spaces and quasi-discrete closure spaces and introduces the notion of
Euclidean and quasi-discrete paths in closure spaces.  \autoref{sec:dsl} briefly recalls  \slcs\ and presents
its extension with the {\em propagation} operator $\ldiff$.  \autoref{sec:CSLCS} introduces the collective
spatial logic \cslcs\ while \autoref{sec:EXAMPLES}  shows some examples of use of the  proposed logics
when interpreted on quasi-discrete closure spaces. In \autoref{sec:model-checking} the model-checking algorithms for
\slcs\  and \cslcs\  interpreted on finite models are presented. In \autoref{sec:discrete-examples} the proof-of-concept
model-checker is shown together with several examples of use. Finally, some conclusions are drawn and lines for future research 
are outlined in \autoref{sec:discussion}. All detailed proofs are provided in the Appendix.

\section{Topological and Closure spaces}

%
%
%

\label{sec:ClosureSpaces}

In this work, we resort to some abstract mathematical structures for the definition of space. The mathematical structure of choice of spatial logics are very often \emph{topological spaces}, possibly enriched with metrics, or other spatial features (see \cite{vBB07}). The use of abstract structures has the advantage to separate logical operators, such as neighbourhood, from the specific nature of space (e.g., the number of dimensions, or the presence or absence of metric features, etc.). However, using topological spaces, it may be difficult to deal with discrete structures, such as finite graphs. In \cite{Gal03}, \emph{closure spaces}, which generalise topological spaces, are proposed as a unifying approach treating
both topological spaces and graphs in a satisfactory way. 
In this section, we recall several definitions and results on 
topological and
closure spaces, most of which are taken from \cite{Gal03}. 


\subsection{Topological spaces}
\label{sec:topological}

We will first provide the basic definitions that are used to relate closure spaces to the more widely known topological spaces. 
The link between topological and closure spaces is deep. In this section we provide a brief introduction to the topic; we refer the reader to, e.g., \cite{Gal03} for more information.
\begin{definition}\label{def:topological-space}
  A \emph{topological space} is a pair $(X,\open)$ of a set $X$ and a collection $\open \subseteq \pow(X)$ of subsets of $X$ called \emph{open sets}, such that $\emptyset, X \in \open$, and subject to closure under arbitrary unions and finite intersections. 
\end{definition}

\begin{definition}\label{def:closed-set}
  In a topological space $(X,\open)$, $A \subseteq X$ is \emph{closed} if its complement is open.
\end{definition}

\begin{definition}\label{def:topological-closure}
 In a topological space $(X,\open)$, the \emph{closure} of $A \subseteq X$ is the \emph{least} closed set containing $A$.
\end{definition}

We remark that closure is well-defined as arbitrary intersections of closed sets are closed, and $X$ itself is both open and closed. An alternative, equivalent formulation of topological spaces is given by the Kuratowski definition.

\begin{definition}\label{def:kuratowski-closure-space}
 According to the Kuratowski definition, a topological space is a pair $(X,\closure)$ where $X$ is a set, and the \emph{closure operator} $\closure : \pow(X) \to \pow(X)$ assigns to each subset of $X$ its \emph{closure}, obeying to the following laws, for all $A,B \subseteq X$:
\begin{enumerate}
 \item \label{def:kuratowski:closure-of_emptyset} $\closure(\emptyset) = \emptyset$;
 \item \label{def:kuratowski:closure-larger} $A \subseteq \closure(A)$;
 \item \label{def:kuratowski:closure-union} $\closure(A \cup B) = \closure(A) \cup \closure(B)$;
 \item \label{def:kuratowski:closure-idempotent}  $\closure(\closure(A))= \closure(A)$.
\end{enumerate}
\end{definition}

%

 The Kuratowski and open sets definitions of a topological space are equivalent. 
The proof can be sketched as follows. To obtain the Kuratowski definition from a topological space defined in terms of open sets, one defines $\closure(A)$ as topological closure (\autoref{def:topological-closure}). The properties of \autoref{def:kuratowski-closure-space} can be shown to hold. For the converse, starting from a Kuratowski topological space $(X,\closure)$, the open sets are defined as those sets $A$ that are equal to their \emph{interior}, that is, 
$A = \overline{\closure{(\overline{A}})}$ where for any $B\subseteq X$ we let $\overline{B}$ denote the \emph{complement} of $B$, i.e. $X\setminus B$.
%
%
%
%

\subsection{Closure spaces}
A \emph{closure space} (also called \emph{\v Cech closure space} or \emph{preclosure space} in the literature), is composed of a set (of points) and a (closure) operator on
subsets (of points), as specified by the following definition:

\begin{definition}\label{def:closure-space}
A \emph{closure space} is a pair $(X,\closure)$ where $X$ is a set, and the \emph{closure operator} $\closure : \pow(X) \to \pow(X)$ assigns to each subset of $X$ its \emph{closure}, obeying to the following laws, for all $A,B \subseteq X$:
\begin{enumerate}
 \item \label{def:closure-space:closure-of_emptyset} $\closure(\emptyset) = \emptyset$;
 \item \label{def:closure-space:closure-larger} $A \subseteq \closure(A)$;
 \item \label{def:closure-space:closure-union} $\closure(A \cup B) = \closure(A) \cup \closure(B)$.
\end{enumerate}
\end{definition}

Closure spaces are a generalisation of \emph{topological spaces}, which is easy to see by comparing \autoref{def:closure-space} with \autoref{def:kuratowski-closure-space}; the difference is that the \emph{idempotency axiom} $\closure(\closure(A)) = \closure(A)$ is not required in closure spaces. Indeed, topological spaces are precisely the subclass of closure spaces where such axiom holds. We shall call a closure space \emph{topological} or \emph{idempotent} or \emph{Kuratowski} in that case. We note in passing that the notion of \emph{continuous function} also extends to closure spaces (see \autoref{def:continuous-function}), making closure spaces a \emph{category} in the sense of category theory, and topological spaces a \emph{full subcategory}.



Below, we consider an example of a closure space, with set of points $X$ in a classical Euclidean space, but exhibiting a non-standard closure operator.

\begin{figure}
\centering
\includegraphics[width=.40\textwidth]{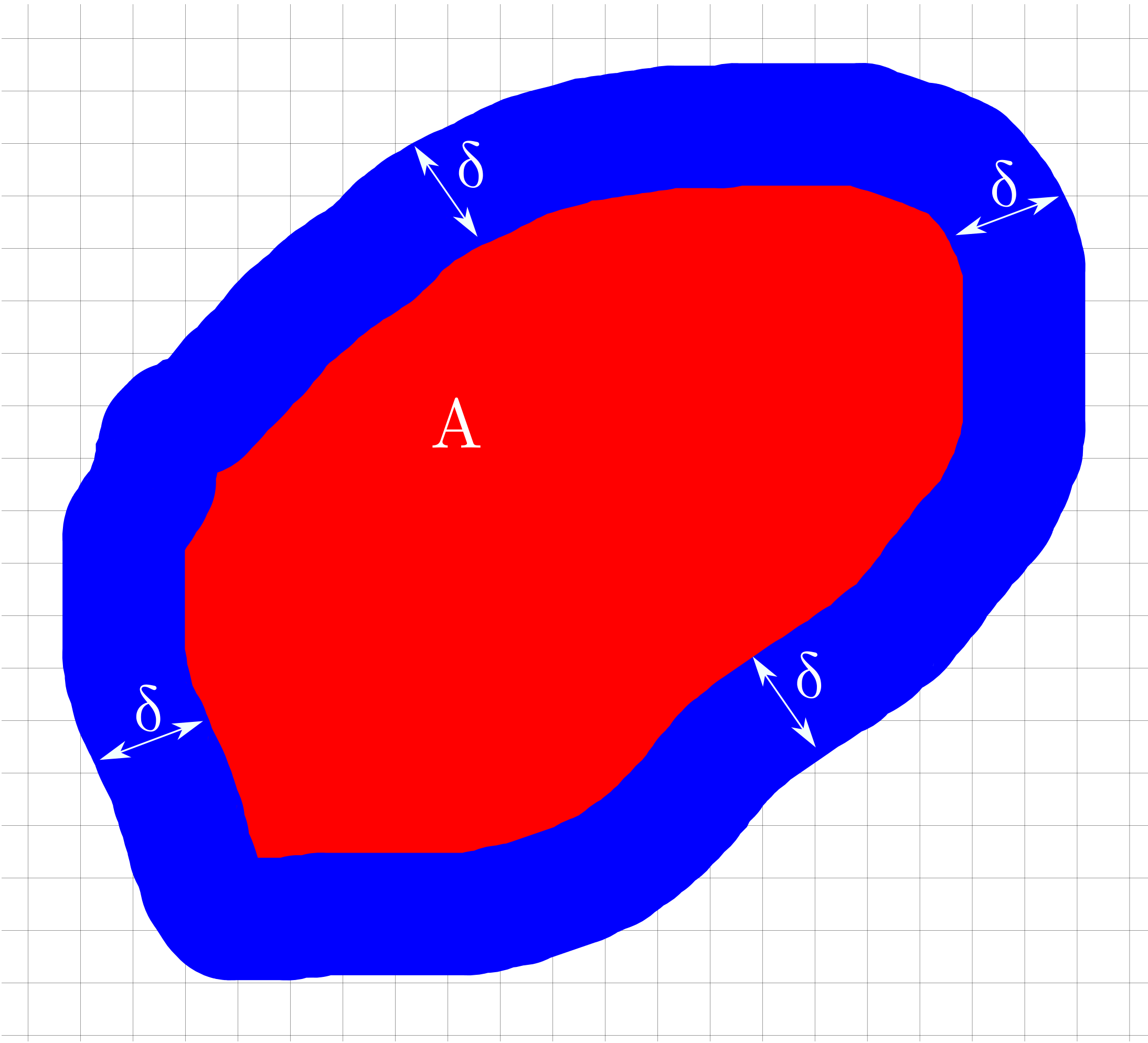}
\caption{\label{fig:potato} A picture of \autoref{ex:r2distance}; the union of the blue and red areas is the closure of the red area}
\end{figure}

\begin{example}\label{ex:r2distance}
Let $\delta\in \reals_{>0}$ and $\closure_{\delta}: \pow(\reals^{2}) \to \pow(\reals^{2})$ be such that:
%
\[
\closure_{\delta}(A)=\{ (x_1,y_1) \in \reals^{2} | \exists (x_2,y_2)\in A. \sqrt{(x_2-x_1)^{2}+(y_2-y_1)^{2}}\leq\delta\}
\]
Function $\closure_{\delta}$ maps each subset $A$ of $\reals^{2}$ to the set of points 
located
in a radius  $\delta$ from a point in $A$ (see \autoref{fig:potato}).
It is easy to see that $\closure_{\delta}$ satisfies all the three conditions of \autoref{def:closure-space} and that
$(\reals^{2},\closure_{\delta})$ is a closure space.
\clsexa
\end{example}
%

\begin{example}
The closure space of \autoref{ex:r2distance} is \emph{not} a \emph{topological} space, as its closure operator is not idempotent.
\clsexa
\end{example}

\begin{definition}\label{def:closure-concepts}
Let $(X,\closure)$ be a closure space; for each $A \subseteq X$:
\begin{enumerate}
\item the \emph{interior} $\interior(A)$ of $A$ is the set $\overline{\closure(\overline A)}$;
\item \label{def:closure-concepts:neighbourhood}$A$ is a \emph{neighbourhood} of $x \in X$ if and only if $x \in \interior(A)$;
\item \label{def:closure-concepts:closed-open}$A$ is \emph{closed} if $A = \closure(A)$ while it is \emph{open} if $A = \interior(A)$.
\end{enumerate} 
\end{definition}


\begin{example}\label{ex:r2distance2}
Let us consider the closure space $(\reals^{2},\closure_{\delta})$, introduced in \autoref{ex:r2distance}, assuming, for simplicity, that $\delta \leq 1$.
Let $A=\{ (x,y)\in\reals^{2} | \sqrt{x^2+y^2}\leq 1\}$. We have that:
\begin{itemize}
\item $\interior(A)=\{ (x,y)\in \reals^{2} | \sqrt{x^2+y^2}\leq 1-\delta \}$; 
\item for any $(x_1,y_1)\in \reals^{2}$, $A$ is a neighbourhood of $(x_1,y_1)$ if and only if:
\[
\{ (x_2,y_2)\in \reals^{2} | \sqrt{(x_2-x_1)^{2}+(y_2-y_1)^{2}}\leq\delta \}\subseteq A
\]
\item the only \emph{closed} set (of the closure operator $\closure_{\delta}$) in $\pow(\reals^{2})$ is $\reals^{2}$, while $\emptyset$ is the only \emph{open} set.
\clsexa
\end{itemize}
\end{example}
 
\noindent The following proposition states a number of general properties of
closure spaces.

\begin{proposition}\label{lem:interior-monotone}
Let $(X,\closure)$ be a closure space, the following properties hold:
\begin{enumerate}
\item \label{lem:interior-monotone-1} $A\subseteq X$ is open if and only if $\overline A$ is closed;
\item \label{lem:interior-monotone-2} closure and interior are monotone operators over the inclusion order, that is:
$A \subseteq B \implies \closure(A) \subseteq \closure(B)\mbox{ and }\interior(A) \subseteq \interior(B)$
\item \label{lem:interior-monotone-3} Finite intersections and arbitrary unions of open sets are open.
 \end{enumerate}
\end{proposition}

Given a closure space $(X,\closure)$, and $A\subseteq X$, we can define the \emph{boundary} of $A$. The latter is only given in terms of closure and interior, and coincides with the definition of boundary in a topological space.
We also provide two similar notions, namely the \emph{interior} and \emph{closure} boundary (the latter is sometimes called \emph{frontier}).

\begin{definition}\label{def:boundary}
 In a closure space $(X,\closure)$, the \emph{boundary} of $A \subseteq X$ is defined as $\boundary(A) = \closure(A) \setminus \interior(A)$. The \emph{interior boundary} is $\iboundary(A) = A \setminus \interior(A)$, and the \emph{closure boundary} is $\cboundary(A) = \closure(A) \setminus A$.
\end{definition}

%

In \cite{Gal99}, a discrete variant of the topological definition of the boundary of a set $A$ is given, for the case where a closure 
operator is derived from a reflexive and symmetric relation (see \autoref{def:closure-operator-of-a-relation} in
the next section). Therein, in Lemma 5, it is proved that the definition of~\cite{Gal99} coincides with the one we provide above. 

\begin{proposition}\label{pro:boundary-properties}
 The following equations hold in a closure space:
 \begin{align}
    \boundary(A) & =  \cboundary(A) \cup \iboundary(A)  \label{eqn:boundary-properties-1}\\
    \cboundary(A) \cap \iboundary(A) & = \emptyset \label{eqn:boundary-properties-2} \\
    \boundary(A) & =  \boundary(\overline A) \label{eqn:boundary-properties-3} \\
    \cboundary(A) & =  \iboundary(\overline A)  \label{eqn:boundary-properties-4} \\
    \cboundary(A) & =  \boundary(A) \cap \overline A \label{eqn:boundary-properties-5} \\
    \iboundary(A) & =  \boundary(A) \cap A \label{eqn:boundary-properties-6} \\
    \boundary(A) & =  \closure(A) \cap \closure(\overline A) \label{eqn:boundary-properties-7}
 \end{align}
\end{proposition}

A closure space can be also obtained by restricting the domain of another space.


\begin{definition}\label{def:subspace}
 Given a closure space $(X,\closure)$ and a subset $Y \subseteq X$, we call \emph{subspace closure} the operation $\closure^{Y}: \pow(Y) \to \pow(Y)$ defined as $\closure^{Y}(A) = \closure(A) \cap Y$. We call $(Y, \closure^{Y})$ the subspace of $(X, \closure)$ \emph{generated by} $Y$.
\end{definition}

%


\begin{proposition}\label{lem:subspace-closure-is-closure}
 The subspace closure is a closure operator.
\end{proposition}

\begin{example}
$(\reals_{\geq 0}^{2},\closure_{\delta}^{\reals_{\geq 0}^{2}})$ 
is a subspace of the closure space $(\reals^{2},\closure_{\delta})$ introduced
in \autoref{ex:r2distance}, generated by $\reals_{\geq 0}^{2}$.
\clsexa
\end{example}

\subsection{Quasi-discrete closure spaces}
\label{sec:quasidiscrete}

A closure space may be derived starting from a \emph{binary relation}, that is, a \emph{graph}.
Such closure spaces may be characterised as {\em quasi-discrete} as
briefly presented in this section. For additional details we refer the interested reader to~\cite{Gal03}.
%

\begin{definition}\label{def:closure-operator-of-a-relation}
Consider a set $X$ and a relation $R \subseteq X \times X$. A closure operator is obtained from $R$ as
$\closure_R(A) = A \cup \{ x \in X \mid \exists a \in A . (a,x) \in R \}$.
\end{definition}

\begin{proposition}\label{pro:closure-space-of-a-relation}
 The pair $(X,\closure_R)$ is a closure space.
\end{proposition}

Closure operators obtained by \autoref{def:closure-operator-of-a-relation} are not necessarily idempotent. Lemma 11 in \cite{Gal03} provides a necessary and sufficient condition, that we rephrase below. We let $R^=$ denote the reflexive closure of $R$, that is, the smallest reflexive relation containing $R$, which is defined as the union of $R$ with the identity relation on the same domain.

\begin{lemma}\label{lem:idempotency-and-transitivity}
 $\closure_R$ is idempotent if and only if $R^=$ is transitive.
\end{lemma}

Note that when $R$ is transitive, so is $R^=$, thus $\closure_R$ is idempotent. The vice-versa is not true. For instance, it may happen that $(x,y) \in R$, and $(y,x) \in R$, but $(x,x) \notin R$. 

%
%

\begin{remark}\label{rem:open-sets-coarse}
In topology, open sets play a fundamental role. However, the situation is different in closure spaces derived from a relation $R$. For example, in a closure space derived from a symmetric relation, whose graph is connected, the only open sets are the whole space, and the empty set.
\end{remark}

\begin{proposition}\label{pro:interior-boundary-in-quasi-discrete}
 Given $R \subseteq X \times X$, in the space $(X,\closure_R)$, we have:
 \begin{align}
 \interior(A) & = \{ x \in A \mid \lnot \exists a \in \overline A . (a,x) \in R \} \label{eqn:boundary-quasi-discrete-1} \\
 \iboundary(A) & = \{ x \in A \mid \exists a \in \overline A . (a,x) \in R\} \label{eqn:boundary-quasi-discrete-2} \\
 \cboundary(A) & = \{ x \in \overline A \mid \exists a \in A . (a,x) \in R \} \label{eqn:boundary-quasi-discrete-3}
 \end{align}
\end{proposition}

Closure spaces derived from a relation can be characterised as \emph{quasi-discrete} spaces (see also Lemma 9 of \cite{Gal03} and the subsequent statements).

\begin{definition}\label{def:quasi-discrete-closure-space}
 A closure space is \emph{quasi-discrete} if and only if one of the following equivalent conditions holds:  
 \begin{enumerate}
  \item[i)] each $x \in X$ has a \emph{minimal neighbourhood}\footnote{A \emph{minimal neighbourhood} of $x$ is a set that is a neighbourhood of $x$ (\autoref{def:closure-concepts}~(\ref{def:closure-concepts:neighbourhood})) and is included in all other neighbourhoods of $x$.} $\mneigh_x$; 
  
  \item[ii)] for each $A \subseteq X$, $\closure(A) = \bigcup_{a \in A} \closure(\{a\})$.
 \end{enumerate}

\end{definition}

\noindent The following is proved as Theorem 1 in \cite{Gal03}.

\begin{theorem}
 A closure space $(X,\closure)$ is quasi-discrete if and only if there is a relation $R \subseteq X \times X$ such that $\closure = \closure_R$.
\end{theorem}

Summing up, whenever one starts from an arbitrary relation $R \subseteq X \times X$, the obtained closure space $(X,\closure_R)$ enjoys minimal neighbourhoods, and the closure of a set $A$ is the union of the closure of the singletons composing $A$. Furthermore, such nice properties are only true in a closure space when there is some $R$ such that the closure operator of the space is derived from $R$. In the remainder of this section, we exemplify some aspects of quasi-discreteness.

\begin{figure}[tbp]
\begin{center}
\begin{tikzpicture}{}
\foreach \i /\j / \fc / \bc in { 
	1/1/yellow!90/yellow!50, 
	2/1/yellow!90/yellow!50, 
	3/1/red!90/red!50,
	4/1/black/white,
	5/1/black/white,
	6/1/black/white,
	7/1/black/white,
	8/1/black/white,
	9/1/black/white,
	1/2/yellow!90/yellow!50, 
	2/2/yellow!90/yellow!50, 
	3/2/red!90/red!50,
	4/2/black/white,	
	5/2/black/white,
	6/2/blue!90/blue!50,
	7/2/blue!90/blue!50,
	8/2/black/white,
	9/2/black/white,
	1/3/red!90/red!50, 
	2/3/red!90/red!50, 
	3/3/black/white,
	4/3/black/white,
	5/3/blue!90/blue!50,
	6/3/green!90/green!50,
	7/3/green!90/green!50,
	8/3/blue!90/blue!50,
	9/3/black/white,
	1/4/black/white, 
	2/4/black/white, 
	3/4/black/white,
	4/4/black/white,
	5/4/blue!90/blue!50,
	6/4/green!90/green!50,
	7/4/green!90/green!50,
	8/4/blue!90/blue!50,
	9/4/black/white,
	1/5/black/white, 
	2/5/black/white, 
	3/5/black/white,
	4/5/black/white,
	5/5/black/white,
	6/5/blue!90/blue!50,
	7/5/blue!90/blue!50,
	8/5/black/white,
	9/5/black/white
	} {
		\node [circle,draw=\fc,fill=\bc,thick,inner sep=0pt,minimum size=4mm] at (\i*0.8,\j*0.8) (node\i\j) {};
}

\foreach \i [evaluate = \j as \ipp using int(\i+1)] in {1,2,3,4,5,6,7,8,9} {
	\foreach \j [evaluate = \j as \jpp using int(\j+1)] in {1,2,3,4,5} {
		\ifnum\j<5
		\draw[<->] (node\i\j) -- (node\i\jpp);
		\fi
		\ifnum\i<9
		\draw[<->] (node\i\j) -- (node\ipp\j);
		\fi
	}
}
\end{tikzpicture}
\end{center}
\caption{\label{fig:qdspace1}A  graph inducing a \emph{quasi-discrete} closure space}
\end{figure}
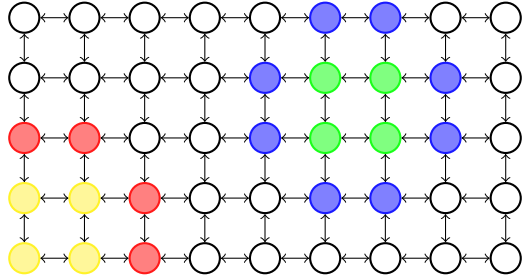
\begin{example}{}
\label{ex:qdspace1}
Every graph induces a \emph{quasi-discrete} closure space. For instance, consider the 
(undirected) graph depicted in \autoref{fig:qdspace1}. Let $R$ be the (symmetric) binary relation 
induced by the graph edges, and let $Y$ and $G$ denote the set 
of \emph{yellow} and \emph{green} nodes, respectively.
The closure $\closure_{R}(Y)$ consists of all \emph{yellow} nodes and \emph{red} nodes, while the closure
$\closure_{R}(G)$ contains all \emph{green} nodes and \emph{blue} nodes.
The interior $\interior(Y)$ of  $Y$ contains a single node, the one located at the 
bottom-left in \autoref{fig:qdspace1}. The \emph{interior} $\interior(G)$ of $G$ is empty.
Indeed, we have that $\boundary(G)=\closure_R(G)$, while $\iboundary(G)=G$ and $\cboundary(G)$ consists
of the \emph{blue} nodes.
\clsexa
\end{example}

\begin{example}\label{ex:r2distance-as-relation}
The closure space of \autoref{ex:r2distance} is a quasi discrete closure space. Indeed, define $R_{\delta}\subseteq \reals^{2}\times\reals^{2}$
as:
\[
R_{\delta}=\{ ( (x_1,y_1) , (x_2,y_2) ) |  \sqrt{(x_2-x_1)^{2}+(y_2-y_1)^{2}}\leq\delta \}
\]
It is easy to prove that $\closure_{\delta}=\closure_{R_{\delta}}$. Note that
$R_{\delta}$ is reflexive but not transitive. So, the closure space is not a topological
space.
\clsexa
\end{example}
%
%

Existence of minimal neighbourhoods does not depend on finiteness of the space; moreover, it is not even required that each point has a finite neighbourhood, as illustrated by the 
following example:

\begin{example}\label{exa:quasi-discrete-rational}
Consider the rational numbers $\rats$, with the relation $\leq$. Such a relation is reflexive and
transitive, thus the closure space $(\rats,\closure_\leq)$ is \emph{topological} and \emph{quasi-discrete} (but not finite). For any $x\in \rats$, we have $\mneigh_x=\{ y\in\rats | y\leq x\}$,
which is not finite.
\clsexa
\end{example}

%

\begin{example}\label{exa:quasi-discrete-not-topological}
Another example of closure space exhibiting minimal neighbourhoods in absence of finite neighbourhoods is the one considered in \autoref{ex:r2distance}. In \autoref{ex:r2distance2} we show that for any $(x_1,y_1)\in \reals^{2}$, $A$ is a neighbourhood of $(x_1,y_1)$ if and only if 
\[
\{ (x_2,y_2)\in \reals^{2} | \sqrt{(x_2-x_1)^{2}+(y_2-y_1)^{2}}\leq\delta \}\subseteq A.
\]
Hence, $\mneigh_{(x_1,y_1)}=\{ (x_2,y_2)\in \reals^{2} | \sqrt{(x_2-x_1)^{2}+(y_2-y_1)^{2}}\leq\delta \}$.
\clsexa
\end{example}

\begin{example}\label{exa:topological-not-quasi-discrete}
 An example of a topological closure space which is not quasi-discrete is the set of real numbers equipped with the Euclidean topology (the topology induced by arbitrary union and finite intersection of open intervals). To see that the space is not quasi-discrete, one applies \autoref{def:quasi-discrete-closure-space}. Consider an open interval $(x,y)$. We have $\closure((x,y)) = [x,y]$, but for each point $z$, we also have $\closure(z) = [z,z] = \{z\}$. Therefore $\bigcup_{ z \in (x,y) } \closure({z}) = \bigcup_{ z \in (x,y) } \{z\} = (x,y) \neq [x,y]$.
\clsexa
\end{example}

We note in passing that {\em any} finite space is trivially a quasi-discrete closure space.
Quasi discrete closure spaces can be used to model spatial structures in $\reals^{n}$, as shown below.

%

\begin{example}
 Let $\mathscr{F}\subseteq \pow(\reals^{n})$ be a partition
of $\reals^{n}$, each element of which is either \emph{open} or \emph{closed}, i.e.
$\cup_{A\in \mathscr{F}} A = \reals^{n}$, 
$\forall A, B\in \mathscr{F}: A\not=B\rightarrow A\cap B=\emptyset$,
$\forall A\in \mathscr{F}: A =\interior(A) \vee A =\closure(A)$.
We let $R^{\mathscr{F}}\subseteq \mathscr{F}\times\mathscr{F}$ be the \emph{connectedness} relation among elements of $\mathscr{F}$, 
formally:
%
\[
R^\mathscr{F}=\{ (A,B) | A, B \mbox{ open and }\closure(A) \cap \closure(B)  \neq \emptyset \}
\]
where $\closure$ is the standard topological closure over $\reals^{n}$. It is easy to see that $(\mathscr{F},\closure_{R^\mathscr{F}})$ is a
quasi discrete closure space.  \autoref{fig:RealQDCS} shows an example
in $\reals^{2}$, where the open sets are shown in pink, while the only
closed set is shown in black.
\begin{figure}
\centering{\includegraphics[width=.3\textwidth]{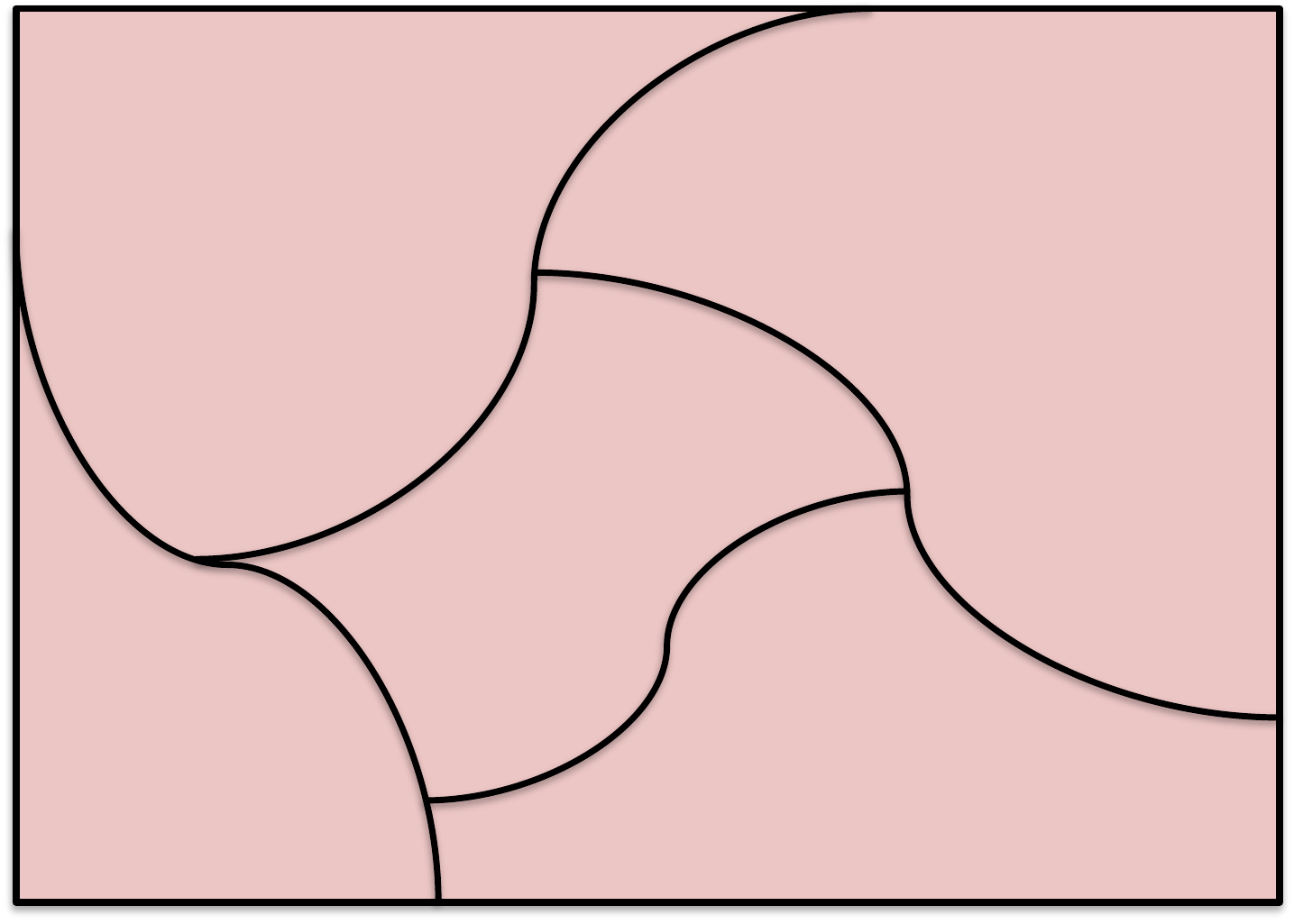}}
\caption{\label{fig:RealQDCS} A quasi-discrete closure space inducing a spatial structure.}
\end{figure}
\clsexa
\end{example}

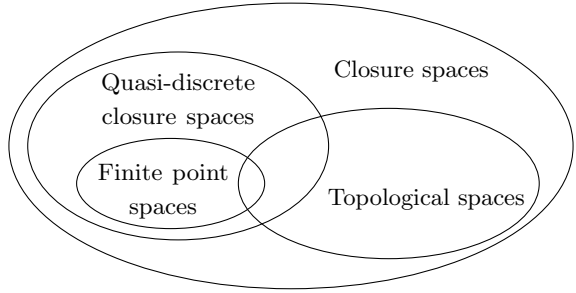
\begin{figure}
\begin{tikzpicture}
\draw (0,0) ellipse (3.75cm and 1.9cm);	
\draw (-1.5,0) ellipse (2cm and 1.25cm); 
\draw (1.3,-0.5) ellipse (2cm and 1cm); 
\draw (-1.6,-0.5) ellipse (1.25cm and 0.6cm); 
\node at (1.6,1) {\footnotesize Closure spaces};
\node at (1.8,-0.7) {\footnotesize Topological spaces};
\node [text width=2.25cm,align=center] at (-1.5,0.6) {\footnotesize Quasi-discrete closure spaces};
\node [text width=1.75cm,align=center] at (-1.7,-0.6) {\footnotesize Finite point spaces};
\end{tikzpicture}
\caption{\label{fig:SpaceHierarchy} The hierarchy of closure spaces.}
\end{figure}

In \autoref{fig:SpaceHierarchy}, the hierarchy of closure spaces with respect to quasi-discreteness is shown. All finite spaces are quasi-discrete, as closure of arbitrary sets is determined by that of the singletons, by the axiom $\closure(A) \cup \closure(B) = \closure(A \cup B)$. Obviously there are quasi-discrete infinite spaces (any infinite graph interpreted as a closure space is an example). A quasi-discrete space which is also topological is the space associated to any complete graph. In this case, for any set, $\closure(A)$ is the whole space, thus closure is idempotent. More precisely, the topology determined by the closure operator associated to a complete graph is the \emph{indiscrete} topology, where the only open sets are the empty set and the whole space. It is obvious that there are topological spaces that are not quasi-discrete, such as Euclidean spaces. Finally there are closure spaces that are neither topological nor quasi discrete. The most obvious example is the \emph{coproduct} (disjoint union) of a topological space 
which is not quasi-discrete (e.g. any Euclidean space), and a quasi-discrete, but not topological, closure space. The disjoint union of two closure spaces is defined below (we omit the proof that it actually obeys to the axioms of a closure space, as it is an easy exercise).

\begin{definition}\label{def:coproduct}
 Given two closure spaces $(X, \closure^X)$ and $(Y,\closure^Y)$, consider the disjoint union of $X$ and $Y$, represented as $X \uplus Y = X' \cup Y'$ with $X' = \{(1,x) \mid x \in X\}$ and $Y' = \{(2,y) \mid y \in Y\}$. In order to equip the set $X \uplus Y$ with a closure operator, for each $A \subseteq X \uplus Y$, let $A^X = \{ x \mid (1,x) \in A \}$ and $A^Y = \{ y \mid (2,y) \in A\}$.  Define $\closure(A) = \{(1,x) \mid x \in \closure^X(A^X)\} \cup \{(2,y) \mid y \in \closure^Y(A^Y)\}$.
\end{definition}

\subsection{Paths and connectedness in closure spaces}\label{sec:paths-new}

In this section we define paths and connectedness for interesting classes of closure spaces. A uniform definition of paths in closure spaces is non-trivial. It is possible, and often done, to borrow the notion of path from topology. However, as we shall see, the extension is not fully satisfactory. For example, the topological definition does not yield graph-theoretical paths in the case of quasi-discrete closure spaces. Our solution is pragmatic. We define paths as it is natural in interesting classes of closure spaces. We leave open the possibility to change this notion, in chosen classes of closure spaces, practically making our theory dependent on such choice. The theoretical question of finding a truly uniform notion of path (e.g., by some form of category-theoretical \emph{universal property} characterising a path-connected class of spaces) is left for future work. First of all we introduce the definition of continuous function, which restricts to topological continuity in the setting of idempotent closure spaces\footnote{Note that in topological spaces one may equivalently use the definition we propose here, based on the Kuratowski axioms, or the definition of continuity using open sets, namely $f$ is continuous whenever for each open set $o$, $f^{-1}(o)$ is open. However, the two definitions do not coincide for arbitrary closure spaces (open sets play a less important role in closure spaces, see \autoref{rem:open-sets-coarse}). }.

\begin{definition}\label{def:continuous-function}
 A \emph{continuous function} $f : (X_1,\closure_1) \to (X_2,\closure_2)$ is a function $f : X_1 \to X_2$ such that, for all $A \subseteq X_1$, we have $f(\closure_1(A)) \subseteq \closure_2(f(A))$.
\end{definition}

\noindent Below, two kinds of paths are introduced: \emph{Euclidean} paths and \emph{quasi-discrete} paths.
 
\begin{definition}\label{def:path}
For each closure space $(X,\closure)$, assume a chosen closure space $\mathfrak{I}$, equipped with a linear order $\leq$ with bottom $0$, and call path a continuous function $p : \mathfrak{I} \to (X,\closure)$.
In particular, call \emph{Euclidean path} any continuous function whose domain is the half-line $\reals_{\geq 0} = \{x \in \reals \mid 0 \leq x \}$, equipped with the Euclidean (topological) closure operator. Call \emph{quasi-discrete path} any continuous function whose domain is the quasi-discrete closure space $(\nats,\closure_\succ)$ where $(n,m) \in \succ \iff m = n+1$. 
Whenever $(X,\closure)$ is an Euclidean topological space (resp. a quasi-discrete closure space), call \emph{path} an Euclidean (resp. quasi-discrete) path whose codomain is $(X,\closure)$.
\end{definition}

Note that in \autoref{def:path} we do not require compatibility conditions between the closure operator and the linear order of $\mathfrak{I}$. Depending on the application context, different orders may be chosen, obtaining different interpretations of logics, or different degrees of compatibility between closure and paths (see e.g. \autoref{thm:other-side-of-simple-discrete-paths-until-S}).
We consider the study of appropriate compatibility conditions, determining a universal notion of path for certain classes of closure spaces, out of scope for the current paper. We can, though, provide a hint about the complexity of such study. One of the major difficulties in finding a unifying notion is that Euclidean paths are not directed, whereas quasi-discrete paths are directed. The examples in this section are also aimed at making this problem more clear. Directed paths in topology are a highly non-trivial topic by themselves, and gave rise to the subject of \emph{directed algebraic topology}  \cite{Gra09}. Generalizing directed algebraic topology to work in the setting of closure spaces could be a relevant strategy to face these issues.
%

As a matter of notation, we call $p$ a path \emph{from} $x$, and write $p : x \pto{}{}$, when $p(0) = x$. We write $y \in p$ whenever there is $i$ such that $p(i) = y$. We also write $p : x \pto{i}{y}$ when $p$ is a path from $x$ and $p(i)=y$. 

The definition of Euclidean path is intuitively similar to the classical topological definition of a path, namely a continuous function from the unit interval $[0,1]$, except that Euclidean paths that we defined are ``open-ended on the right'' (note that the open interval $[0,1)$ and $\reals^+$ are continuously isomorphic). The definition of quasi-discrete path, on the other hand, mimics the classical definition of infinite path in a graph. Simply adopting Euclidean paths in quasi-discrete spaces yields counter-intuitive results, as shown below.

\begin{example}\label{exa:continuous-path-and-directed-graphs}
 Consider the quasi-discrete closure space obtained from the graph $G=(\{a,b\},\{(b,a)\})$ having two nodes $a$, $b$, and only one edge, from $b$ to $a$. Note that there is no graph-theoretical path from $a$ to $b$. However, consider the function $p : \reals_{\geq 0} \to \{a,b\}$, defined by $p(0) = a$, and $p(i) = b$ for $i \neq 0$. This function is continuous, thus it is an Euclidean path starting from $a$ and traversing $b$. To see this, choose any subset $J$ of the half-line.
\begin{itemize}
  \item If $J = \emptyset$, the thesis is trivially obtained; otherwise, assuming $J \neq \emptyset$:

  \item  if $J = \{0\}$, then $p(\closure(J))= p(\{0\}) = p(J) \subseteq \closure(p(J))$; otherwise, assuming $J \neq \emptyset$ and $J \neq \{0\}$, necessarily $b \in p(J)$, and:

  \item if $0 \notin J$ and $0 \notin \closure(J)$, then $p(\closure(J)) = p(J) = \{b\} \subseteq \closure(p(J))$;

  \item if $0 \notin J$ and $0 \in \closure(J)$, then $p(\closure(J))=\{a,b\} = \closure(\{b\})  = \closure(p(J))$;
  
  \item if $0 \in J$, then $p(\closure(J)) \subseteq \{a,b\} = \closure(p(J))$.
\end{itemize}\clsexa
\end{example}


We saw that Euclidean paths may not yield the expected results in quasi-discrete closure spaces. On the other hand, graph-theoretical and quasi-discrete paths coincide.

\begin{lemma}\label{lem:paths-are-paths}
 Given a (quasi-discrete) path $p$ in a quasi-discrete space $(X, \closure_R)$, for all $i \in \nats$ with $p(i) \neq p(i+1)$, we have $(p(i),p(i+1)) \in R$, i.e., the image of $p$ is a (graph theoretical, countably infinite) path in the graph of $R$. Conversely, each countable path in the graph of $R$ uniquely determines a quasi-discrete path.
\end{lemma}

Note that, in particular, in \autoref{exa:continuous-path-and-directed-graphs} there is no \emph{quasi-discrete} path rooted in $a$ and passing by $b$, whereas there are quasi-discrete paths rooted in $b$ and passing by $a$ (for example, the path defined by $p(0)=b$ and $p(i>0) = a$).
Let us introduce the notion of connectedness that we use in this work.

\begin{definition}\label{def:connected-path}
 Given a closure space $(X,\closure)$, set $A \subseteq X$ is \emph{path-connected} if and only if for each $x,y \in A$ there is a path $p$ and an index $i$ such that $p(0) = x$, $p(i) = y$ and, for all $j \leq i$,   $p(j) \in A$. 
\end{definition}

Note that, for quasi-discrete closure spaces, by \autoref{lem:paths-are-paths}, \autoref{def:connected-path} coincides with the usual notion of \emph{strong connectedness} in graph theory.


\begin{remark}
It is worth mentioning that connectedness can be also borrowed from topology, resorting to the notion of \emph{separation}. Formally, let $(X,\closure)$ be a closure space. Two sets $A_1,A_2 \subseteq X$ are \emph{separated} if and only if $\closure(A_1) \cap A_2 = \emptyset = A_1 \cap \closure(A_2)$. Note that separated sets are also disjoint, since for all sets $A$, we have $A \subseteq \closure(A)$. Thus, there is no explicit requirement that $A_1$ and $A_2$ are disjoint. Set $A \subseteq X$ is \emph{connected} if and only if there are no non-empty, separated sets $A_1, A_2 \subseteq X$ such that $A = A_1 \cup A_2$.
In the case of topological spaces, the difference between this definition and path connectedness is widely known. There is a difference also in quasi-discrete closure spaces. A quasi-discrete closure space which is connected, but not path-connected is the space $(\{1,2,3\}, \closure_R)$, where $R = \{(1,2),(3,2)\}$. By \autoref{lem:paths-are-paths} there is no path from $1$ to $3$; however, it is not possible to find two non-empty, separated sets $A_1,A_2$ with $X = A_1 \cup A_2$. The only possible choices, recalling that separated sets must be disjoint, are $A_1 = \{1,2\}, A_2 = \{3\}$, with $\closure(A_2) \cap A_1 = \{2\}$, $A_1 = \{1\}, A_2 = \{2,3\}$, with $\closure(A_1) \cap A_2 = \{2\}$, and $A_1 = \{1,3\}, A_2 = \{2\}$ with $\closure(A_1) \cap A_2 = \{2\}$.
\end{remark}

%
%
%
%

%
%

\section{Spatial logics for closure spaces}
\label{sec:dsl}

In this section we present \slcs: a \emph{Spatial Logic for Closure Spaces}, that we first proposed in \cite{Ci+14b}. The logic is meant to assign to formulas a \emph{local} meaning; for each point, formulas may predicate both on the possibility of reaching other points satisfying specific properties, or of being reached from them, along paths of the space. 
In \cite{Ci+14b}, \slcs\ is equipped with two \emph{spatial operators}:
a ``one step'' modality, called ``near'' and denoted by $\lnear$, turning the closure operator $\closure$ into a logical operator,
and a binary \emph{spatial until} 
operator $\mathcal{U}$, which is a spatial counterpart of the temporal \emph{until} operator.
In the present paper we extend \slcs\  with an additional binary operator, 
$\ldiff$, used to model  \emph{propagation}, and propose a new interpretation for $\mathcal{U}$, based on the notion of paths that we introduced in \autoref{sec:paths-new}. In order to avoid confusion, we call the newly defined connective \emph{surrounded}, and use the symbol $\lsurr$. Operator $\lsurr$ coincides with $\mathcal{U}$ in the case of quasi-discrete closure spaces, and enhances it by also providing an intuitively meaningful interpretation in the case of continuous (e.g. Euclidean) spaces. 
The proposed spatial logic combines these new operators with 
standard boolean operators. Assume a finite or countable set $\props$ of \emph{atomic propositions}.

\begin{definition}\label{def:logic-syntax}
 The syntax of \slcs\ is defined by the grammar in \autoref{fig:slcs-syntax}, where $a$ ranges over 
$\props$.
\end{definition}

\begin{figure}
 $$\begin{array}{l c l l}
\Phi & ::=  & a & \mbox{\sc{[Atomic proposition]}} \\
     & \mid & \top & \mbox{\sc{[True]}} \\	
     & \mid & \lnot \Phi & \mbox{\sc{[Not]}}\\
     & \mid & \Phi \land \Phi & \mbox{\sc{[And]}} \\
     & \mid & \lnear \Phi & \mbox{\sc{[Near]}} \\
     & \mid & \Phi \lsurr \Phi & \mbox{\sc{[Surrounded]}} \\
     & \mid & \Phi \ldiff \Phi & \mbox{\sc{[Propagation]}} \\
\end{array}$$
%
\caption{\label{fig:slcs-syntax}\slcs\ syntax}
\end{figure}


In \autoref{fig:slcs-syntax}, $\top$ denotes the truth value \emph{true}, $\lnot$ is negation, $\land$ is conjunction, 
$\lnear$ is the \emph{closure} operator,
$\lsurr$ is the \emph{surrounded} operator, and
$\ldiff$ is the \emph{propagation} operator. 
From now on, with a small overload of notation, we let $\Phi$ denote the set of \slcs\ formulas. We shall now define the interpretation of formulas. 


\begin{definition}\label{def:model}
 A \emph{closure model} is a pair $\model = ((X,\closure),\eval)$ consisting of a closure space $(X,\closure)$ and a valuation $\eval : \props \to 2^X$, assigning to each atomic proposition the set of points where it holds. 
\end{definition}

\begin{definition}\label{def:closure-semantics}
 Satisfaction $\model,x \models \phi$ of formula $\phi \in \Phi$ at point $x \in X$ in model $\model = ((X,\closure),\eval)$ is defined by induction on the structure of terms, by the equations in \autoref{fig:closure-semantics}.
\end{definition}
%

\begin{figure}[htbp]
 \[\begin{array}{rclcl}
   \model,x & \models & a \in AP & \iff & x \in \eval(a) \\
   \model,x & \models & \top & \iff & \mathit{true} \\
   \model,x & \models & \lnot \phi & \iff & \model,x \not \models \phi \\
   \model,x & \models & \phi_1 \land \phi_2 & \iff & \model,x \models \phi_1 \text{ and } \model, x \models \phi_2 \\
   \model,x & \models & \lnear \phi & \iff & x \in \closure(\{ y \in X | \model, y \models \phi \}) \\
   \model,x & \models & \phi_1 \lsurr \phi_2 & \iff & \model, x \models \phi_1 \land \forall p : x \pto {}{} . \forall l .  \model, p(l) \models \lnot \phi_1 \\ & & & & \qquad  \implies \exists k . 0 < k \leq l . \model,p(k) \models \phi_2 \\
   \model,x & \models & \phi_1\ldiff \phi_2 & \iff & \model,x \models \phi_2 \land \exists y . \exists p : y \pto {l}{x} . \model, y \models \phi_1 \land \\ & & & & \qquad \forall i . 0 < i < l \implies \model, p(i) \models \phi_2
 \end{array}\]
\caption{\label{fig:closure-semantics}\slcs\ semantics}
\end{figure}


Atomic propositions and boolean connectives have the expected meaning. For formulas of the form $\phi_1 \lsurr \phi_2$, the basic idea is that point $x$ satisfies $\phi_1 \lsurr \phi_2$ whenever 
there is ``no way out'' from 
$\phi_1$
unless passing by a point that satisfies $\phi_2$. 
For instance, if we consider the model of \autoref{fig:qdspace1}, \emph{yellow} nodes should satisfy $yellow \lsurr red$
while \emph{green} nodes should satisfy $green \lsurr blue$. 
%
%
A point $x$ satisfies $\phi_1\ldiff \phi_2$ if it satisfies $\phi_2$ and it is reachable from a point satisfying $\phi_1$ via a path such that all of its points, except possibly the starting point, satisfy $\phi_2$.
For instance, if we consider again the model of \autoref{fig:qdspace1}, \emph{blue}, \emph{green} and \emph{white} nodes satisfy $green \ldiff \neg red$ while the same formula is not satisfied by \emph{yellow} nodes. 

In \autoref{fig:derivableoperators}, we present some derived operators. Besides standard logical connectives, the logic can express the 
\emph{interior} ($\linterior\phi$), the \emph{boundary} ($\lboundary \phi$),
the \emph{interior boundary} ($\liboundary \phi$) and the \emph{closure boundary} ($\lcboundary \phi$) 
of the set of points satisfying formula $\phi$. Moreover, by appropriately using the \emph{surrounded} operator, operators concerning
\emph{reachability} ($\phi_1 \ldualuntil \phi_2$),  \emph{global satisfaction} ($ \leverywhere \phi$,
{\em everywhere} $\phi$) and \emph{possible satisfaction} 
($\lsomewhere \phi$, {\em somewhere} $\phi$) can be derived. Finally we define the $\ldualdiff$ connective, expressing that $\phi_2$ keeps $x$ ``apart'' from $\phi_1$. More explanation is provided below.


\begin{proposition}
\label{remark:duals}
We have that:
\begin{enumerate}
\item $\model,x \models \phi_1 \ldualuntil \phi_2$  if and only if there is $p : x \pto {}{}$ and $k$ such that $\model, p(k) \models \phi_2$ and for each $j$ with $0 < j \leq k$, we have $\model, p(j) \models \phi_1$;
\item  $\model,x \models \phi_1 \ldualdiff \phi_2$, if and only if $\model, x \models \phi_2$ or for any $y$ such that 
$\model, y \models \phi_1$, and for any 
$p : y \pto {l}{x}$, there exists $i$ such that $0 < i < l$ and $\model, p(i) \models \phi_2$.

\item $\model,x \models \leverywhere \phi_1$  if and only if for each $p : x \pto {}{}$ and $i\in \nats$, $\model, p(i) \models \phi_1$; 
\item $\model,x \models \lsomewhere \phi_1$  if and only if there is $p : x \pto {}{}$ and $i\in \nats$ such that $\model, p(i) \models \phi_1$.
\end{enumerate} 
\end{proposition}

\newcommand{\calT}{\,\mathcal{T}\,}

Note that point $x$ 
satisfies $\phi_1 \ldualuntil \phi_2$ if and only if either $\phi_2$ is satisfied by $x$ or there exists a sequence of
points after $x$, all satisfying $\phi_1$, leading to a point satisfying both $\phi_2$ and $\phi_1$. In the second case,
it is not required that $x$ itself satisfies $\phi_1$. 
For instance, both \emph{red} and \emph{green} nodes in \autoref{fig:qdspace1}
satisfy $(\mathit{white}\vee \mathit{blue})\ldualuntil \mathit{blue}$, as well as the white and blue nodes.
The formula is not satisfied by the yellow nodes. This is so because the
first node of a path leading to a blue node is not required to satisfy
white or blue. It is easy to strengthen the notion of reachability when
we want to identify all white nodes from which a blue node can be
reached by requiring in addition that the first node of the path has to
be white. We can define this notion as a derived operator as follows:
$$\phi_1 \calT \phi_2 \triangleq \phi_1 \land ((\phi_1 \lor \phi_2) \ldualuntil \phi_2)$$ 
Note
also that $\phi_2$ is occurring also in the first argument of $\ldualuntil$. This is
because satisfaction of $\phi_1\ldualuntil \phi_2$ requires that the final node on the path satisfies
both 
$\phi_1$ and $\phi_2$.

A point $x$ satisfies  $\model,x \models \phi_1 \ldualdiff \phi_2$ if 
it satisfies $\phi_2$ or every path from a point $y$ satisfying $\phi_1$ to $x$
passes by a point satisfying $\phi_2$, located
between $y$ and $x$.
For instance, with reference to \autoref{fig:qdspace1}, let us consider
$\mathit{yellow} \ldualdiff \mathit{red}$, that is
$ \neg(\mathit{yellow} \ldiff \neg \mathit{red})$. Note that
$\mathit{yellow} \ldiff \neg \mathit{red}$ is satisfied by the yellow points in the figure: for each yellow point $x$, let $y$ be any yellow point (even $x$ itself) and $p$ a path starting from $y$ and passing by $x$ staying in the yellow area. Furthermore, points that are not yellow do not satisfy $yellow \ldiff \neg \mathit{red}$ by definition of $\ldiff$.
Therefore, $\mathit{yellow} \ldualdiff \mathit{red}$ is satisfied by all other points in the figure,
including the red ones. Furthermore, all white nodes in the figure
satisfy both $\mathit{yellow} \ldualdiff \mathit{red}$ and
$\mathit{green} \ldualdiff \mathit{blue}$.
%
%

It is worth noting that in some situations, operators dealing with paths in opposite directions may be inter-expressible. However, an appropriate formalisation of such kinds of axioms, and the study of the associated classes of closure models, is left for future work.




\begin{figure}[tbp]
 \[\begin{array}{lclclcl}
    \bot & \triangleq & \lnot \top & \qquad \qquad &
    \phi_1 \lor \phi_2 & \triangleq & \lnot (\lnot \phi_1 \land \lnot \phi_2) 
    \\
    \linterior \phi & \triangleq & \lnot (\lnear \lnot \phi) & 
    &
    \lboundary \phi & \triangleq & (\lnear \phi) \land (\lnot \linterior \phi) 
    \\
    \liboundary \phi & \triangleq & \phi \land (\lnot \linterior \phi) & 
    &
    \lcboundary \phi & \triangleq & (\lnear \phi) \land (\lnot \phi) 
    \\
    \phi_1 \ldualuntil \phi_2 & \triangleq & \neg ( (\neg \phi_2) \lsurr (\neg \phi_1) ) & 
    & \leverywhere \phi & \triangleq & \phi \lsurr \bot 
    \\  
    \lsomewhere \phi & \triangleq & \neg \leverywhere (\neg \phi) & & 
        \phi_1 \ldualdiff \phi_2 & \triangleq & \neg ( \phi_1 \ldiff (\neg \phi_2) ) 
 \end{array}\]
\caption{\label{fig:derivableoperators}Some \slcs\ derived operators}
\end{figure}


%
We conclude this section by restricting our attention to \emph{quasi discrete closure models}, i.e. \emph{closure models} that are originated from \emph{quasi discrete closure spaces}, in order to compare \autoref{def:closure-semantics} with the interpretation of $\lsurr$ studied in \cite{Ci+14b}.


\begin{definition}\label{def:qdcmodel}
 A \emph{quasi discrete closure model} is a pair $\model = ((X,\closure),\eval)$ consisting of a quasi discrete closure space $(X,\closure)$ and a valuation $\eval : \props \to 2^X$, assigning to each atomic proposition the set of points where it holds.
\end{definition} 


\begin{example}
\label{ex:imagesascs}
For $k,h \in \nats$, let $\nats^2_{k,h}$ be the set $\{ (i,j) \in \nats \times \nats \mid i \in [1,k] \land j \in [1,h] \}$. A \emph{digital image} of size $k \times h$, on finite set of \emph{colours} $C$, is a function $f : \nats^2_{k,h} \to C$, assigning a colour to each point of a finite rectangle in $\nats^2$. Such an image gives rise to the quasi-discrete closure space $(\nats_{k,h},\closure_{4adj})$, where $$ ((x_1,y_1),(x_2,y_2)) \in 4adj \iff (x_1 - x_2)^2 + (y_1 - y_2)^2 = 1$$
 Furthermore, we also define the closure model $((\nats_{k,h},\closure_{4adj}),\eval)$ with atomic propositions in $C$, where $\eval(c \in C) = \{ (i,j) \in \nats^2_{k,h} \mid f(i,j) = c\}$.

In words, such closure model is based on a regular grid, where each pixel, except those on the borders, has four neighbours, corresponding to the directions right, left, up and down.  On top of this space, atomic propositions are interpreted as the colours of pixels.
\clsexa
\end{example}

In \cite{Ci+14b}, we introduced the \emph{spatial until} operator $\phi_1 \mathcal{U} \phi_2$, with a similar intended meaning as $\lsurr$. The main difference is that the definition of $\mathcal{U}$ requires existence of a set of points satisfying $\phi_1$, having closure boundary satisfying $\phi_2$. The definitions of $\lsurr$ and $\mathcal{U}$ coincide in the case of quasi-discrete spaces (see \autoref{thm:other-side-of-simple-discrete-paths-until-S}). As we will see, the definition using paths behaves in a more natural way for topological spaces. First, we compare the interpretation of $\lsurr$ given in \cite{Ci+14b} with \autoref{def:closure-semantics}.

\begin{theorem}\label{thm:other-side-of-simple-discrete-paths-until-S}
 In a \emph{quasi-discrete} closure model $\model$: 
 $\model,x \models \phi_1 \lsurr \phi_2$  according to \autoref{def:closure-semantics} if and only if $\model, x \models \phi_1 \mathcal{U} \phi_2$ according to \cite{Ci+14b}, namely, there is $A \subseteq X$ such that $x \in A$, and $\forall y \in A . \model, y \models \phi_1$, and $\forall z \in \cboundary(A) . \model, z \models \phi_2$.
\end{theorem}

We conclude this section by showing two examples where the definition of \cite{Ci+14b} behaves in a counter-intuitive way, whereas the definition using paths works as expected.

\begin{figure}
\centering{\includegraphics[width=.7\textwidth]{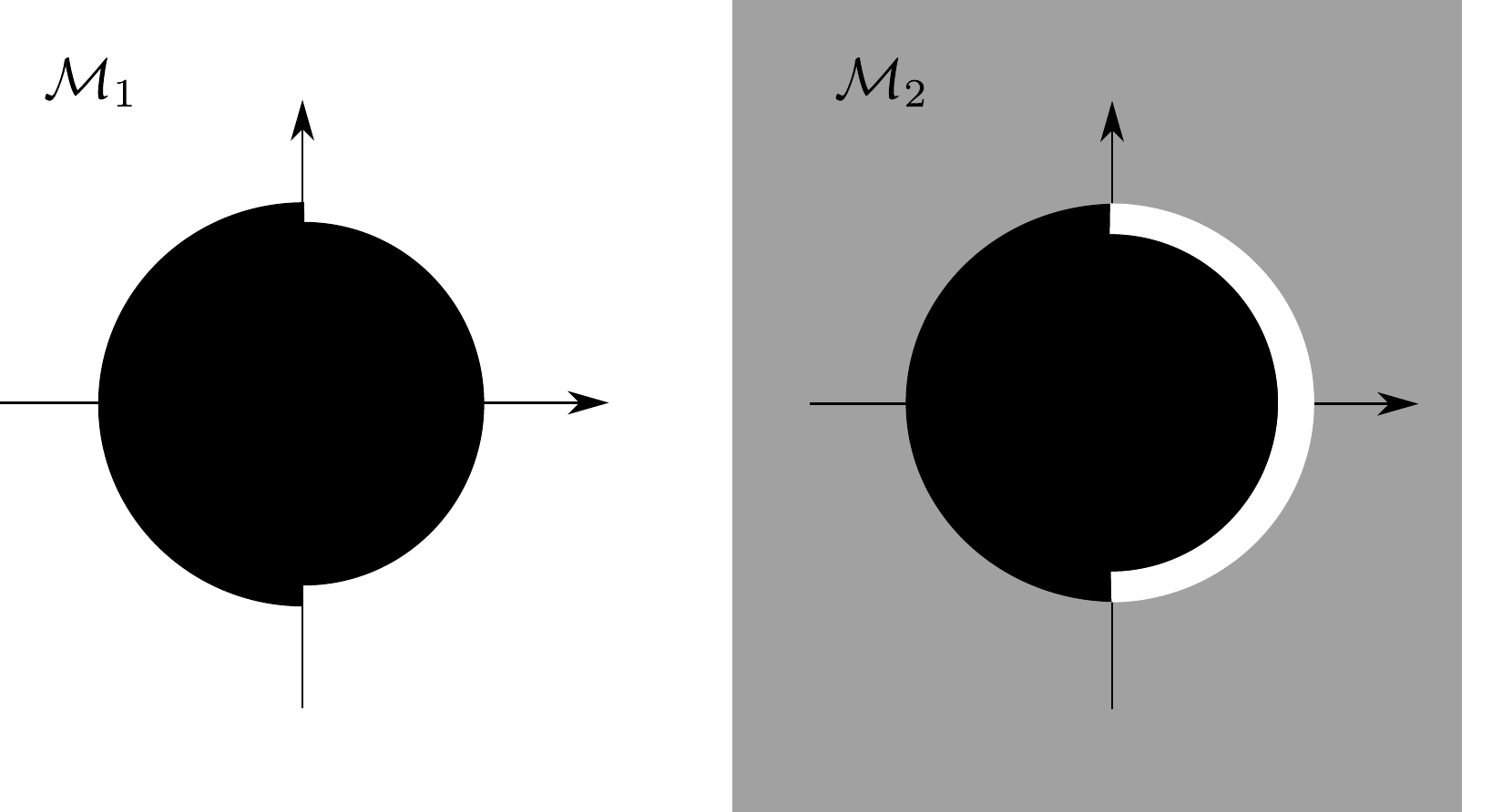}}
\caption{\label{fig:example-iboundary-cboundary-not-enough} Two continuous closure models (boundaries are deliberately represented as very thick, but the reader should think of them as infinitely thin).}
\end{figure}

\begin{example}
We define two models based on the Euclidean topology over $\reals^2$, seen as a closure space $(\reals^2,\closure)$. We use propositions $b,w,g$, depicted in \autoref{fig:example-iboundary-cboundary-not-enough} 
as black, white and grey areas, respectively. Consider the sets $H = \{ (x,y) | x^2 + y^2 < 1 \}$, $H^< = \{ (x,y) | x^2 + y^2 = 1 \land x < 0\}$, $H^{\geq} = \{ (x,y) | x^2 + y^2 = 1 \land x \geq 0 \}$. Let $\model_i = ((\reals^2,\closure), \eval_i)$, for $i \in \{1,2\}$. Fix valuations as follows: $\eval_1(b) = H \cup H^<$, $\eval_1(w) = \reals^2 \setminus \eval_1(b)$,  $\eval_1(g) = \emptyset$, 
$\eval_2(b) = \eval_1(b)$,   
$\eval_2(w) = H^\geq$,  
$\eval_2(g) = \reals_2 \setminus (H \cup H^< \cup H^\geq)$. Let $x \in H$. Clearly, we have $\model_1, x \models b \lsurr w$, and $\model_2 , x \nmodels b \lsurr w$, as there are paths starting at a black point in $\model_2$ and reaching a grey point, which does not satisfy $b$, without passing by white points. The expectation is that $b\, \mathcal{U} w$ holds at $x$ in $\model_1$, which is true by the choice $A = H \cup H^{<}$, but note that $\cboundary(A) = H^{\geq}$. 
For this reason, we also have 
$\model_2, x \models b \, \mathcal{U} w$ by the choice $A = H \cup H^{<}$, which is not what one would expect when thinking of the area $H$ being ``surrounded'' by white points.
\end{example}

\section{The collective spatial logic \cslcs}
\label{sec:CSLCS}

%
%

So far, the properties expressed by our logic refer to points in space, when considered \emph{individually}. However, when looking at space, it is also natural to formulate properties of \emph{sets} of points, considered as a \emph{collective} entity. As we shall see, our notion of collectivity is that of a set of points that are inter-reachable by paths in the whole space. Therefore, not only connected sets are of interest to our logic, but also sets of isolated points or components, that are subsets of  path-connected sets satisfying given properties. Other logics predicating on sets of points include the family of \emph{region calculi} (see \cite{KKWZ07} for a comprehensive overview), describing properties of \emph{regular} sets, and using mereotopological boolean connectives (e.g., ``part of'', ``boundary'', and so on). Such logics characterise \emph{regions} of space. We explicitly divert from this research line, because we aim at characterizing \emph{local} properties of \emph{points}, fitting in the tradition of modal logics, and relating individuals to the collectivity they live in.  Our choice of collective operators is driven by this principle, and is modulated by the requirement of a computationally feasible model checking procedure. 
%


Getting into detail, given a closure model $\model = ((X, \closure), \eval)$, one may introduce ``collective'' formulas $\psi$ (whose syntax and semantics will be clarified in the sequel) equipped with a \emph{collective interpretation}, assigning a boolean valuation to the problem $\model, A \models \psi$ for each \emph{set} of points $A \subseteq X$. 
We define the \emph{collective spatial logic of closure spaces} \cslcs, which is interpreted on closure models. The logic has a \emph{collective} fragment and an \emph{individual} fragment. The collective fragment is evaluated on \emph{subsets} of the set of points of the space. The individual fragment, which is evaluated on single points, is the logic \slcs\ defined in \autoref{sec:dsl}. 

%



\begin{definition}\label{def:set-based:logic-syntax} 
Fix a set $\props$ of \emph{atomic propositions}. The syntax of formulas is defined by the grammar in \autoref{fig:cslcs-syntax}, where 
$a$ ranges over $\props$.
$\mbox{ }$\closenv
\end{definition}




\begin{figure}
\centering
$\begin{array}{l c l l c l c l l}
\multicolumn{4}{c}{\text{\sc{{\bf Collective formulas}}}} & & \multicolumn{4}{c}{\text{\sc{{\bf Individual formulas}}}}\\ \\
\Psi & ::=  & \top & \mbox{\sc{[True]}} & \qquad & \Phi & ::=  & a & \mbox{\sc{[Atomic proposition]}} \\
     & \mid & \lnot \Psi & \mbox{\sc{[Not]}} & & & \mid & \top & \mbox{\sc{[True]}} \\	
     & \mid & \Psi \land \Psi & \mbox{\sc{[And]}} & & & \mid & \lnot \Phi & \mbox{\sc{[Not]}}\\
     & \mid &  \Phi \lShare \Psi & \mbox{\sc{[Share]}} & & & \mid & \Phi \land \Phi & \mbox{\sc{[And]}} \\
     & \mid & \lGroup \Phi & \mbox{\sc{[Group]}} & & & \mid & \lnear \Phi & \mbox{\sc{[Near]}} \\
     &      &              &                     & & & \mid & \Phi \ldiff \Phi & \mbox{\sc{[Propagation]}} \\
     &      &              &                     & & & \mid & \Phi \lsurr \Phi & \mbox{\sc{[Surrounded]}} \\
\end{array}$
\caption{\label{fig:cslcs-syntax} \cslcs\ syntax.}
\end{figure}

We deliberately use the same syntax for boolean connectives both in the individual and the collective fragment, as usage of either fragment is always clear from the context.  Boolean operators are standard. The novel operators we propose are the \emph{share} connective and the \emph{group} connective. Let $\phi$ be an individual formula, and $\psi$ a collective formula. Informally, $\phi \lShare \psi$ (read: $\phi \text{ \emph{share} }\psi$) is satisfied by set $A$ when the subset of points of $A$ satisfying the individual property $\phi$ also satisfies the collective property $\psi$. Formula $\lGroup \phi$ holds on set $A$ when its elements belong to a \emph{group}, that is, a possibly larger, path-connected set of points, all satisfying the individual formula $\phi$.


The satisfaction relation of the logic for each collective formula $\psi$ is given in the form $\model,A \models_C \psi$, where $\model$ is a closure model (see \autoref{def:model}), and $A \subseteq X$ is a set of points.

\begin{definition}\label{def:set-based:satisfaction} 
Given a model $\model = ((X,\closure), \eval))$, and $A \subseteq X$, collective satisfaction $\models_C$ is given by the inductive definition below, where $\models$ is the individual satisfaction relation of \autoref{def:closure-semantics}:

$$
\begin{array}{r c l c l}
\model, A & \models_C & \top \\
\model, A & \models_C & \lnot \psi & \iff &  \model, A \nmodels_C \psi \\
\model, A & \models_C & \psi_1 \land \psi_2 & \iff & \model, A \models_C \psi_1 \text{ and } \model, A \models_C \psi_2 \\
\model, A & \models_C & \phi \lShare \psi & \iff & \model, \{ x \in A \mid \model, x \models \phi \}  \models_C \psi \\
\model, A & \models_C & \lGroup \phi & \iff & \exists B \subseteq X . A \subseteq B \land B \text{ is path-connected } \land \\ & & & & \forall z \in B . \model, z \models \phi\\
\end{array}
$$
\end{definition}

The definition of $\lGroup$ requires the existence of a set $B$ which is possibly larger than $A$. The intuition is that the elements of $A$ are part of a larger ``collective'', consisting of elements satisfying $\phi$.
We consider variants of connectedness as the most basic forms of \emph{collective} and \emph{spatial} property. In particular, we use path-connectedness, in line with the path-based interpretation of \slcs. Connectedness is ``collective'' in the sense that it is not merely determined by a property of the singletons composing a set, and it is not even preserved in subsets of a connected set. On the other hand, even though one could imagine all sorts of collective predicates on a model, we focus on (path-)connectedness, as it is completely determined by the structure of a closure space. For this reason, we consider it a fundamental collective property, deserving special treatment in the field of spatial logics, akin to the notion of transition in models of modal logics. 
Due to the restrictions that we introduce (mainly the strict layering of the collective and individual fragments) the logic \cslcs\ can be automatically verified at a computational cost which is comparable to that of \slcs. Using \cslcs\ one is able to check that given individuals 
are
in the \emph{same} area of space, and they share specific properties. Informally (and depending on the chosen closure model), this idea can be interpreted, for example, as: the fact that certain individuals are able to connect and act as a group; that they may follow the same route to reach a goal; that they are located all together in a protected environment; etc. Below, we develop this concept by the means of some derived operators. In \autoref{sec:discrete-examples} we provide some examples.

\begin{definition}\label{def:collective-derived-operators}
The following derived operators may be defined, where $\psi_1$ and $\psi_2$ are 
collective formulas, and $\phi$ is an \slcs\ formula:
$$
\begin{array}{l c l l}
\bot & \triangleq & \lnot \top& \mbox{\sc{[False]}}\\
\psi_1 \lor \psi_2 & \triangleq & \lnot  ((\lnot \psi_1) \land (\lnot \psi_2))& \mbox{\sc{[Or]}} \\
\lForall \phi & \triangleq & \lnot \phi \lShare \lGroup \bot & \mbox{\sc{[Forall, Individually]}} \\
\lExists \phi & \triangleq & \lnot (\lForall \lnot \phi) & \mbox{\sc[Exists]}\\
\emptyset & \triangleq & \lForall \bot & \mbox{\sc[Empty]}
\end{array}
$$ 
\end{definition}

\noindent The definition of $\lForall$ uses the fact that the only set $A$ such that $\model, A \models \lGroup \bot$ is the empty set, which is trivially path-connected. This is made formal by the following lemma.

\begin{lemma}\label{lem:collective-derived-operators}
 We have:
 \begin{enumerate}
  \item \label{lem:individually} $\model, A \models_C \lForall \phi$ if and only if $\forall x \in A . \model, x \models \phi$;  
  \item \label{lem:exists} $\model, A \models_C \lExists \phi$ if and only if $\exists x \in A . \model, x \models \phi$;
  \item \label{lem:empty} $\model, A \models_C \emptyset$ if and only if $A = \emptyset$.
 \end{enumerate}
\end{lemma}

The $\lForall$ and $\lExists$ connectives also exist in the classical topological logic $\logicSfourU$ (see \cite{KKWZ07}); additionally, \cslcs\ provides the possibility to classify subsets, instead of whole models. However, \emph{global satisfaction}, defined on models, is obtained as a side effect.


\begin{definition}\label{def:global-satisfaction}
\emph{Global} satisfaction is defined for each model $\model = ((X,\closure), \eval)$ and collective formula $\psi$ as $\model \models_G \psi \iff \model, X \models_C \psi$.
\end{definition}

From now on, we will sometimes omit the subscripts $C$ and $G$ from the satisfaction relation, when clear from the context. Apart from the usual derived connectives, such as disjunction or logical implication, \cslcs\ can express some 
useful 
derived operators.

\begin{definition}\label{def:collective-more-derived-operators}
Define the following collective derived operators:

$$\begin{array}{l c l l}
\phi_1 \lSSurr \phi_2  & \triangleq & \lGroup (\lnot \phi_2 \land (\phi_1 \lsurr \phi_2)) & \mbox{\sc[Collectively surrounded]} \\
\phi_1 \lPartitioned \phi_2  & \triangleq & \lForall ((\phi_1 \lor \phi_2) \land \lnot (\phi_1 \land \phi_2)) \land   & \mbox{\sc[Collectively partitioned]} \\ && (\phi_1 \lShare (\phi_1 \lSSurr \phi_2)) \land (\phi_2 \lShare (\phi_2 \lSSurr \phi_1))
\end{array}
$$ 
\end{definition}

%



A set $A$ satisfies $\model, A \models \phi_1 \lSSurr \phi_2$ if and only if the points in $A$ satisfy $\phi_1$, and are ``collectively'' surrounded by a set of points satisfying $\phi_2$. More precisely, using the connective $\lGroup$, it is required that a path-connected set $B$ including $A$ exists, with all points of $B$ satisfying $\phi_1 \lsurr \phi_2$, but not $\phi_2$. Not only there can be no path rooted in $B$ and leaving $\phi_1$ without passing by $\phi_2$, but also, noting that all the elements of $B$ satisfy $\phi_1 \land \lnot \phi_2$, such set $B$ must be a path-connected component of $\lnot \phi_2$, the elements of which are surrounded in the sense of \slcs\ by points satisfying $\phi_2$.



\begin{figure}
\begin{center}
\begin{tikzpicture}{}
\node [circle,draw=black,fill=blue,thick,inner sep=0pt,minimum size=4mm] at (-0.15,0) (node0) {};
\node [circle,draw=black,fill=blue,thick,inner sep=0pt,minimum size=4mm] at (0.5,0.5) (node1) {};
\node [circle,draw=black,fill=blue,thick,inner sep=0pt,minimum size=4mm] at (0.5,-0.5) (node2) {};
\node [circle,draw=black,fill=red,thick,inner sep=0pt,minimum size=4mm] at (1.25,0.5) (node3) {};
\node [circle,draw=black,fill=red,thick,inner sep=0pt,minimum size=4mm] at (1.25,-0.5) (node4) {};
\node [circle,draw=black,fill=red,thick,inner sep=0pt,minimum size=4mm] at (1.90,.0) (node5) {};
\draw[<->] (node0) -- (node1);
\draw[<->] (node0) -- (node2);
\draw[<->] (node2) -- (node1);
\draw[<->] (node1) -- (node3);
\draw[<->] (node2) -- (node4);
\draw[<->] (node3) -- (node4);
\draw[<->] (node4) -- (node5);
\draw[<->] (node5) -- (node3);
\end{tikzpicture}
\qquad\qquad
\begin{tikzpicture}{}
\node [circle,draw=black,fill=red,thick,inner sep=0pt,minimum size=4mm] at (-0.15,0) (node0) {};
\node [circle,draw=black,fill=red,thick,inner sep=0pt,minimum size=4mm] at (0.5,0.5) (node1) {};
\node [circle,draw=black,fill=blue,thick,inner sep=0pt,minimum size=4mm] at (0.5,-0.5) (node2) {};
\node [circle,draw=black,fill=blue,thick,inner sep=0pt,minimum size=4mm] at (1.25,0.5) (node3) {};
\node [circle,draw=black,fill=blue,thick,inner sep=0pt,minimum size=4mm] at (1.25,-0.5) (node4) {};
\node [circle,draw=black,fill=red,thick,inner sep=0pt,minimum size=4mm] at (1.90,.0) (node5) {};
\draw[<->] (node0) -- (node1);
\draw[<->] (node0) -- (node2);
\draw[<->] (node2) -- (node1);
\draw[<->] (node1) -- (node3);
\draw[<->] (node2) -- (node4);
\draw[<->] (node3) -- (node4);
\draw[<->] (node4) -- (node5);
\draw[<->] (node5) -- (node3);
\end{tikzpicture}
\end{center}

\caption{\label{fig:Partitioning} The model on the left satisfies $red \lPartitioned blue$;  the one on the right does not.}
\end{figure}
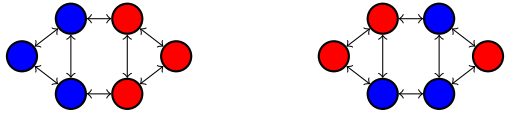

For the $\lPartitioned$ connective, we look at its global interpretation. The statement $\model \models \phi_1 \lPartitioned \phi_2$ expresses that all the points of the space satisfy either $\phi_1$ or $\phi_2$, that \emph{all} the points satisfying $\phi_1$ can be connected to each other, forming a set of points satisfying $\phi_1$ and surrounded by points satisfying $\phi_2$, and vice-versa. The sets of points satisfying $\phi_1$ is path-connected, and so is the set satisfying $\phi_2$. For example, the model in the left-hand side of \autoref{fig:Partitioning}
satisfies $red \lPartitioned blue$ while the model in the right-hand-side of the figure
does not satisfy the same formula.

%

%


\section{Example: emergency evacuation}
\label{sec:EXAMPLES}

In this section we show some examples of interpreting \slcs\ and \cslcs\ on quasi-discrete closure spaces. First, starting from our running example, let us define a closure space to provide a simple model of short-range communication.

\begin{example}\label{ex:r2distance2-b}
Let us consider again the closure space presented in \autoref{ex:r2distance}. This closure space can 
be used to model a network of agents distributed over a two-dimensional physical space, that communicate via wireless devices having fixed communication radius $\delta$.
In the left hand side of \autoref{fig:sensors} a graphical representation of such model is provided. There \emph{green}, \emph{purple}
and \emph{blue} dots identify different kinds of agents located in the space. Let us consider the colours as atomic propositions.

The set $\closure_\delta( green \cup purple \cup blue )$ consists of points
in $\reals^{2}$ that are in the communication range of at least one agent, represented by the pink area in the right-hand side 
of \autoref{fig:sensors}.
Suppose that the \emph{green} agent of our example is the source of some relevant information, which is meant to be transmitted from the \emph{green} device to the other devices that are reachable after some \emph{hops}.
The set of devices that can receive the information sent by the \emph{green} device is characterised, using the propagation operator, by the formula $green \ldiff (purple \cup blue)$, satisfied by the black points in \autoref{fig:sensors_flooding}.
\clsexa
\end{example}

\newcommand{\sensornode}[5]{
\node[circle,white,fill=#4,thick,inner sep=0pt,minimum size=0.15cm] at (#1,#2) {};
}

\newcommand{\communicationrange}[5]{
\node[circle,white,draw=#5!50,fill=#5!10,thick,inner sep=0pt,minimum size=#3] at (#1,#2) {};
}

\newcommand{\communicationborder}[5]{
\node[circle,white,draw=#5!50,thick,inner sep=0pt,minimum size=#3] at (#1,#2) {};
}

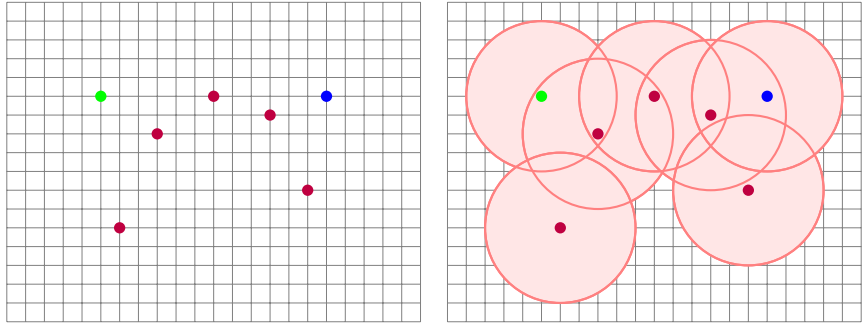
\begin{figure}[tbp]
\begin{center}
\begin{tabular}{cc}
\begin{tikzpicture}
\draw[step=0.25cm,gray,very thin] (-2.25,-2.5) grid (3.25,1.75);
\sensornode{-1.0}{0.5}{2.5cm}{green}{red!10};
\sensornode{-.25}{0}{2.5cm}{purple}{red!10};
\sensornode{0.5}{0.5}{1.5cm}{purple}{red!10};
\sensornode{1.25}{0.25}{1.5cm}{purple}{red!10};
\sensornode{1.75}{-0.75}{1.5cm}{purple}{red!10};
\sensornode{-0.75}{-1.25}{1.5cm}{purple}{red!10};
\sensornode{2.0}{0.5}{1.5cm}{blue}{red!10};
\end{tikzpicture} & 
\begin{tikzpicture}
\draw[step=0.25cm,gray,very thin] (-2.25,-2.5) grid (3.25,1.75);
\communicationrange{-1.0}{0.5}{2.0cm}{green}{red};
\communicationrange{-.25}{0}{2.0cm}{purple}{red};
\communicationrange{0.5}{0.5}{2.0cm}{purple}{red};
\communicationrange{1.25}{0.25}{2.0cm}{purple}{red};
\communicationrange{1.75}{-0.75}{2.0cm}{purple}{red};
\communicationrange{-.75}{-1.25}{2.0cm}{purple}{red};
\communicationrange{2.0}{0.5}{2.0cm}{blue}{red};
\sensornode{-1.0}{0.5}{2.5cm}{green}{red!10};
\sensornode{-.25}{0}{2.5cm}{purple}{red!10};
\sensornode{0.5}{0.5}{1.5cm}{purple}{red!10};
\sensornode{1.25}{0.25}{1.5cm}{purple}{red!10};
\sensornode{1.75}{-0.75}{1.5cm}{purple}{red!10};
\sensornode{-0.75}{-1.25}{1.5cm}{purple}{red!10};
\sensornode{2.0}{0.5}{1.5cm}{blue}{red!10};
\communicationborder{-1.0}{0.5}{2.0cm}{green}{red};
\communicationborder{-.25}{0}{2.0cm}{purple}{red};
\communicationborder{0.5}{0.5}{2.0cm}{purple}{red};
\communicationborder{1.25}{0.25}{2.0cm}{purple}{red};
\communicationborder{1.75}{-0.75}{2.0cm}{purple}{red};
\communicationborder{-.75}{-1.25}{2.0cm}{purple}{red};
\communicationborder{2.0}{0.5}{2.0cm}{blue}{red};
\end{tikzpicture}
\end{tabular}
\caption{A graphical representation of \autoref{ex:r2distance2-b}.}
\label{fig:sensors}
\end{center}
\end{figure}

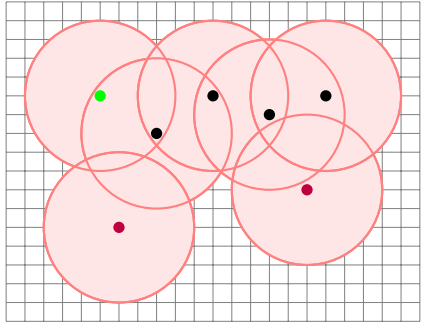
\begin{figure}[tbp]
\begin{center}
\begin{tabular}{cc}
\begin{tikzpicture}
\draw[step=0.25cm,gray,very thin] (-2.25,-2.5) grid (3.25,1.75);
\communicationrange{-1.0}{0.5}{2.0cm}{green}{red};
\communicationrange{-.25}{0}{2.0cm}{purple}{red};
\communicationrange{0.5}{0.5}{2.0cm}{purple}{red};
\communicationrange{1.25}{0.25}{2.0cm}{purple}{red};
\communicationrange{1.75}{-0.75}{2.0cm}{purple}{red};
\communicationrange{-.75}{-1.25}{2.0cm}{purple}{red};
\communicationrange{2.0}{0.5}{2.0cm}{blue}{red};
\sensornode{-1.0}{0.5}{2.5cm}{green}{red!10};
\sensornode{-.25}{0}{2.5cm}{black}{red!10};
\sensornode{0.5}{0.5}{1.5cm}{black}{red!10};
\sensornode{1.25}{0.25}{1.5cm}{black}{red!10};
\sensornode{1.75}{-0.75}{1.5cm}{purple}{red!10};
\sensornode{-0.75}{-1.25}{1.5cm}{purple}{red!10};
\sensornode{2.0}{0.5}{1.5cm}{black}{red!10};
\communicationborder{-1.0}{0.5}{2.0cm}{green}{red};
\communicationborder{-.25}{0}{2.0cm}{purple}{red};
\communicationborder{0.5}{0.5}{2.0cm}{purple}{red};
\communicationborder{1.25}{0.25}{2.0cm}{purple}{red};
\communicationborder{1.75}{-0.75}{2.0cm}{purple}{red};
\communicationborder{-.75}{-1.25}{2.0cm}{purple}{red};
\communicationborder{2.0}{0.5}{2.0cm}{blue}{red};
\end{tikzpicture} 
\end{tabular}
\caption{In \emph{black} the devices that can receive data from the \emph{green} device.}
\label{fig:sensors_flooding}
\end{center}
\end{figure}

%


%

Taking advantage of both \autoref{ex:imagesascs} (interpreting digital images as closure models) and \autoref{ex:r2distance2-b}, we will now set up a more complex closure space, comprising a communication layer, with closure determined by communication ranges, and a physical layer, with closure determined by the structure of a regular grid. The two layers are linked by a binary relation. On top of this set-up, we will discuss the interpretation of some example properties, assuming that a set of agents (modelled by appropriate atomic propositions) is distributed in the physical layer.

\begin{figure}[tbp]
\centering
\includegraphics[width=.7\textwidth]{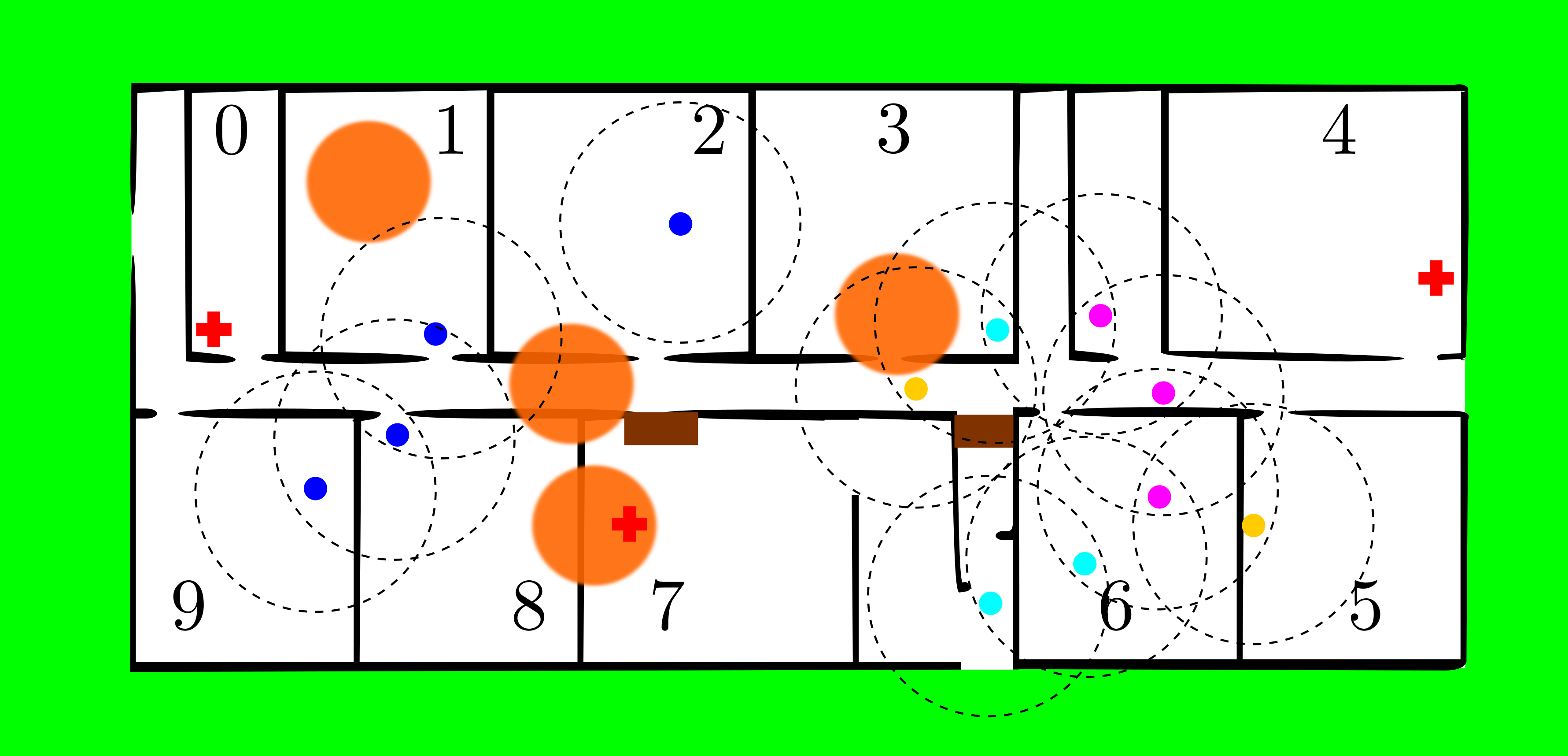}
\caption{\label{fig:escape}A representation of agents in a building in an emergency condition.}
\end{figure}

\begin{example}\label{exa:escape}
Recall from \autoref{ex:imagesascs} that digital images can be treated as finite quasi-discrete closure models. Consider one such model, with underlying space $(X,\closure_{4adj})$, with $X \subset \nats^2$. In this example, we will use a digital image representing a portion of a two-dimensional physical space; therefore, each point of the image is also mapped to a \emph{position}, or \emph{coordinate}, in the Euclidean space $\reals^2$, giving rise to a function
$map : X \to \reals^2$. Let $Y$ be the finite image of the function $map$. Assume $X$ and $Y$ are disjoint, for simplicity.  Let $pos$ be the graph of the function $map$, that is, the set of pairs $\{(x,y) \in X \times Y \mid map(x) = y\}$. In a similar way as in \autoref{ex:r2distance2-b}, fix a communication range $\delta$, and introduce the relation $R_\delta \subseteq \reals^2$ from \autoref{ex:r2distance-as-relation}. Then, let $R_\delta'=R_\delta \cap Y^2$ be the restriction of $R_\delta$ to the image of the function $map$.
%
Consider the set $Z = X \cup Y$.  Define the quasi-discrete closure space $(Z,\closure_R)$ using the relation $$R \triangleq 4adj \cup pos \cup R_\delta'$$ 

The closure space $(Z,\closure_R)$ can be thought of as ``two-layered''. One layer is the digital image, the other one is a finite subset of $\reals^2$ equipped with the closure $\closure_\delta$ restricted to $Y$, in a similar way to \autoref{ex:r2distance} . The two layers are linked by the relation $pos$; note that each position in $Y$ is thus ``close'', in the sense of the operator $\lnear$, to a point of the digital image. By this, as we shall see, logic formulas can simultaneously predicate on proximity in the image, acting as a ``physical'' layer, where proximity means adjacency in space, and in Euclidean coordinates, acting as a ``communication'' layer, where proximity is based on distance. We will also consider a set of agents, first-aid facilities, obstructions, and dangerous areas, formalised as atomic propositions, giving rise to a quasi-discrete closure \emph{model}. 

Before making this idea formal, we look at a picture of an instance of such construction, in \autoref{fig:escape}. The digital image in the background, the points of which form the set $X$, represents the map of a building at a specific instant in time, where an emergency situation occurs (note that rooms have been numbered for reader's convenience, but we are not considering numbers, graphically, as part of the underlying map). The \emph{white} points form the areas where agents can walk. Some of the white points, however, are covered by obstructions, painted in brown, or are in the range of some source of hazard. Hazardous areas are painted in semi-transparent orange. The green points are a safe area, accessible via exit doors. Some white points are also part of areas where first aid is available, which are represented by a red cross. The walls are painted in black.  Coloured (blue, cyan, purple, yellow) dots represent agents, with their communication range (dashed circles). The set $Y$ is the set of actual coordinates of the points in space denoted by pixels of the digital image.

We define a valuation function $\eval$, obtaining the quasi-discrete model $((Z,\closure_R),\eval)$. Atomic propositions are the colours $white$, $black$, $green$, $red$, $blue$, $cyan$, $purple$, $yellow$, $brown$, $agent$, $danger$, and $coord$. Function $\eval$ is such that each point in the image satisfies its own colour. Proposition $danger$ is true only at points in the image under the orange semi-transparent circles. Each point may satisfy more than one atomic proposition; in particular, points under the orange circles also satisfy other atomic propositions. Agents are represented by additionally colouring points of $X$ in blue, cyan, purple, or yellow. Points that satisfy $red$ or $brown$ also satisfy $white$, as in principle these are areas where it is possible to walk, even though there is an obstruction in the current situation.   Points of $Y$ satisfy just one predicate, namely $coord$, and are not represented in \autoref{fig:escape}. In addition, no other point in $Z$ satisfies predicate $coord$. Finally, define the short-hands $obstacle \triangleq black \lor brown \lor danger$,  $agent \triangleq blue \lor cyan \lor purple \lor yellow$, and $\mathit{safe} \triangleq white \land \lnot obstacle$.
%
%

In this situation, we suppose that groups of agents of the same colour are expected to address an emergency situation together. Agents must be able to reach both first-aid points and exit doors without passing by dangerous areas. Agents belonging to the same group should reach a first-aid point and the exit together with other members of the group; in case an agent is isolated from her group, an agent of another group must be able to reach first aid, and then rescue her. \clsexa
\end{example}

We remark that, for simplicity, when dealing with paths concerning agents, we do not consider the cases in which an agent may exit and re-enter the building through a different access, passing by the green area\footnote{Depending on the application domain, one may take into account agents that exit and re-enter the building by using another set of logical properties (the logic easily distinguishes between these two different kinds of paths).}.
In the remainder of this section, we present some example properties, and their interpretation in the situation of \autoref{fig:escape}. First, recall the definition of the derived operator $\phi_1 \calT \phi_2 \triangleq \phi_1 \land ((\phi_1 \lor \phi_2) \ldualuntil \phi_2)$. Point $x$ satisfies $\phi_1 \calT \phi_2$ whenever it satisfies $\phi_1$ and there is a path $p$, and an index $i$, with $p(0) = x$, such that, for all $j \in (0,i)$, point $p(j)$ satisfies $\phi_1 \lor \phi_2$, and point $p(i)$ satisfies $\phi_2$. Informally speaking, we may say that $\calT$ expresses reachability in space from a point satisfying formula $\phi_1$ to a point satisfying $\phi_2$, only passing by points satisfying $\phi_1$ or $\phi_2$.

\begin{example}\label{exa:escape-locked-room}
 There may be safe points, with no escape route. This is defined as the formula $$\phi_1 \triangleq \mathit{safe} \lsurr obstacle$$ satisfied by the white points in Room $3$.  \clsexa
\end{example}

\begin{example}\label{exa:safety-route}
The walking areas, from which a first-aid point can be safely reached, are classified by the derived operator $\ldualuntil$. Consider the formula:
$$\phi_2 \triangleq \mathit{safe} \calT (red \land \mathit{safe})$$ 
Points satisfying formula $\phi_2$ are required to be $\mathit{safe}$, and furthermore, to be at the start of a path of $\mathit{safe}$ points, leading to a point which is $red$ and $\mathit{safe}$. 
In \autoref{fig:escape}, $\phi_2$ is satisfied, among other points, by all the positions of agents, except those in rooms 3 and 7. That is, $\phi_2$ is satisfied by those white points that are the start of a path that avoids obstacles (including dangerous areas), leading to safe first-aid facilities, while only traversing $white$ points.  
Similarly, the points from which an exit may be reached are characterised by the formula
$$\phi_3 \triangleq \mathit{safe} \calT \mathit{green}$$
which is satisfied by the blue, yellow, and violet points, but not by any cyan point (note that we are not considering the possibility of passing by the green area and re-enter the building, as we explained earlier). The points where first-aid facilities are located, and from where it is possible to safely reach an exit (all the red points in \autoref{fig:escape}), satisfy the formula
$$\phi_4 \triangleq (red \land \mathit{safe}) \land (\mathit{safe} \calT \mathit{green})$$
 Combining $\phi_2$ and $\phi_4$ one is then able to define the set of points from which one can safely walk to a first aid point and then to the exit. These points are identified by the formula 
 $$\phi_5 \triangleq \mathit{safe} \calT \phi_4$$ 
For instance, the white points in Room 8, but not those in Room 7, satisfy $\phi_5$.
 \clsexa
\end{example}

We shall now introduce some collective formulas, that for simplicity are evaluated under the global interpretation of \autoref{def:global-satisfaction}.

\begin{example}
  We can define a collective formula, parametrised by a colour, that is true whenever all agents of the given colour are connected in the communication layer of the model. 
  $$ \phi_6(colour) = (coord \land \lnear colour) \lShare \lGroup (coord \land \lnear agent)$$
  In the definition of $\phi_6$, note that $\lnear colour$ denotes the set of points that are \emph{near} to a point satisfying $colour$. Such set is the union of the points in the digital image where the agents of the given colour are located, their neighbours in the digital image, and their coordinates in the communication layer. Therefore, when $colour$ is the colour of an agent, the sub-formula $coord \land \lnear colour$ precisely identifies the coordinates in $Y$ that are positions of agents in the group identified by $colour$. Such coordinates are required to be part of a larger set of points, which are \emph{connected} in the communication layer, and also are positions of arbitrary agents, so that the communication flow required by the formula may also include agents of \emph{different} colours. 
  In the model of \autoref{fig:escape}, $\phi_6(colour)$ holds for all the colours of agents, except blue. \clsexa
\end{example}

\begin{example}
Agents of the same colour should be able to reach a first aid point, and then an exit, all together. We leave the colour as a parameter of the formula.
 $$\phi_7(colour) = colour \lShare \lGroup \phi_5$$
 In \autoref{fig:escape}, $\phi_7(colour)$ holds for colours yellow and purple, but not cyan and blue.
\end{example}

\begin{example}
 We shall now deal with rescuing of agents. An agent of a given colour can be rescued if there is an agent of a different colour that can reach her, after passing by a first-aid point, and the two can safely reach an exit. 
 First consider formula $\phi_8(colour)$, describing first-aid points that can be reached by an agent of a different colour than the given one (this is achieved by the sub-formula $agent \land (\lnot colour)$ below), by a safe route:
 $$\phi_8(colour) \triangleq (red \land \lnot obstacle) \land (\lnear ((agent \land (\lnot colour))\ldiff \mathit{safe}))$$
 Points satisfying $\phi_8(colour)$ are $red$ and not an obstacle, that is, they are safe first-aid locations. Furthermore, the definition of $\phi_8(colour)$ also uses the $\ldiff$ operator in order to guarantee that such points are directly connected (operator $\lnear$) to points that \emph{can be reached}\footnote{Since the model of our example is symmetric, reachability in opposite directions may not make an actual difference. However, models similar to the one we are depicting may feature e.g., one way doors. We are not adding one-way links in our model, as we do not deem it necessary for illustrating the connectives of the logic, and it makes the formal definition of the underlying closure space less readable.} from a point where an agent of a different colour is located, passing only through safe points.
 Thus, agents of a specific colour that can be rescued satisfy the formula $$\phi_9(colour) \triangleq agent \land \phi_2 \land \lnear (\lnot obstacle \land ( \phi_8(colour) \ldiff \mathit{safe} )) $$ 
We can also define a collective formula expressing that, for a given colour, either $\phi_7(colour)$ holds, or all agents can be rescued: $$\phi_{10}(colour) \triangleq \phi_7(colour) \lor \lForall (colour \lShare \phi_9(colour))$$
In our example model, $\phi_{10}(blue)$ is true, whereas $\phi_{10}(cyan)$ is false.
 \clsexa
\end{example}

\section{Spatial model checking}\label{sec:model-checking}

In this section we describe a model checking algorithm for \slcs\ and \cslcs. The algorithm is composed of two 
procedures, one for individual formulas, that is, the logic \slcs, and one for 
collective formulas, making use of the procedure for individual formulas. As we shall see, the procedure for individual formulas is a \emph{global} model checking procedure for 
\slcs. Given model $\model = ((X,\closure_{R}), \eval)$ and formula $\phi$, 
the procedure returns the set $\{x \in X \mid \model, x \models \phi\}$. The procedure for 
collective formulas, on the other hand, is a \emph{local} model checking algorithm, that 
is, given model $\model$, formula $\psi$ and set of points $A$, it returns the 
boolean satisfaction value of $\model, A \models \psi$. We choose a local algorithm for the collective fragment, since enumeration of a set of subsets is 
a problem of inherent exponential complexity. Merely returning a result for a global model checking procedure would require some kind of symbolic description, which is left for future investigation. 

\begin{figure}[tbp]
\small
\begin{tabular}{p{.45\textwidth}p{.45\textwidth}}
\begin{algorithm}[H] 
\myfun{$\check(\model,\phi)$}{
  
  \KwIn{Finite, quasi-discrete closure model $\model = ((X,\closure_R),\eval)$, formula $\phi$}
  \KwOut{Set of points $\{ x \in X \mid \model, x \models \phi\}$}
  
  \Match{$\phi$}
  {
     \lCase{$\top:$}{\Return $X$}
     \lCase{$p:$}{\Return $\eval(p)$}
     \Case{$\lnot \phi_1:$}{
	\Let $P = \check(\model,\phi_1)$ \In\; 
	\Return $X \setminus P$
      }
     \Case{$\phi_1 \land \phi_2:$}{
	\Let $P = \check(\model,\phi_1)$ \In\;
	\Let $Q = \check(\model,\phi_2)$ \In\;
	\Return $P \cap Q$
      }
     \Case{$ \phi_1\ldiff\phi_2:$}{
        \Return \checkDiff($\model$,$\phi_1$,$\phi_2$)     
      }
  
     \Case{$\phi_1 \lsurr \phi_2:$}{
        \Return \checkUntil($\model$,$\phi_1$,$\phi_2$)     
     }
   }   
  }
\end{algorithm}
&
\begin{algorithm}[H] 
        \myfun{\checkUntil($\model$,$\phi_1$,$\phi_2$)} {
        \KwIn{Finite, quasi-discrete closure model $\model = ((X,\closure_R),\eval)$, 
formulas $\phi_1, \phi_2$}
	\KwOut{Set of points $\{ x \in X \mid \model, x \models \phi_1 \lsurr 
\phi_2\}$}
	
	\Var $V := \check(\model,\phi_1)$\; 
	\Let $Q = \check(\model,\phi_2)$ \In\;
 	\Var $T$ := $\cboundary(V \cup Q)$ \;
  
	\While{$T \neq \emptyset$}{
        \Var $T'$ := $\emptyset$\;
		\For{$x\in T$}{
              \Let $N = pre(x) \cap V$ \In\; 
              $V$ := $V \setminus N$\;
              $T'$ := $T'\cup (N \setminus Q)$\;
		}                
        $T$ := $T'$;
        }
  	\Return $V$

     }
\end{algorithm} 
\\
\begin{algorithm}[H]\caption{\label{alg:decision-procedure}Decision procedure 
for the model checking problem of \slcs.}\end{algorithm}
&
\begin{algorithm}[H]\caption{\label{alg:check-until-quasi-discrete}Checking 
\emph{surrounded} formulas in a quasi-discrete closure space.}\end{algorithm}\\
\end{tabular}
\end{figure}
%
%
\begin{figure}[tbp]
\small
\begin{tabular}{p{.90\textwidth}p{.45\textwidth}}
\begin{algorithm}[H] 
        \myfun{\checkDiff($\model$,$\phi_1$,$\phi_2$)} {
        \KwIn{Finite, quasi-discrete closure model $\model = ((X,\closure_R),\eval)$, 
formulas $\phi_1, \phi_2$}
	\KwOut{Set of points $\{ x \in X \mid \model, x \models \phi_1 \ldiff 
\phi_2\}$}
	
	\Var $V := \check(\model,\phi_1)$\; 
	\Var $Q = \check(\model,\phi_2)$ \In\;
	\Var $T$ := $\closure_{R}(V) \cap Q$ \;
	\Var $R$ := $T$ \;
	\Var $Q$ := $Q \setminus T$ \;
	\While{$T \neq \emptyset$}{
        		\Var $T'$ := $\emptyset$\;
		\For{$x\in T$}{
			$T'$ := $T' \cup (Q \cap post(x))$ \;
		} 
		$Q$ := $Q \setminus T'$ \;
		$R$ := $R\cup T'$\;
	        $T$ := $T'$ \;
	}                
  	\Return $R$
     }

\end{algorithm} 
\\
\begin{algorithm}[H]\caption{\label{alg:check-diff-quasi-discrete}Checking 
\emph{propagation} formulas in a quasi-discrete closure space.}\end{algorithm}
\end{tabular}
\end{figure}

Function \check, computed by \autoref{alg:decision-procedure}, implements 
the model checker for \slcs. The function takes as input a finite, quasi-discrete model 
$\model=((X,\closure_{R}),\eval)$ and a \slcs\ formula $\phi$, 
and returns the set of all points  in $X$ satisfying $\phi$. 
The function is inductively defined on the structure of $\phi$ and, 
following a bottom-up approach, computes the resulting set via an
appropriate combination of the recursive invocations of \check on the subformulas of $\phi$.
When $\phi$ is of the form $\top$, $p$, $\lnot \phi_1$ or $\phi_1 \land 
\phi_2$, the definition of $\check(\model,\phi)$ is straightforward. 
To compute the set of points satisfying $\lnear \phi_1$, the closure
operator $\closure$ of the space is applied to the set of points satisfying 
$\phi_1$. When $\phi$ is of the form $\phi_1 \lsurr \phi_2$, function \check relies on
the function \checkUntil defined in \autoref{alg:check-until-quasi-discrete}. When $\phi$ is of the form $\phi_1 \ldiff \phi_2$, function \check relies on
the function \checkDiff defined in \autoref{alg:check-diff-quasi-discrete}.

\begin{figure}[tbp]

\begin{center}
\begin{tabular}{ccc}
\scalebox{0.75}{\begin{tikzpicture}{}
\node [circle,draw=yellow!50,fill=yellow!20,,thick,inner sep=0pt,minimum size=4mm] at (0,0) (node0) {0};
\node [circle,draw=yellow!50,fill=yellow!20,,thick,inner sep=0pt,minimum size=4mm] at (1,0) (node1) {1};
\node [circle,draw=yellow!50,fill=yellow!20,,thick,inner sep=0pt,minimum size=4mm] at (1,1) (node2) {2};
\node [circle,draw=red!50,fill=red!20,thick,inner sep=0pt,minimum size=4mm] at (1,2) (node3) {3};
\node [circle,draw=red!50,fill=red!20,thick,inner sep=0pt,minimum size=4mm] at (2.5,1) (node4) {4};
\node [circle,draw=black,fill=white,thick,inner sep=0pt,minimum size=4mm] at (2.5,2) (node5) {5};
\node [circle,draw=black,fill=white,thick,inner sep=0pt,minimum size=4mm] at (1.75,3) (node6) {6};
\node [circle,draw=black,fill=white,thick,inner sep=0pt,minimum size=4mm] at (3.25,3) (node7) {7};
\node [circle,draw=yellow!50,fill=yellow!20,thick,inner sep=0pt,minimum size=4mm] at (4.0,1.5) (node8) {8};
\node [circle,draw=yellow!50,fill=yellow!20,thick,inner sep=0pt,minimum size=4mm] at (4.0,0.5) (node9) {9};
\draw[<->] (node0) -- (node1);
\draw[<->] (node0) -- (node2);
\draw[<->] (node2) -- (node1);
\draw[<->] (node2) -- (node3);
\draw[<->] (node3) -- (node5);
\draw[<->] (node3) -- (node6);
\draw[<->] (node5) -- (node6);
\draw[<->] (node7) -- (node6);
\draw[<->] (node7) -- (node5);
\draw[<->] (node5) -- (node4);
\draw[<->] (node2) -- (node4);
\draw[<->] (node8) -- (node4);
\draw[<->] (node8) -- (node4);
\draw[<->] (node8) -- (node9);
\draw[<->] (node4) -- (node9);
\draw[<->] (node5) -- (node8);
\end{tikzpicture}} &
\scalebox{0.75}{\begin{tikzpicture}{}
\node [circle,draw=yellow!50,fill=yellow!20,,thick,inner sep=0pt,minimum size=4mm] at (0,0) (node0) {0};
\node [circle,draw=yellow!50,fill=yellow!20,,thick,inner sep=0pt,minimum size=4mm] at (1,0) (node1) {1};
\node [circle,draw=yellow!50,fill=yellow!20,,thick,inner sep=0pt,minimum size=4mm] at (1,1) (node2) {2};
\node [circle,draw=red!50,fill=red!20,thick,inner sep=0pt,minimum size=4mm] at (1,2) (node3) {3};
\node [circle,draw=red!50,fill=red!20,thick,inner sep=0pt,minimum size=4mm] at (2.5,1) (node4) {4};
\node [circle,draw=black,fill=black,thick,inner sep=0pt,minimum size=4mm] at (2.5,2) (node5) {5};
\node [circle,draw=black,fill=black,thick,inner sep=0pt,minimum size=4mm] at (1.75,3) (node6) {6};
\node [circle,draw=black,fill=white,thick,inner sep=0pt,minimum size=4mm] at (3.25,3) (node7) {7};
\node [circle,draw=yellow!50,fill=yellow!20,thick,inner sep=0pt,minimum size=4mm] at (4.0,1.5) (node8) {8};
\node [circle,draw=yellow!50,fill=yellow!20,thick,inner sep=0pt,minimum size=4mm] at (4.0,0.5) (node9) {9};
\draw[<->] (node0) -- (node1);
\draw[<->] (node0) -- (node2);
\draw[<->] (node2) -- (node1);
\draw[<->] (node2) -- (node3);
\draw[<->] (node3) -- (node5);
\draw[<->] (node3) -- (node6);
\draw[<->] (node5) -- (node6);
\draw[<->] (node7) -- (node6);
\draw[<->] (node7) -- (node5);
\draw[<->] (node5) -- (node4);
\draw[<->] (node2) -- (node4);
\draw[<->] (node8) -- (node4);
\draw[<->] (node8) -- (node4);
\draw[<->] (node8) -- (node9);
\draw[<->] (node4) -- (node9);
\draw[<->] (node5) -- (node8);
\end{tikzpicture}} &
\scalebox{0.75}{
\begin{tikzpicture}{}
\node [circle,draw=yellow!50,fill=yellow!20,,thick,inner sep=0pt,minimum size=4mm] at (0,0) (node0) {0};
\node [circle,draw=yellow!50,fill=yellow!20,,thick,inner sep=0pt,minimum size=4mm] at (1,0) (node1) {1};
\node [circle,draw=yellow!50,fill=yellow!20,,thick,inner sep=0pt,minimum size=4mm] at (1,1) (node2) {2};
\node [circle,draw=red!50,fill=red!20,thick,inner sep=0pt,minimum size=4mm] at (1,2) (node3) {3};
\node [circle,draw=red!50,fill=red!20,thick,inner sep=0pt,minimum size=4mm] at (2.5,1) (node4) {4};
\node [circle,draw=black,fill=black,thick,inner sep=0pt,minimum size=4mm] at (2.5,2) (node5) {5};
\node [circle,draw=black,fill=black,thick,inner sep=0pt,minimum size=4mm] at (1.75,3) (node6) {6};
\node [circle,draw=black,fill=white,thick,inner sep=0pt,minimum size=4mm] at (3.25,3) (node7) {7};
\node [circle,draw=black,fill=black!20,thick,inner sep=0pt,minimum size=4mm] at (4.0,1.5) (node8) {8};
\node [circle,draw=yellow!50,fill=yellow!20,thick,inner sep=0pt,minimum size=4mm] at (4.0,0.5) (node9) {9};
\draw[<->] (node0) -- (node1);
\draw[<->] (node0) -- (node2);
\draw[<->] (node2) -- (node1);
\draw[<->] (node2) -- (node3);
\draw[<->] (node3) -- (node5);
\draw[<->] (node3) -- (node6);
\draw[<->] (node5) -- (node6);
\draw[<->] (node7) -- (node6);
\draw[<->] (node7) -- (node5);
\draw[<->] (node5) -- (node4);
\draw[<->] (node2) -- (node4);
\draw[<->] (node8) -- (node4);
\draw[<->] (node8) -- (node4);
\draw[<->] (node8) -- (node9);
\draw[<->] (node4) -- (node9);
\draw[<->] (node5) -- (node8);
\end{tikzpicture}}\\
(a) & (b) & (c) \\[.25cm]
\scalebox{0.75}{
\begin{tikzpicture}{}
\node [circle,draw=yellow!50,fill=yellow!20,,thick,inner sep=0pt,minimum size=4mm] at (0,0) (node0) {0};
\node [circle,draw=yellow!50,fill=yellow!20,,thick,inner sep=0pt,minimum size=4mm] at (1,0) (node1) {1};
\node [circle,draw=yellow!50,fill=yellow!20,,thick,inner sep=0pt,minimum size=4mm] at (1,1) (node2) {2};
\node [circle,draw=red!50,fill=red!20,thick,inner sep=0pt,minimum size=4mm] at (1,2) (node3) {3};
\node [circle,draw=red!50,fill=red!20,thick,inner sep=0pt,minimum size=4mm] at (2.5,1) (node4) {4};
\node [circle,draw=black,fill=black,thick,inner sep=0pt,minimum size=4mm] at (2.5,2) (node5) {5};
\node [circle,draw=black,fill=black,thick,inner sep=0pt,minimum size=4mm] at (1.75,3) (node6) {6};
\node [circle,draw=black,fill=white,thick,inner sep=0pt,minimum size=4mm] at (3.25,3) (node7) {7};
\node [circle,draw=black,fill=black,thick,inner sep=0pt,minimum size=4mm] at (4.0,1.5) (node8) {8};
\node [circle,draw=yellow!50,fill=yellow!20,thick,inner sep=0pt,minimum size=4mm] at (4.0,0.5) (node9) {9};
\draw[<->] (node0) -- (node1);
\draw[<->] (node0) -- (node2);
\draw[<->] (node2) -- (node1);
\draw[<->] (node2) -- (node3);
\draw[<->] (node3) -- (node5);
\draw[<->] (node3) -- (node6);
\draw[<->] (node5) -- (node6);
\draw[<->] (node7) -- (node6);
\draw[<->] (node7) -- (node5);
\draw[<->] (node5) -- (node4);
\draw[<->] (node2) -- (node4);
\draw[<->] (node8) -- (node4);
\draw[<->] (node8) -- (node4);
\draw[<->] (node8) -- (node9);
\draw[<->] (node4) -- (node9);
\draw[<->] (node5) -- (node8);
\end{tikzpicture}} &
\scalebox{0.75}{
\begin{tikzpicture}{}
\node [circle,draw=yellow!50,fill=yellow!20,,thick,inner sep=0pt,minimum size=4mm] at (0,0) (node0) {0};
\node [circle,draw=yellow!50,fill=yellow!20,,thick,inner sep=0pt,minimum size=4mm] at (1,0) (node1) {1};
\node [circle,draw=yellow!50,fill=yellow!20,,thick,inner sep=0pt,minimum size=4mm] at (1,1) (node2) {2};
\node [circle,draw=red!50,fill=red!20,thick,inner sep=0pt,minimum size=4mm] at (1,2) (node3) {3};
\node [circle,draw=red!50,fill=red!20,thick,inner sep=0pt,minimum size=4mm] at (2.5,1) (node4) {4};
\node [circle,draw=black,fill=black,thick,inner sep=0pt,minimum size=4mm] at (2.5,2) (node5) {5};
\node [circle,draw=black,fill=black,thick,inner sep=0pt,minimum size=4mm] at (1.75,3) (node6) {6};
\node [circle,draw=black,fill=white,thick,inner sep=0pt,minimum size=4mm] at (3.25,3) (node7) {7};
\node [circle,draw=black,fill=black,thick,inner sep=0pt,minimum size=4mm] at (4.0,1.5) (node8) {8};
\node [circle,draw=black,fill=black!20,thick,inner sep=0pt,minimum size=4mm] at (4.0,0.5) (node9) {9};
\draw[<->] (node0) -- (node1);
\draw[<->] (node0) -- (node2);
\draw[<->] (node2) -- (node1);
\draw[<->] (node2) -- (node3);
\draw[<->] (node3) -- (node5);
\draw[<->] (node3) -- (node6);
\draw[<->] (node5) -- (node6);
\draw[<->] (node7) -- (node6);
\draw[<->] (node7) -- (node5);
\draw[<->] (node5) -- (node4);
\draw[<->] (node2) -- (node4);
\draw[<->] (node8) -- (node4);
\draw[<->] (node8) -- (node4);
\draw[<->] (node8) -- (node9);
\draw[<->] (node4) -- (node9);
\draw[<->] (node5) -- (node8);
\end{tikzpicture}} &
\scalebox{0.75}{
\begin{tikzpicture}{}
\node [circle,draw=yellow!50,fill=yellow!20,,thick,inner sep=0pt,minimum size=4mm] at (0,0) (node0) {0};
\node [circle,draw=yellow!50,fill=yellow!20,,thick,inner sep=0pt,minimum size=4mm] at (1,0) (node1) {1};
\node [circle,draw=yellow!50,fill=yellow!20,,thick,inner sep=0pt,minimum size=4mm] at (1,1) (node2) {2};
\node [circle,draw=red!50,fill=red!20,thick,inner sep=0pt,minimum size=4mm] at (1,2) (node3) {3};
\node [circle,draw=red!50,fill=red!20,thick,inner sep=0pt,minimum size=4mm] at (2.5,1) (node4) {4};
\node [circle,draw=black,fill=black,thick,inner sep=0pt,minimum size=4mm] at (2.5,2) (node5) {5};
\node [circle,draw=black,fill=black,thick,inner sep=0pt,minimum size=4mm] at (1.75,3) (node6) {6};
\node [circle,draw=black,fill=white,thick,inner sep=0pt,minimum size=4mm] at (3.25,3) (node7) {7};
\node [circle,draw=black,fill=black,thick,inner sep=0pt,minimum size=4mm] at (4.0,1.5) (node8) {8};
\node [circle,draw=black,fill=black,thick,inner sep=0pt,minimum size=4mm] at (4.0,0.5) (node9) {9};
\draw[<->] (node0) -- (node1);
\draw[<->] (node0) -- (node2);
\draw[<->] (node2) -- (node1);
\draw[<->] (node2) -- (node3);
\draw[<->] (node3) -- (node5);
\draw[<->] (node3) -- (node6);
\draw[<->] (node5) -- (node6);
\draw[<->] (node7) -- (node6);
\draw[<->] (node7) -- (node5);
\draw[<->] (node5) -- (node4);
\draw[<->] (node2) -- (node4);
\draw[<->] (node8) -- (node4);
\draw[<->] (node8) -- (node4);
\draw[<->] (node8) -- (node9);
\draw[<->] (node4) -- (node9);
\draw[<->] (node5) -- (node8);
\end{tikzpicture}}\\
(d) & (e) & (f) 
\end{tabular}
\end{center}
\caption{\label{fig:mcalg}Model-checking \emph{yellow}\;$\lsurr$\;\emph{red}}
\end{figure}
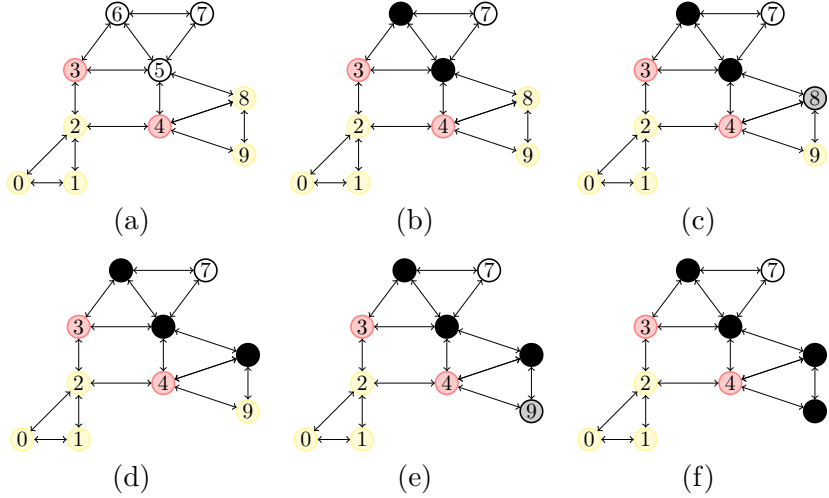

Function \checkUntil takes as parameters a finite, quasi-discrete closure model $\model$, and two \slcs\ formulas 
$\phi_1$ and $\phi_2$. The function  
computes the set of points in $\model$ satisfying $\phi_1 \lsurr \phi_2$. This 
is performed iteratively by removing from $V=\check(\model,\phi_1)$
points that we may intuitively call \emph{bad}. More precisely, a point is \emph{bad} if, in the underlying relation of the quasi-discrete closure model, there is a path rooted in it, reaching a point 
satisfying $\neg\phi_1$, without crossing any point satisfying $\phi_2$.
Let $Q=\check(\model,\phi_2)$ be the set of points in $\model$ satisfying 
$\phi_2$. 
To identify the \emph{bad} points in $V$ the function \checkUntil performs a \emph{backward search} from 
$T=\cboundary(V\cup Q)$. Note that any path leaving $V\cup Q$ must  
pass through points
in $T$. Moreover, $T$ only contains points that satisfy  neither
$\phi_1$ nor $\phi_2$. 
Until $T$ is empty, function \checkUntil first picks an element $x$ in $T$ and then removes from $V$ 
the set of (bad) points $N$ that can reach  $x$ in \emph{one step}.  
To compute the set $N$ we use the function $pre(x) = \{ y \in X \mid  (y,x) \in R \}$ 
$=\{y \in X \mid x \in \closure_{R}(\{y\})\}$.
At the end of each iteration the variable $T$ is updated by considering the set of 
newly 
discovered \emph{bad points}. Note that such new bad points do not include ``candidate bad points'' that also satisfy $\phi_2$. This is because any such point $x$ satisfies both formulas, thus every path starting from $x$ and reaching $\lnot \phi_1$ also passes (trivially) by a point satisfying $\phi_2$.
%
The evolution of this algorithm is illustrated in an informal way in~\autoref{fig:mcalg} for the formula \emph{yellow}\;$\lsurr$\;\emph{red}. At the beginning we have that $V=\{0,1,2,8,9\}$ and $Q=\{3,4\}$. Variable $T$ is initialized to the external boundary of $V\cup Q$, that is the set $\{ 5, 6 \}$. Points in $T$ are the black ones in \autoref{fig:mcalg}~(b). In the next step, yellow points that are neighbours of black ones (coloured in grey in \autoref{fig:mcalg}~(c)) are removed from $V$ and included in $T$ (see \autoref{fig:mcalg}~(d)). This ``refinement'' step is iterated until a fixed point is reached. The remaining yellow points are those satisfying \emph{yellow}\;$\lsurr$\;\emph{red} (see \autoref{fig:mcalg}~(f)).

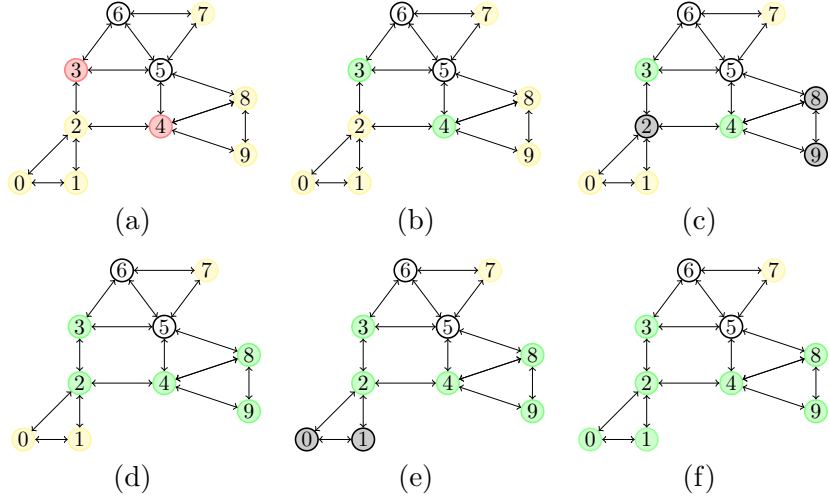
\begin{figure}[tbp]
\begin{center}
\begin{tabular}{ccc}
\scalebox{0.75}{\begin{tikzpicture}{}
\node [circle,draw=yellow!50,fill=yellow!20,,thick,inner sep=0pt,minimum size=4mm] at (0,0) (node0) {0};
\node [circle,draw=yellow!50,fill=yellow!20,,thick,inner sep=0pt,minimum size=4mm] at (1,0) (node1) {1};
\node [circle,draw=yellow!50,fill=yellow!20,,thick,inner sep=0pt,minimum size=4mm] at (1,1) (node2) {2};
\node [circle,draw=red!50,fill=red!20,thick,inner sep=0pt,minimum size=4mm] at (1,2) (node3) {3};
\node [circle,draw=red!50,fill=red!20,thick,inner sep=0pt,minimum size=4mm] at (2.5,1) (node4) {4};
\node [circle,draw=black,fill=white,thick,inner sep=0pt,minimum size=4mm] at (2.5,2) (node5) {5};
\node [circle,draw=black,fill=white,thick,inner sep=0pt,minimum size=4mm] at (1.75,3) (node6) {6};
\node [circle,draw=yellow!40,fill=yellow!20,thick,inner sep=0pt,minimum size=4mm] at (3.25,3) (node7) {7};
\node [circle,draw=yellow!50,fill=yellow!20,thick,inner sep=0pt,minimum size=4mm] at (4.0,1.5) (node8) {8};
\node [circle,draw=yellow!50,fill=yellow!20,thick,inner sep=0pt,minimum size=4mm] at (4.0,0.5) (node9) {9};
\draw[<->] (node0) -- (node1);
\draw[<->] (node0) -- (node2);
\draw[<->] (node2) -- (node1);
\draw[<->] (node2) -- (node3);
\draw[<->] (node3) -- (node5);
\draw[<->] (node3) -- (node6);
\draw[<->] (node5) -- (node6);
\draw[<->] (node7) -- (node6);
\draw[<->] (node7) -- (node5);
\draw[<->] (node5) -- (node4);
\draw[<->] (node2) -- (node4);
\draw[<->] (node8) -- (node4);
\draw[<->] (node8) -- (node4);
\draw[<->] (node8) -- (node9);
\draw[<->] (node4) -- (node9);
\draw[<->] (node5) -- (node8);
\end{tikzpicture}} &
\scalebox{0.75}{\begin{tikzpicture}{}
\node [circle,draw=yellow!50,fill=yellow!20,,thick,inner sep=0pt,minimum size=4mm] at (0,0) (node0) {0};
\node [circle,draw=yellow!50,fill=yellow!20,,thick,inner sep=0pt,minimum size=4mm] at (1,0) (node1) {1};
\node [circle,draw=yellow!50,fill=yellow!20,,thick,inner sep=0pt,minimum size=4mm] at (1,1) (node2) {2};
\node [circle,draw=green!50,fill=green!20,thick,inner sep=0pt,minimum size=4mm] at (1,2) (node3) {3};
\node [circle,draw=green!50,fill=green!20,thick,inner sep=0pt,minimum size=4mm] at (2.5,1) (node4) {4};
\node [circle,draw=black,fill=white,thick,inner sep=0pt,minimum size=4mm] at (2.5,2) (node5) {5};
\node [circle,draw=black,fill=white,thick,inner sep=0pt,minimum size=4mm] at (1.75,3) (node6) {6};
\node [circle,draw=yellow!40,fill=yellow!20,thick,inner sep=0pt,minimum size=4mm] at (3.25,3) (node7) {7};
\node [circle,draw=yellow!50,fill=yellow!20,thick,inner sep=0pt,minimum size=4mm] at (4.0,1.5) (node8) {8};
\node [circle,draw=yellow!50,fill=yellow!20,thick,inner sep=0pt,minimum size=4mm] at (4.0,0.5) (node9) {9};
\draw[<->] (node0) -- (node1);
\draw[<->] (node0) -- (node2);
\draw[<->] (node2) -- (node1);
\draw[<->] (node2) -- (node3);
\draw[<->] (node3) -- (node5);
\draw[<->] (node3) -- (node6);
\draw[<->] (node5) -- (node6);
\draw[<->] (node7) -- (node6);
\draw[<->] (node7) -- (node5);
\draw[<->] (node5) -- (node4);
\draw[<->] (node2) -- (node4);
\draw[<->] (node8) -- (node4);
\draw[<->] (node8) -- (node4);
\draw[<->] (node8) -- (node9);
\draw[<->] (node4) -- (node9);
\draw[<->] (node5) -- (node8);
\end{tikzpicture}} &
\scalebox{0.75}{
\begin{tikzpicture}{}
\node [circle,draw=yellow!50,fill=yellow!20,,thick,inner sep=0pt,minimum size=4mm] at (0,0) (node0) {0};
\node [circle,draw=yellow!50,fill=yellow!20,,thick,inner sep=0pt,minimum size=4mm] at (1,0) (node1) {1};
\node [circle,draw=black,fill=black!20,,thick,inner sep=0pt,minimum size=4mm] at (1,1) (node2) {2};
\node [circle,draw=green!50,fill=green!20,thick,inner sep=0pt,minimum size=4mm] at (1,2) (node3) {3};
\node [circle,draw=green!50,fill=green!20,thick,inner sep=0pt,minimum size=4mm] at (2.5,1) (node4) {4};
\node [circle,draw=black,fill=white,thick,inner sep=0pt,minimum size=4mm] at (2.5,2) (node5) {5};
\node [circle,draw=black,fill=white,thick,inner sep=0pt,minimum size=4mm] at (1.75,3) (node6) {6};
\node [circle,draw=yellow!40,fill=yellow!20,thick,inner sep=0pt,minimum size=4mm] at (3.25,3) (node7) {7};
\node [circle,draw=black,fill=black!20,thick,inner sep=0pt,minimum size=4mm] at (4.0,1.5) (node8) {8};
\node [circle,draw=black,fill=black!20,thick,inner sep=0pt,minimum size=4mm] at (4.0,0.5) (node9) {9};
\draw[<->] (node0) -- (node1);
\draw[<->] (node0) -- (node2);
\draw[<->] (node2) -- (node1);
\draw[<->] (node2) -- (node3);
\draw[<->] (node3) -- (node5);
\draw[<->] (node3) -- (node6);
\draw[<->] (node5) -- (node6);
\draw[<->] (node7) -- (node6);
\draw[<->] (node7) -- (node5);
\draw[<->] (node5) -- (node4);
\draw[<->] (node2) -- (node4);
\draw[<->] (node8) -- (node4);
\draw[<->] (node8) -- (node4);
\draw[<->] (node8) -- (node9);
\draw[<->] (node4) -- (node9);
\draw[<->] (node5) -- (node8);
\end{tikzpicture}}\\
(a) & (b) & (c) \\[.25cm]
\scalebox{0.75}{
\begin{tikzpicture}{}
\node [circle,draw=yellow!50,fill=yellow!20,,thick,inner sep=0pt,minimum size=4mm] at (0,0) (node0) {0};
\node [circle,draw=yellow!50,fill=yellow!20,,thick,inner sep=0pt,minimum size=4mm] at (1,0) (node1) {1};
\node [circle,draw=green!50,fill=green!20,,thick,inner sep=0pt,minimum size=4mm] at (1,1) (node2) {2};
\node [circle,draw=green!50,fill=green!20,thick,inner sep=0pt,minimum size=4mm] at (1,2) (node3) {3};
\node [circle,draw=green!50,fill=green!20,thick,inner sep=0pt,minimum size=4mm] at (2.5,1) (node4) {4};
\node [circle,draw=black,fill=white,thick,inner sep=0pt,minimum size=4mm] at (2.5,2) (node5) {5};
\node [circle,draw=black,fill=white,thick,inner sep=0pt,minimum size=4mm] at (1.75,3) (node6) {6};
\node [circle,draw=yellow!40,fill=yellow!20,thick,inner sep=0pt,minimum size=4mm] at (3.25,3) (node7) {7};
\node [circle,draw=green!50,fill=green!20,thick,inner sep=0pt,minimum size=4mm] at (4.0,1.5) (node8) {8};
\node [circle,draw=green!50,fill=green!20,thick,inner sep=0pt,minimum size=4mm] at (4.0,0.5) (node9) {9};
\draw[<->] (node0) -- (node1);
\draw[<->] (node0) -- (node2);
\draw[<->] (node2) -- (node1);
\draw[<->] (node2) -- (node3);
\draw[<->] (node3) -- (node5);
\draw[<->] (node3) -- (node6);
\draw[<->] (node5) -- (node6);
\draw[<->] (node7) -- (node6);
\draw[<->] (node7) -- (node5);
\draw[<->] (node5) -- (node4);
\draw[<->] (node2) -- (node4);
\draw[<->] (node8) -- (node4);
\draw[<->] (node8) -- (node4);
\draw[<->] (node8) -- (node9);
\draw[<->] (node4) -- (node9);
\draw[<->] (node5) -- (node8);
\end{tikzpicture}} &
\scalebox{0.75}{
\begin{tikzpicture}{}
\node [circle,draw=black,fill=black!20,,thick,inner sep=0pt,minimum size=4mm] at (0,0) (node0) {0};
\node [circle,draw=black,fill=black!20,,thick,inner sep=0pt,minimum size=4mm] at (1,0) (node1) {1};
\node [circle,draw=green!50,fill=green!20,,thick,inner sep=0pt,minimum size=4mm] at (1,1) (node2) {2};
\node [circle,draw=green!50,fill=green!20,thick,inner sep=0pt,minimum size=4mm] at (1,2) (node3) {3};
\node [circle,draw=green!50,fill=green!20,thick,inner sep=0pt,minimum size=4mm] at (2.5,1) (node4) {4};
\node [circle,draw=black,fill=white,thick,inner sep=0pt,minimum size=4mm] at (2.5,2) (node5) {5};
\node [circle,draw=black,fill=white,thick,inner sep=0pt,minimum size=4mm] at (1.75,3) (node6) {6};
\node [circle,draw=yellow!40,fill=yellow!20,thick,inner sep=0pt,minimum size=4mm] at (3.25,3) (node7) {7};
\node [circle,draw=green!50,fill=green!20,thick,inner sep=0pt,minimum size=4mm] at (4.0,1.5) (node8) {8};
\node [circle,draw=green!50,fill=green!20,thick,inner sep=0pt,minimum size=4mm] at (4.0,0.5) (node9) {9};
\draw[<->] (node0) -- (node1);
\draw[<->] (node0) -- (node2);
\draw[<->] (node2) -- (node1);
\draw[<->] (node2) -- (node3);
\draw[<->] (node3) -- (node5);
\draw[<->] (node3) -- (node6);
\draw[<->] (node5) -- (node6);
\draw[<->] (node7) -- (node6);
\draw[<->] (node7) -- (node5);
\draw[<->] (node5) -- (node4);
\draw[<->] (node2) -- (node4);
\draw[<->] (node8) -- (node4);
\draw[<->] (node8) -- (node4);
\draw[<->] (node8) -- (node9);
\draw[<->] (node4) -- (node9);
\draw[<->] (node5) -- (node8);
\end{tikzpicture}} &
\scalebox{0.75}{
\begin{tikzpicture}{}
\node [circle,draw=green!40,fill=green!20,,thick,inner sep=0pt,minimum size=4mm] at (0,0) (node0) {0};
\node [circle,draw=green!40,fill=green!20,,thick,inner sep=0pt,minimum size=4mm] at (1,0) (node1) {1};
\node [circle,draw=green!50,fill=green!20,,thick,inner sep=0pt,minimum size=4mm] at (1,1) (node2) {2};
\node [circle,draw=green!50,fill=green!20,thick,inner sep=0pt,minimum size=4mm] at (1,2) (node3) {3};
\node [circle,draw=green!50,fill=green!20,thick,inner sep=0pt,minimum size=4mm] at (2.5,1) (node4) {4};
\node [circle,draw=black,fill=white,thick,inner sep=0pt,minimum size=4mm] at (2.5,2) (node5) {5};
\node [circle,draw=black,fill=white,thick,inner sep=0pt,minimum size=4mm] at (1.75,3) (node6) {6};
\node [circle,draw=yellow!40,fill=yellow!20,thick,inner sep=0pt,minimum size=4mm] at (3.25,3) (node7) {7};
\node [circle,draw=green!50,fill=green!20,thick,inner sep=0pt,minimum size=4mm] at (4.0,1.5) (node8) {8};
\node [circle,draw=green!50,fill=green!20,thick,inner sep=0pt,minimum size=4mm] at (4.0,0.5) (node9) {9};\draw[<->] (node0) -- (node1);
\draw[<->] (node0) -- (node2);
\draw[<->] (node2) -- (node1);
\draw[<->] (node2) -- (node3);
\draw[<->] (node3) -- (node5);
\draw[<->] (node3) -- (node6);
\draw[<->] (node5) -- (node6);
\draw[<->] (node7) -- (node6);
\draw[<->] (node7) -- (node5);
\draw[<->] (node5) -- (node4);
\draw[<->] (node2) -- (node4);
\draw[<->] (node8) -- (node4);
\draw[<->] (node8) -- (node4);
\draw[<->] (node8) -- (node9);
\draw[<->] (node4) -- (node9);
\draw[<->] (node5) -- (node8);
\end{tikzpicture}}\\
(d) & (e) & (f) 
\end{tabular}
\end{center}
\caption{\label{fig:mcalprop}Model-checking \emph{red}\;$\ldiff$\;\emph{yellow}}
\end{figure}

Function \checkDiff takes as parameters a finite, quasi-discrete closure model $\model$, and two \slcs\ formulas 
$\phi_1$ and $\phi_2$. The function computes the set of points in $\model$ satisfying $\phi_1 \ldiff \phi_2$. Such computation is performed iteratively via a breadth-first search that starts from all the points in $V=\check(\model,\phi_1)$ and that traverses only points that are in $Q=\check(\model,\phi_2)$. To select points at the next level, the function $post(x) = \{y | (x,y)\in R\} = \closure_{R}(\{x\})$ is used.
%
The evolution of this algorithm is illustrated in an informal way in~\autoref{fig:mcalprop} for the formula \emph{red}\;$\ldiff$\;\emph{yellow}. First, all \emph{red} points -- which satisfy \emph{red}\;$\ldiff$\;\emph{yellow} -- are included in the set $R$ (the green points in \autoref{fig:mcalprop}~(b)). In the next step, the algorithm selects all the yellow points that are neighbours of an element in $R$ (\autoref{fig:mcalprop}~(c) and \autoref{fig:mcalprop}~(e)). These points are added to the set $R$ until a fixed point is reached (see \autoref{fig:mcalprop}~(f)). When the algorithm terminates, the points in $R$ are exactly the ones satisfying the considered formula. These are the green points in \autoref{fig:mcalprop}~(f).

\newcommand{\Sat}{{\tt Sat}}
\newcommand{\CC}{{\tt CC}}
\newcommand{\Flood}{{\tt Flood}}

The local model checking algorithm for \cslcs\ formulas is given in 
\autoref{alg:model-checking}. Function $\Sat_C$ takes as input a 
finite, quasi-discrete model $\model = ((X,\closure),\eval)$, a subset $A$ of 
$X$ and a 
collective formula $\psi$, and returns the truth value of $\model, A \models \psi$. 
The definition uses function $\Sat$ as defined 
above. The implementation of boolean operators is straightforward.  The case for $\phi \lShare \psi$ uses the global model checker $\Sat$ for individual formulas to compute the set of points satisfying $\phi$, and recursively checks if the intersection of such set with $A$ satisfies $\psi$. The case for $\lGroup \phi$ first performs some checks for corner cases of the definition, namely when $A$ is the empty set (then $\lGroup \phi$ is true), and when $A$ is not included in the set of points satisfying $\phi$ (then $\lGroup \phi$ is false). After this, a variant of the classical Tarjan's algorithm \cite{Tar72} for computing strongly connected components is executed on 
 the underlying graph of the space, starting from an 
arbitrary point of $A$.   The pseudo-code for such procedure is reported in \autoref{alg:check-group}.

More specifically, the difference between our algorithm and the classical procedure by Tarjan is that we only visit nodes in $B$ (that is, the semantics of $\phi$), and reachable from a chosen element $x$ of $A$, whereas the classical procedure visits all the nodes of the graph. This choice is motivated by the fact that we do not need to collect all the strongly connected components, but only to determine whether there is a strongly connected component, in the subgraph determined by $B$, that contains $A$. In the algorithm, $s$ is a stack; for simplicity we assume an operation $\mathit{popUntil(s,x)}$ that removes from a stack the most recently inserted elements including $x$, and returns the set of all such elements. Note that such set is only needed to compare it with $A$; this check can be efficiently implemented with a while loop that pops elements out of the stack and checks whether such elements belong to $A$, thus avoiding to store an additional set of possibly large size. Furthermore, $ll$ is a \emph{map} (the ``low link'' array of Tarjan's algorithm), indexed by elements of $X$. We omit the details of its implementation; clearly, if $X$ is enumerated by a contiguous subset of the natural numbers, a standard array can be used.

In order to address termination, complexity and correctness of our algorithms, we first
define the notion of \emph{size} of a formula.

\begin{definition}
 For $\phi$ a \slcs\ formula, let $size(\phi)$ be inductively defined as follows:
\begin{itemize}
\item $size(\top)=size(p)=1$
\item $size(\neg\phi)=size(\lnear\phi)=1+size(\phi)$
\item $size(\phi_1\wedge\phi_2)=size(\phi_1\lsurr 
\phi_2)=1+size(\phi_1)+size(\phi_2)$
\end{itemize}
 For $\psi$ a \cslcs\ formula, let $size(\psi)$ be inductively defined as follows:
\begin{itemize}
\item $size(\top)=1$
\item $size(\neg\psi)=size(\lGroup\psi)=1+size(\psi)$
\item $size(\psi_1\wedge\psi_2)= 1+size(\phi_1)+size(\phi_2)$ 
\item $size(\phi \lShare \psi)=1 + size(\phi) + size(\psi)$
\end{itemize}

\end{definition}

\begin{lemma}
\label{lemma:termination}
 For any finite quasi-discrete model $\model=((X,\closure_{R}),\eval)$ and 
\slcs\ formula $\phi$ of size $k$, 
 \check terminates  in $\mathcal{O}(k\cdot(|X|+|R|))$ steps.
\end{lemma}

\begin{theorem}\label{thm:sound-compl}
For any finite quasi-discrete closure model $\model=((X,\closure),\eval)$ and 
\slcs\ formula $\phi$, $x\in \check(\model,\phi)$
if and only if $\model,x\models \phi$.
\end{theorem}

%
%
%

%

\begin{theorem}\label{thm:check-collective}
For any finite, quasi-discrete closure model $\model = ((X,\closure_R), \eval)$, formula $\psi$ with $size(\psi) = k$, and $A \subseteq X$, we have $\check_C(\model, A, \psi) = True$ if and only if $\model, A 
\models \psi$, taking in the worst case $\mathcal{O}(k \cdot (|X| + |R|))$ steps.
\end{theorem}


\begin{figure}
\begin{tabular}{p{.48\textwidth}p{.48\textwidth}}
\small
\begin{algorithm}[H] 
\myfun{$\Sat_C(\model, A,\psi)$}{
  
  \KwIn{Finite, quasi-discrete closure model $\model = ((X,\closure_R),\eval)$, Set of 
points $A$, collective formula $\psi$}
  \KwOut{Truth value of $\model, A \models \psi$}
  
  \Match{$\psi$}
  {
     \lCase{$\top:$}{\Return $True$}
     \Case{$\lnot \psi:$}{
	\Let $R = \Sat_C(\model, A,\psi)$ \In\; 
	\Return \Neg $R$
      }
     \Case{$\psi_1 \land \psi_2:$}{
	\Let $R = \Sat_C(\model,A,\psi_1)$ \In\;
	\Let $S = \Sat_C(\model,A,\psi_2)$ \In\;
	\Return $R$ \Conj $S$
      }
     \Case{$\phi \lShare \psi_1:$}{
	\Let $B = \Sat(\phi) \cap A$ \In\;
	\Return $\Sat_C(\model,B,\psi_1)$	
      }
      \Case{$\lGroup \phi:$}{
	\lIf{$(A = \emptyset)	$}{\Return $\mathit{True}$}
        \Let $B = \Sat(\phi)$ \In\;
        \lIf{$(A \nsubseteq B)$}{\Return $\mathit{False}$}
        \Let $x \in A$ \In\;        
          \Let $t = \mathit{newCounter}()$ \In\;
          \Let $s = \mathit{newStack}()$ \In\;
          \Let $ll = \mathit{newMap}(X,\mathit{undefined})$ \In\;
          \Return $\visit(\model$,$t$,$s$,$ll$,$A$,$B$,$x)$
      } 
   }   
  }
\end{algorithm}
&
\small
\begin{algorithm}[H] 
        \myfun{\visit$(\model$,$t$,$s$,$ll$,$A$,$B$,$x)$} {
        \KwIn{Finite, quasi-discrete closure model $\model = ((X,\closure_R),\eval)$, counter $t$,
stack $s$, vector $ll$, sets of points $A$, $B$, point $x$ }
	\KwOut{Truth value or \emph{undefined}, depending on the progress of the algorithm when this auxiliary function is called.}
        \Var $\mathit{isRoot} := \mathit{True}$\;        
        \Var $r := \mathit{undefined}$\;
        $\mathit{push}(s,x)$\;
        $ll[x] := increment(t)$\;
        \For{$y \in post(x) \cap B$}{
	\If{$(ll[y] = \mathit{undefined})$}{$\,r:=$\visit$(\model$,$t$,$s$,$ll$,$A$,$B$,$y)$}
	\lIf{$(r \neq \mathit{undefined})$}{\Return $r$}
	\If{$(ll[x] > ll[y])$}{
	  $ll[x] := ll[y]$\;
	  $\mathit{isRoot} := \mathit{False}$
	}
 	}
	\If{$(\mathit{isRoot})$}{
	  \Let $C = \mathit{popUntil}(s,x)$ \;
	  \lIf{$(A \cap C \neq \emptyset)$}{$r := (A \subseteq C)$}
       }
       \Return $r$
     }
\end{algorithm} \\
\begin{algorithm}[H]\caption{\label{alg:model-checking}Algorithm for the model 
checking problem of 
CSLCS}
\end{algorithm}&
\begin{algorithm}[H]\caption{\label{alg:check-group}Checking 
\emph{group} 
formulas in a quasi-discrete closure space.}\end{algorithm}
\end{tabular}
\end{figure}

%
%

\section{A model checker for \slcs\ and \cslcs}\label{sec:discrete-examples}

The algorithms described in \autoref{sec:model-checking} are available as a proof-of-concept tool\footnote{Web site: \url{http://www.github.com/vincenzoml/topochecker}.}. The tool is implemented in OCaml\footnote{See \url{http://ocaml.org}.}, and can be invoked both as a global model checker for SLCS, or as a local model checker for CSLCS.

In the following we discuss a few examples showing how the tool can be used for identifying and analysing regions of interest of a digital image (e.g., a map, a medical image, a picture etc.), using spatial formulas. In this section, digital images are treated as finite, quasi-discrete models in the plane $\nats \times \nats$, equipped with the closure operator $4adj$ of \autoref{ex:imagesascs}.
Other topologies can be readily implemented in the tool. In the case of images, the tool accepts as atomic propositions expressions that denote sets of colours, so that each point $(x,y)$ satisfies precisely those expressions whose semantics includes the colour of the pixel at coordinates $(x,y)$. The \slcs\ model checker, which implements a global algorithm, accepts a formula $\phi$, a colour $c$ and a digital image, and colours with $c$ the points of the image satisfying $\phi$. The \cslcs\ model checker, which is a local algorithm, implements both \autoref{def:set-based:satisfaction} and \autoref{def:global-satisfaction}, accepting a collective formula $\psi$, and optionally\footnote{If no points are specified, the whole space is considered for global satisfaction.}a set of points, and returning a boolean answer. 

\begin{figure}
  \includegraphics[width=.80\textwidth,trim=-100 0 0 0]{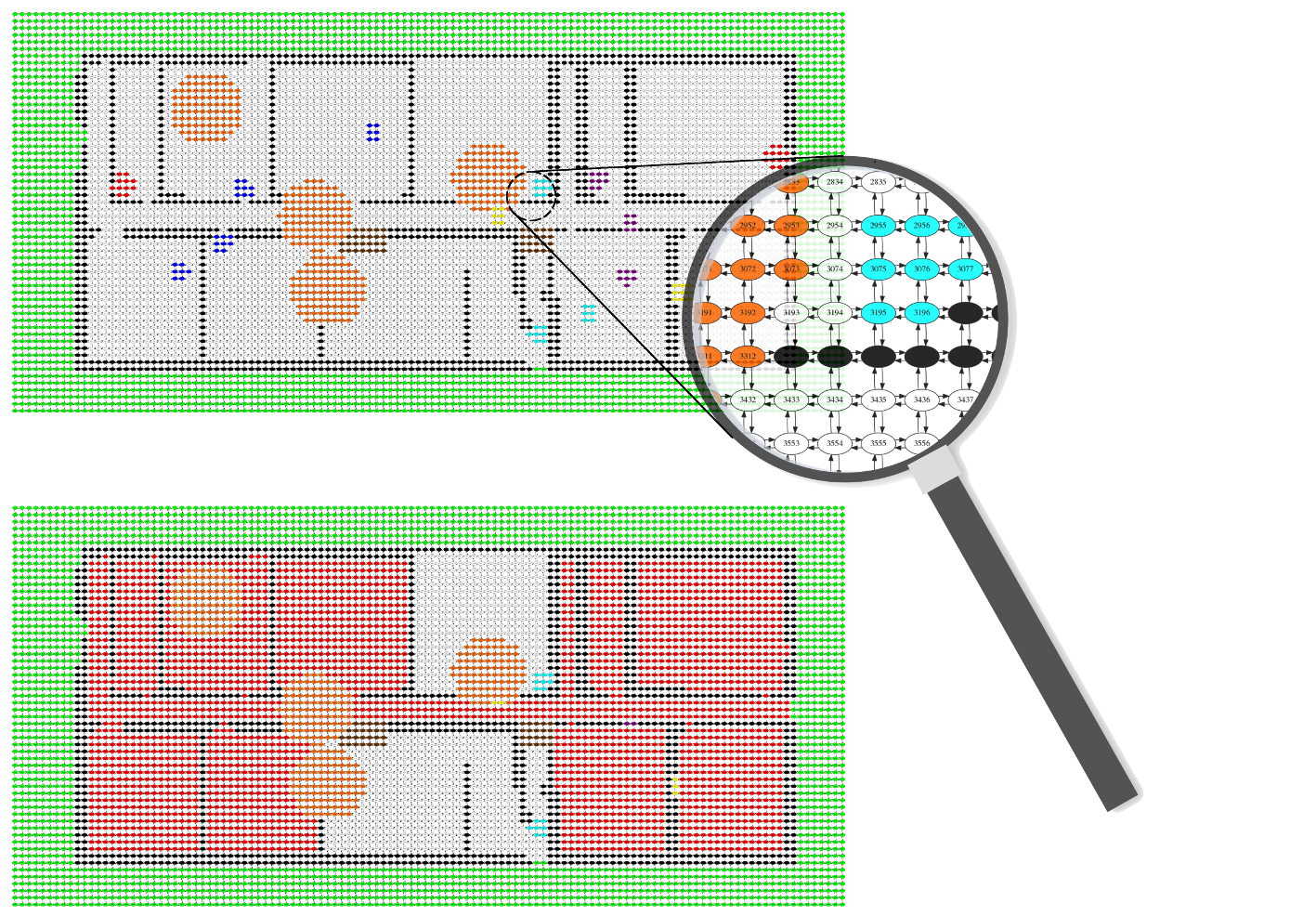}
 \caption{\label{fig:big-example}The model of \autoref{sec:EXAMPLES} rendered as a graph, and the result of model checking formula $\phi_5$ of \autoref{exa:safety-route}.}
\end{figure}

\begin{figure}[!!!tbp]
\centering
  \includegraphics[width=.4\textwidth]{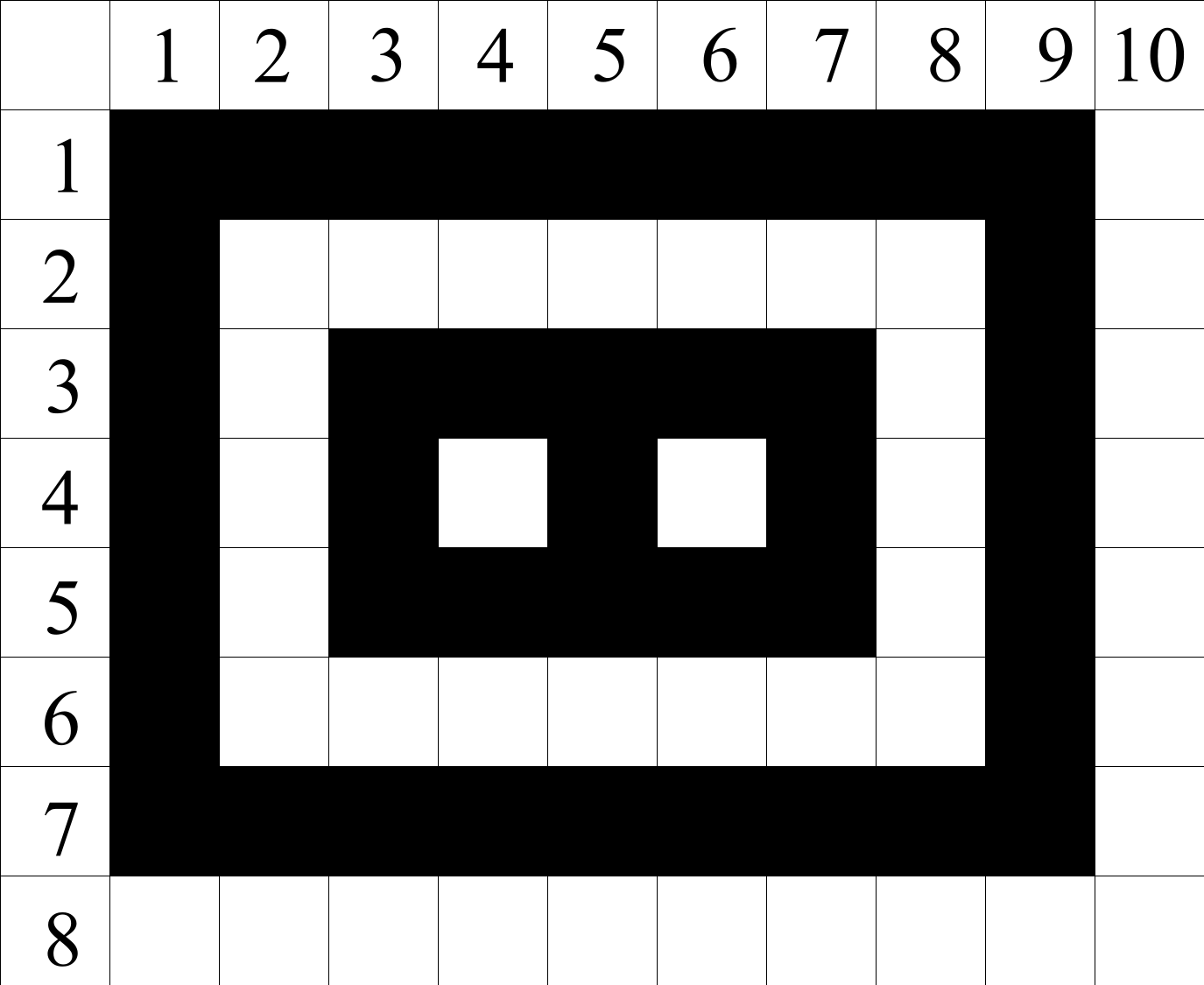}
\caption{\label{fig:collective-surrounded}A test case for the property of being \emph{collectively surrounded} (see \autoref{exa:model-checking-collectively-surrounded}). } 
\end{figure}

\begin{example}
 Finite, quasi-discrete models can be encoded as graphs. By this, the \cslcs\ model checker is able to load also examples with a complex specification. In \autoref{fig:big-example} we show a picture 
 coming from the analysis of the model of \autoref{sec:EXAMPLES} (above), and the output of the tool (below), colouring in red the nodes that satisfy $\phi_5$ from \autoref{exa:safety-route}. In particular, the model is based on a discrete version of a vectorial illustration. Execution times for checking the formulas presented in \autoref{sec:EXAMPLES} depend, indeed, on the resolution of the discrete image. Even though we do not aim at providing benchmarks in this work, just as a hint on execution times, we remark that when the number of points in the image is around one million, verification of formula $\phi_5$ takes around two seconds on a standard (at the time of writing) laptop with 8 gigabytes of main memory.
\end{example}

\begin{example}\label{exa:model-checking-collectively-surrounded}
In \autoref{fig:collective-surrounded} we provide a small, black and white image. 
For $A$ an arbitrary set of points, consider the informal statement ``$A$ is located in a \emph{white} area, and it is \emph{collectively surrounded} by a \emph{black} area''. The intuition here is that all the points of $A$ should be immersed in \emph{the same} white area. However, the meaning of \emph{same} is not thoroughly specified. 
%
 A very liberal interpretation of \emph{collectively surrounded} could let the formula be true at any set $A$ such that all points in $A$ individually satisfy $white \lsurr black$. This is expressed by the \cslcs\ formula $\lForall (white \lsurr  black)$. Here ``the same area'' means ``the same subset''. This notion can be refined. For example,  in \autoref{fig:collective-surrounded}, let $A = \{(4,4),(6,4)\}$ and $B = \{(4,6),(6,6)\}$ (the first coordinate is the horizontal one). It is also sensible to let ``collectively surrounded'' tell $A$ and $B$ apart, as $B$ lays in a \emph{connected} white area surrounded by black points, whereas $A$ does not enjoy such property.  In this case, ``the same area'' is defined as ``the same connected white area''. The derived connective $\lSSurr$ from \autoref{def:collective-more-derived-operators} is designed to do this.
 
The \cslcs\ model checker can be used to verify these two properties on some subsets of the space. First, we verify whether points at coordinates $(4,4),(4,6),(6,4),(6,6)$ individually satisfy $white \lsurr black$. This is checked by expanding the definition of the $\lForall$ connective, from \autoref{def:collective-derived-operators}. Indeed, the model checker answers \emph{true} to this query. The next step is to tell apart different sets of points using the definition of the $\lSSurr$ connective. The definition of $black\, \lSSurr \, white$ is checked on three different sets. The answer is \emph{true} on sets $\{(4,4)\}$ and $\{(4,6),(6,6)\}$ and \emph{false} on the sets $\{(4,4),(6,4)\}$ and $\{(4,4),(4,6)\}$.
\clsexa
\end{example}

\begin{figure}[!!!tbp]
\begin{tabular}{cc}
\small
 \begin{minipage}{.45\textwidth}
 \centering
  \includegraphics[width=.6\textwidth]{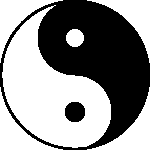}
 \end{minipage}
 &
 \begin{minipage}{.45\textwidth}
 \centering
  \includegraphics[width=.6\textwidth]{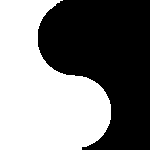}
 \end{minipage}
 \\
\begin{minipage}{.45\textwidth}
\caption{\label{fig:tao}This image is not partitioned by properties \emph{black} and \emph{white}.} 
\end{minipage}
&
\begin{minipage}{.45\textwidth}
\caption{\label{fig:not-tao}This image is partitioned by properties \emph{black} and \emph{white}.}
\end{minipage}
\end{tabular}
\end{figure}

\begin{example}
 Using \cslcs\ it is possible to check whether a given space is \emph{partitioned}, that is, each atomic property lies in a separate area of the image without mixing. We provided a formal definition of such property in \autoref{def:collective-more-derived-operators}, by the means of the $\lPartitioned$ connective. For example, we consider two digital images having only black and white pixels. The \cslcs\ model checker returns \emph{false} on \autoref{fig:tao}, and \emph{true} on \autoref{fig:not-tao}, when requested to verify that $white \lPartitioned black$ is globally satisfied according to \autoref{def:global-satisfaction}.
 %
\clsexa
\end{example}



\begin{figure}[!!!tbp]
\begin{tabular}{cc}
 \begin{minipage}{.45\textwidth}
 \centering
    \includegraphics[width=.8\textwidth]{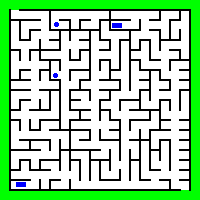}
 \end{minipage} & 
 \begin{minipage}{.45\textwidth}
 \centering
    \includegraphics[width=.8\textwidth]{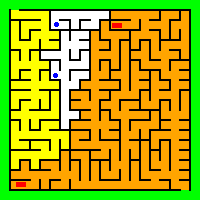}
 \end{minipage} \\
 \begin{minipage}{.45\textwidth}
 \caption{\label{fig:maze} A maze. }
 \end{minipage} &
 \begin{minipage}{.45\textwidth}
 \caption{\label{fig:maze-out} Model checker output. }
 \end{minipage}
\end{tabular}
\end{figure}
%

\begin{example}\label{exa:maze-individual}
In \autoref{fig:maze} we present another example of how \slcs\ can be used for classifying points in a digital image. We use a digital image representing a maze. The green area is the exit. The blue areas are starting points. Three formulas are used to identify interesting areas. Such formulas implicitly make use of the surrounded operator, by the means of the derived operators $\ldualuntil$ and $\mathcal{T}$ (see \autoref{sec:dsl}). 

$$\mathit{toExit} = white \,\mathcal{T}\, green$$
$$\mathit{fromStartToExit} = \mathit{toExit} \land (white \,\mathcal{T}\, blue)$$
$$\mathit{startCanExit}= blue \,\mathcal{T}\, \mathit{fromStartToExit}$$

The output of the tool is in \autoref{fig:maze-out}. The red colour denotes points satisfying $\mathit{startCanExit}$, that is, starting points from which the exit can be reached (for the sake of readability, we have depicted these areas in a rectangular shape, but the tool is obviously not aware of the difference in shape).  Orange and yellow indicate the two regions through which the exit can be reached (formula $toExit$). The orange region includes moreover a start point (formula $\mathit{fromStartToExit}$).
\clsexa
\end{example}

\begin{figure}[!!!!tbp]
\begin{tabular}{cc}
 \begin{minipage}{.45\textwidth}
 \centering
    \includegraphics[width=.8\textwidth]{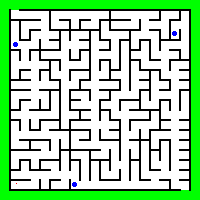}
 \end{minipage} &
 \begin{minipage}{.45\textwidth}
 \centering
    \includegraphics[width=.8\textwidth]{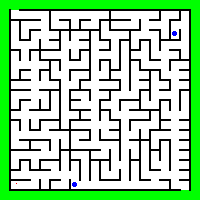}
 \end{minipage}
 \\
 \begin{minipage}{.45\textwidth}
 \caption{\label{fig:maze-disconnected} Blue circles are not able to reach the \emph{same} exit.}
 \end{minipage} &
\begin{minipage}{.45\textwidth}
 \caption{\label{fig:maze-connected} Blue circles are able to reach the \emph{same} exit.}    
 \end{minipage}
\end{tabular}
\end{figure}

\begin{example}\label{exa:maze-collective}
 We continue from \autoref{exa:maze-individual} to show how collective formulas can easily distinguish models having similar individual properties. In \autoref{fig:maze-disconnected}, the three blue circles in the maze can all reach the exit; however, they cannot ``collectively'' do so, as they cannot join and get out through the same exit. In \autoref{fig:maze-connected}, on the other hand, the blue circles \emph{can} get out through the same exit. Importing definitions from \autoref{exa:maze-individual}, the model checker is able to tell the difference between these two models. When invoked on the formula $blue \lShare (\lGroup ((blue \lor white) \,\mathcal{T}\, green))$, the tool returns $false$ in the first model, and $true$ in the second one.
%
%
 The given formula, which is interpreted globally (in the sense of \autoref{def:global-satisfaction}), asserts that all the blue points are part of a strongly connected component of points that can reach the (green) exit passing by points that are either blue or white.
\clsexa
\end{example}

%
%

\begin{figure}[!!tbp]

\begin{center}
\begin{minipage}{.45\textwidth}
\centering
 \includegraphics[width=.8\textwidth]{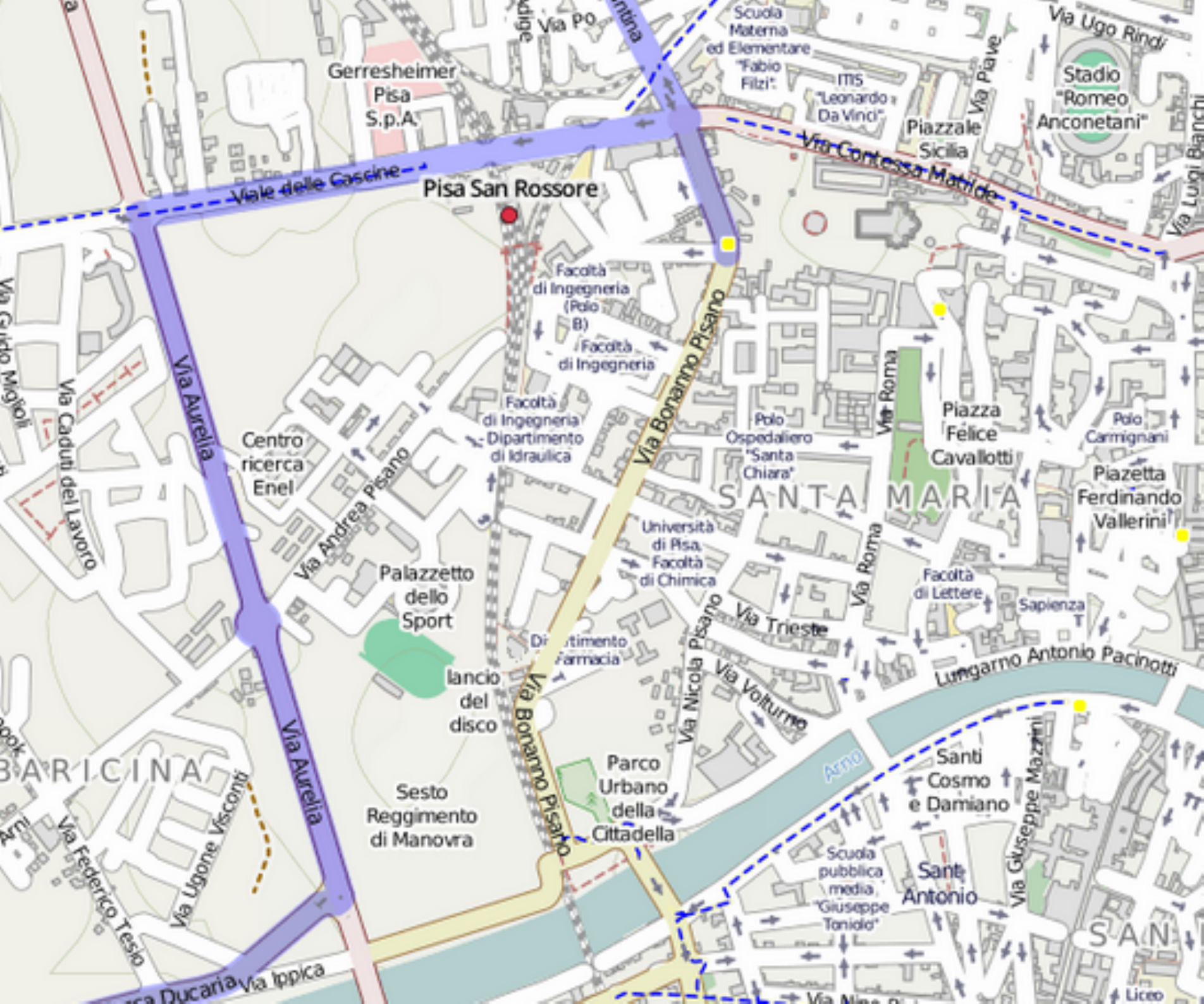}
 \caption{\label{fig:openstreetmap-pisa} Input: the map of a town.}
\end{minipage}
\begin{minipage}{.45\textwidth}
\centering
 \includegraphics[width=.8\textwidth]{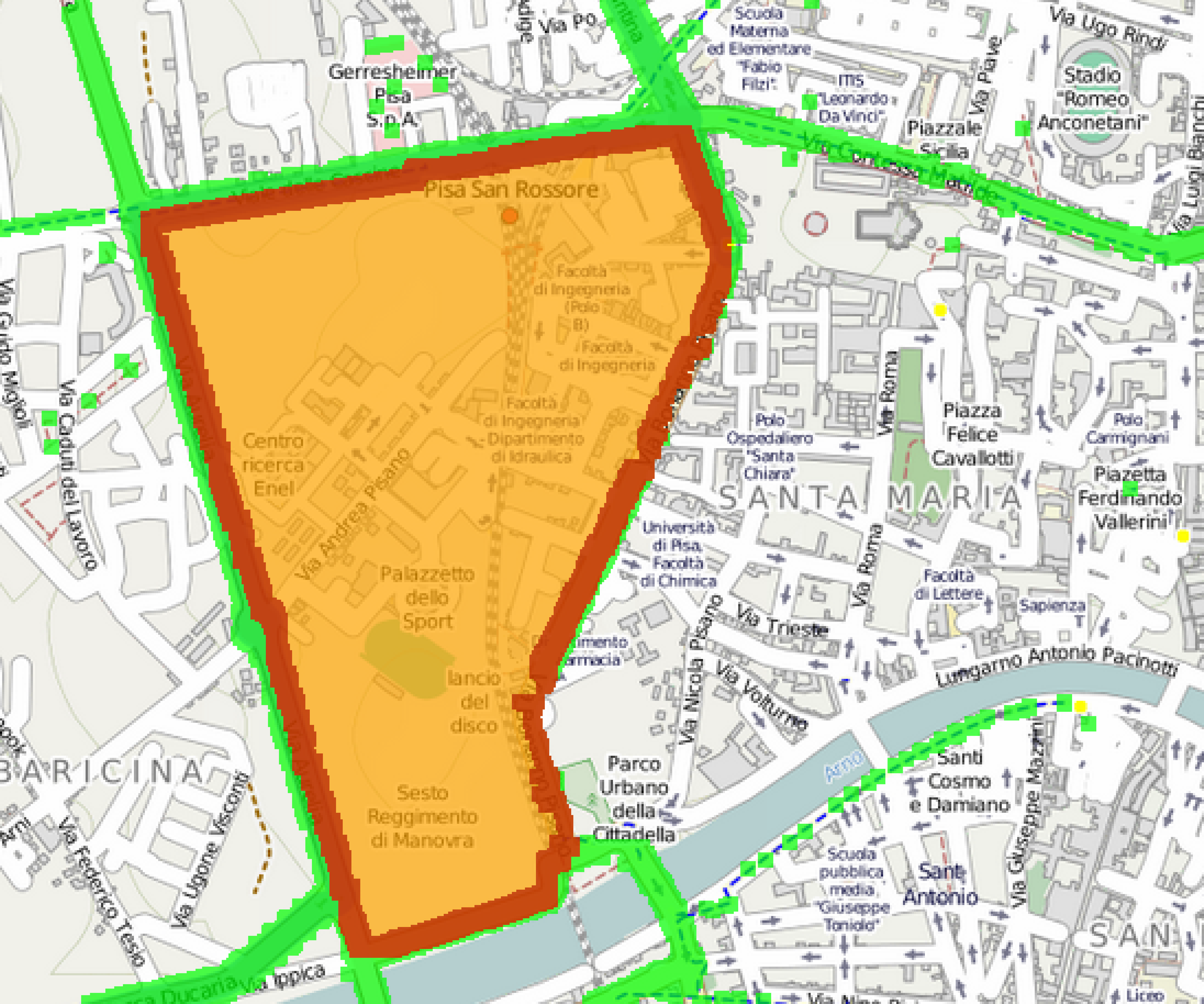}
 \caption{\label{fig:openstreetmap-pisa-out} Output of the tool.}
\end{minipage}
\end{center}
\end{figure}

\begin{example}\label{exa:openstreetmap-pisa}
In \autoref{fig:openstreetmap-pisa} we show a digital image\footnote{\copyright \emph{OpenStreetMap contributors} -- \url{http://www.openstreetmap.org/copyright}.} depicting a portion of the map of Pisa, featuring a red circle which denotes a train station. Streets of different importance are painted with different colours in the map. The \cslcs\ model checker is used to identify and colour the area surrounding the station which is delimited by main streets, 
including 
the delimiting main streets. The output of the tool is shown in \autoref{fig:openstreetmap-pisa-out}, where the station area is coloured in orange, the surrounding main streets are red, and other main streets are in green. 
\clsexa
\end{example}

%

%

\section{Conclusions and Future Work}\label{sec:discussion}

Spatial logics have been studied extensively in the past as a spatial interpretation of modal logics~\cite{HBSL}, with particular emphasis on descriptive languages and aspects such as completeness, decidability, complexity, that are very relevant for mathematical logics. In this paper we have developed this approach in a different direction, namely that of formal and automatic verification and in particular that of spatial model-checking. This focus required us to take several constraints into consideration. On one hand, our aim was to remain as general as possible, in such a way that the developed spatial model checking algorithms can be applied on a wide variety of spatial representations, including forms of continuous space, discrete space, directed and undirected graphs, possibly extended with metric spaces. On the other hand, efficient and effective model checking procedures require finite structures. To this purpose the theoretical framework of closure spaces (a generalisation of topological spaces) has been explored. This framework provides a set of useful basic abstract spatial operators (closure, interior, boundary and many derived ones) that provide a structured way to define higher level spatial logic operators. Moreover, we have shown that they are also suitable for the development of efficient spatial model checking algorithms in which these same closure space based operators play a role as well.

%
In particular, in~\cite{Ci+14b} we have defined the spatial logic \slcs, stemming from the tradition of topological interpretations of 
modal logics, dating back to earlier logicians such as Tarski, where modalities describe neighbourhood. 
The topological definitions have been lifted to a more general setting, also encompassing discrete, graph-based structures. 
In the present paper an alternative, path-based, definition of the logic has been provided which is more general than that presented in~\cite{Ci+14b} and is shown to coincide with the latter in the case of quasi-discrete closure spaces. In addition, 
the framework has been extended with the {\em propagation} operator. This operator captures the notion of spatial propagation;
intuitively the formula $\phi \, \ldiff \, \psi$ describes a situation in which the points satisfying $\psi$ can be reached 
by paths rooted in points satisfying $\phi$ and, for the rest, composed only of points satisfying $\psi$.

Furthermore, we have introduced a \emph{collective} logic, which borrows from the spatial logics tradition, but introduces properties that characterise ``collective'', spatial features of \emph{sets}, rather than individuals. For both logics, an efficient model-checking algorithm has been defined and implemented, operating on finite, quasi-discrete closure models.

Future work aims at considering temporal reasoning in addition to spatial verification in order to address system evolution and dynamics within a single logic. Both the theoretical nature of this problem, and the efficiency of model checking algorithms, should be investigated.  In \cite{KKWZ07}, ``snapshot'' models are considered, consisting of a temporal model (e.g., a Kripke frame) where each state is in turn a spatial model, and atomic formulas of the temporal fragment are replaced by spatial formulas. The various possible combinations of temporal and spatial operators, in linear and branching time, are examined therein, for the case of topological models, and basic modal formulas. First results on the extension of snapshot models based on \emph{closure} spaces, and the study of spatio-temporal surrounded operators, has led to an extension of \slcs\ with the branching time logic CTL (Computation Tree Logic ~\cite{ClE82}) and is presented in~\cite{CGLLM15,CLMP15,CLLM16}. It provides spatio-temporal reasoning and model checking. However, the automated verification of snapshot models is susceptible to state-space explosion problems as spatial formulas need to be recomputed at every state. We will therefore also study how to exploit the fact that changes of space
over time are typically incremental and local in nature. Metrics and distance functions can be added in an orthogonal way providing further spatial richness. The theoretical approach pursued in the present paper is starting to find its way to applications such as the detection and analysis of emergent spatial patterns~\cite{NBCLM15} in behaviour modelled as reaction-diffusion equations, such as those involved in the emergence of patterns in animal fur first studied by Turing. In ~\cite{NBCLM15} the closure space based model checking algorithms have been extended with metric spaces and signal temporal logic leading to monitoring algorithms for a linear time spatio-temporal logic. The logic has qualitative and quantitative semantics, and monitoring algorithms have been designed
and implemented. It can be used to verify interesting spatial-temporal properties such as the robustness of patterns to perturbations.
Other ongoing applications of spatio-temporal model checking are the analysis of emergent spatio-temporal phenomena, such as 
the phenomenon of \emph{clumping} (that is, buses with too short headway) in public urban bus transportation systems~\cite{CGLLM14} and the formation of spatial clusters of full stations in bike sharing systems~\cite{CLMP15}. The latter has been also analysed in \cite{ISOLA16}, where statistical spatio-temporal model checking has been used to infer quantitative information like, for example, the probability of cluster formation. In a completely different domain, preliminary work showed very interesting results, combining spatial model checking with \emph{texture analysis} to segment tumour and oedema in medical images, which is of immediate relevance for automatic contouring applications used in radiotherapy~\cite{BCLM16}.

Further promising ideas are presented both in \cite{Gal03}, where principles of ``continuous change'' are proposed in the setting of closure spaces, and in \cite{KM07} where spatio-temporal models are generated by locally-scoped update functions, in order to describe dynamic systems. Another interesting alternative approach to describe spatial properties is based on so-called quad trees. Such trees are constructed by recursively partitioning images into quadrants. The spatio-temporal logic SpaTeL~\cite{bartocci2015} is based on such a spatial superposition logic. 

In the setting of collective adaptive systems, it would be relevant to extend the basic framework we presented here with aspects related to distances or metrics (e.g., distance-bounded variants of the surrounded and propagation operators) and probabilistic aspects, using, e.g., atomic propositions that are probability distributions. 
%
In this work we have considered connectedness and related properties as the most basic forms of collective properties. Indeed, such properties give the logic a ``global'' flavour, witnessed by our \autoref{def:global-satisfaction}. In this respect, \cslcs\ is similar to $\logicSfourU$, even though connectedness can not be represented in the latter. Connectedness may be added as a predicate to spatial logics (see \cite{KPWZ10}). An in-depth comparison between $\logicSfourU$, spatial logics with connectedness, and \cslcs\ will be considered in future work,  possibly taking into account the work of \cite{Sla09} on connectedness in closure spaces. Other logics that consider sets of points rather than properties of individual points are those based on adjacency spaces such as the region calculus studied in discrete mereotopology~\cite{Gal99,Gal14}. In the context of collective properties, one could also consider arbitrary nesting of collective and individual formulas; however, such enhancements translate to inefficient algorithms in the classical exhaustive model checking procedures, as one should enumerate all subsets of the considered set of points. In order to overcome such issues, a \emph{symbolic model checking} approach could be used to represent solution sets without explicit enumeration.
%

A further challenge in spatial and spatio-temporal reasoning is posed by recursive spatial formulas, \emph{a la} $\mu$-calculus, especially on infinite structures with relatively straightforward generating functions (think of fractals, or fluid flow analysis of continuous structures). Such infinite structures could be described by topologically enhanced variants of $\omega$-automata; more generally speaking, the automata-theoretic approach to logics and verification is certainly of interest also in the field of spatial logics. Classes of automata exist living in specific topological structures; an example is given by \emph{nominal automata} (see e.g., \cite{BKL11,GC11,KST12}), that can be defined using presheaf toposes \cite{fs06}, although retaining finite, computationally efficient representations \cite{CM10}.
 This standpoint could be enhanced with notions of neighbourhood coming from closure spaces, with the aim of developing a unifying theory of languages and automata describing physical spaces, graphs, and process calculi with resources. Finally, a more profound study of the generalisation of the notion of paths in closure spaces could lead to further interesting theoretical results.
 
\section{Acknowledgements} This research has been partially funded by the EU FET Proactive project QUANTICOL (nr. 600708). The authors also wish to thank the anonymous reviewers for their valuable comments and suggestions.

\bibliographystyle{alpha}
\bibliography{all.bib}

\appendix

\section{Proofs}\label{sec:proofs}

The proofs of \autoref{lem:interior-monotone}, \autoref{pro:boundary-properties}, and \autoref{lem:subspace-closure-is-closure} are straightforward, and have been omitted from this paper. Full proofs are available in \cite{TRQC062014}.

\begin{proof}(of \autoref{pro:closure-space-of-a-relation})
\medskip

  Axiom \ref{def:closure-space:closure-of_emptyset}:
 $\der \closure_R(\emptyset) = \emptyset \cup \{ x \in X \mid \exists a \in \emptyset . (a,x) \in R\} = \emptyset$
 \medskip
 
 Axiom \ref{def:closure-space:closure-larger}:
 $\der A 
  \jstp \subseteq {A \subseteq A \cup B} \closure_R(A)$
 \medskip
 
 Axiom \ref{def:closure-space:closure-union}:
 $\der \closure_R(A \cup B) 
  \stp = A \cup B \cup \{ x \in X \mid \exists c \in A \cup B . (c,x) \in R \} 
  \jstp = { c \in A \cup B \iff c \in A \lor c \in B }
         A \cup B \cup \{ x \in X \mid \exists c \in A . (c,x) \in R \} \cup \{ x \in X \mid \exists c \in B . (c,x) \in R \} 
  \stp =  \closure_R(A) \cup \closure_R(B)
 $
\end{proof}

\begin{proof}(of \autoref{pro:interior-boundary-in-quasi-discrete})

 \autoref{eqn:boundary-quasi-discrete-1}:
 $
 \der \interior(A) 
 \stp = \overline{\closure_R(\overline A)}
 \stp = \overline{\overline A \cup \{x \in X \mid \exists a \in \overline A . (a,x) \in R \}}
 \stp = A \cap \{ x \in X \mid  \lnot \exists a \in \overline A . (a,x) \in R\} 
 \stp = \{ x \in A \mid \lnot \exists a \in \overline A . (a,x) \in R \}
 $
 
 \autoref{eqn:boundary-quasi-discrete-2}:
 $
 \der \iboundary(A) 
 \stp = A \setminus \interior(A) 
 \stp = A \setminus \{ x \in A \mid \lnot \exists a \in \overline A . (a,x) \in R \}
 \stp = A \cap \{ x \in A \mid \exists a \in \overline A . (a,x) \in R \}
 \stp = \{ x \in A \mid \exists a \in \overline A . (a,x) \in R \}
 $
 
 \autoref{eqn:boundary-quasi-discrete-3}:
 $
 \der \cboundary(A) 
 \stp = \closure(A) \setminus A 
 \stp = (A \cup \{ x \in X \mid \exists a \in A . (a,x) \in R \} ) \setminus A 
 \stp = (A \cup \{ x \in X \mid \exists a \in A . (a,x) \in R \}) \cap \overline A
 \stp = (A \cap \overline A) \cup (\{ x \in X \mid \exists a \in A . (a,x) \in R \} \cap \overline A) 
 \stp = \{ x \in \overline A \mid \exists a \in A . (a,x) \in R \}
 $
\end{proof}

\begin{proof}(of \autoref{lem:paths-are-paths})
 For one direction of the proof, assume $p$ is a continuous function. Importing definitions from \autoref{def:path} and the statement of \autoref{lem:paths-are-paths}, we have 
 $\der
 (i,i+1) \in \succ 
 \stp \implies i + 1 \in \closure_\succ(\{i\})
 \jstp \implies{ p $ continuous$} p(i+1) \in \closure_R(p(\{i\})) 
 \stp \iff p(i+1) \in \closure_R(\{p(i)\}) 
 \stp \iff p(i+1) \in \{ p(i) \} \cup \{ x \mid (p(i),x) \in R \}
 \stp \iff p(i+1) = p(i) \lor (p(i),p(i+1)) \in R
 $
 
 ~ \\
 \noindent For the other direction, given a path $x_{i \in I}$ $R$, define $p(i) = x_i$. Continuity of $p$ is straightforward.
\end{proof}

\begin{proof}(of \autoref{remark:duals})
\begin{enumerate}
\item 
$
\model,x \models \phi_1 \ldualuntil \phi_2
\jstp \iff {\mbox{Definition of $\ldualuntil$}}
 \model,x \models \neg (\neg\phi_2 \lsurr \neg\phi_1)
\jstp \iff {\mbox{Definition of $\lsurr$}}
\neg (\model,x\models \neg\phi_2 \mbox{ and } \crstp \forall p : x \pto {}{}\forall l\in \nats: \model, p(l) \models \lnot \neg\phi_2 \Rightarrow \exists k\in \{1, \ldots, l\}: \model,p(k) \models \neg\phi_1)
\stp \iff
\neg (\model,x\models \neg\phi_2 \mbox{ and } \crstp \forall p : x \pto {}{}\forall l\in \nats:~ \neg(\model, p(l) \models \phi_2) \vee (\exists k\in \{1, \ldots, l\}: \model,p(k) \models \neg\phi_1))
\stp \iff
\model,x\models \phi_2 \mbox{ or } \crstp \exists p : x \pto {}{}\exists l\in \nats:~ \model, p(l) \models \phi_2 \wedge \neg(\exists k\in \{1, \ldots, l\}: \model,p(k) \models \neg\phi_1)
\stp \iff
\exists p : x \pto {}{}\exists l\in \nats:~ \model, p(l) \models \phi_2 \wedge \forall k \in \{1, \ldots, l\}: \model,p(k) \models \phi_1
$
\item 
$
\model,x \models \phi_1 \ldualdiff \phi_2
\jstp \iff {\mbox{Definition of $\ldualdiff$}}
 \model,x \models \neg (\phi_1 \ldiff \neg\phi_2)
\jstp \iff {\mbox{Definition of $\ldiff$}}
\neg (\model,x\models \neg\phi_2 \mbox{ and } \crstp \exists y: \model,y\models \phi_1 \mbox{ and }\exists p : y \pto {l}{x}:\forall i\in \{ 1,\ldots,l-1\}: \model, p(i) \models \neg\phi_2)
\stp \iff
\model,x\models \phi_2 \mbox{ or } \crstp \neg(\exists y: \model,y\models \phi_1 \mbox{ and }\exists p : y \pto {l}{x}:\forall i\in \{ 1,\ldots,l-1\}: \model, p(i) \models \neg\phi_2)
\stp \iff
\model,x\models \phi_2 \mbox{ or } \crstp \forall y: \neg(\model,y\models \phi_1 \mbox{ and }\exists p : y \pto {l}{x}:\forall i\in \{ 1,\ldots,l-1\}: \model, p(i) \models \neg\phi_2)
\stp \iff
\model,x\models \phi_2 \mbox{ or } \crstp \forall y: \neg\model,y\models \phi_1 \mbox{ or } \neg(\exists p : y \pto {l}{x}:\forall i\in \{ 1,\ldots,l-1\}: \model, p(i) \models \neg\phi_2)
\stp \iff
\model,x\models \phi_2 \mbox{ or } \crstp \forall y: \neg\model,y\models \phi_1 \mbox{ or } \forall p : y \pto {l}{x}:\neg(\forall i\in \{ 1,\ldots,l-1\}: \model, p(i) \models \neg\phi_2)
\stp \iff
\model,x\models \phi_2 \mbox{ or } \crstp \forall y: \neg\model,y\models \phi_1 \mbox{ or } \forall p : y \pto {l}{x}:\exists i\in \{ 1,\ldots,l-1\}: \neg\model, p(i) \models \neg\phi_2)
\stp \iff
\model,x\models \phi_2 \mbox{ or } \crstp \forall y: \neg\model,y\models \phi_1 \mbox{ or } \forall p : y \pto {l}{x}:\exists i\in \{ 1,\ldots,l-1\}: \model, p(i) \models \phi_2)
\stp \iff
\model,x\models \phi_2 \mbox{ or } \crstp \forall y: \model,y\models \phi_1 \Rightarrow \forall p : y \pto {l}{x}:\exists i\in \{ 1,\ldots,l-1\}: \model, p(i) \models \phi_2)
\stp \iff
\model,x\models \phi_2 \mbox{ or } \crstp \forall y: \model,y\models \phi_1: \forall p : y \pto {l}{x}:\exists i\in \{ 1,\ldots,l-1\}: \model, p(i) \models \phi_2)
$
\item 
$
\model,x \models \leverywhere \phi
\jstp \iff {\mbox{Definition of $\leverywhere$}}
\model,x \models \phi \lsurr \bot
\jstp \iff {\mbox{Definition of $\lsurr$}}
\forall p : x \pto {}{}\forall l\in \nats: \model, p(l) \models \lnot \phi \Rightarrow \exists k \in \{1, \ldots, l\}: \model,p(k) \models \bot
\stp \iff
\forall p : x \pto {}{}\forall l\in \nats: \model, p(l) \models \phi
$  
\item 
$
\model,x \models \lsomewhere \phi
\jstp \iff {\mbox{Definition of $\lsomewhere$}}
\model,x \models \neg \leverywhere \neg\phi
\jstp \iff {\mbox{\autoref{remark:duals}~(3)}}
\neg (\forall p : x \pto {}{}\forall l\in \nats: \model, p(l) \models \neg\phi)
\stp \iff
\exists p : x \pto {}{}\exists l\in \nats: \model, p(l) \models \phi
$\vspace{-19.5 pt}
\end{enumerate}
\end{proof}

\begin{proof}(of \autoref{thm:other-side-of-simple-discrete-paths-until-S})
Consider a quasi-discrete closure model $\model = ((X,\closure),\eval)$ and suppose $\model, x \models \phi_1 \mathcal{U} \phi_2$ as defined in \cite{Ci+14b}, that is, suppose there is a set $A$ with $x \in A$, $\forall y \in A . \model, y \models \phi_1$, and $\forall z \in \cboundary(A) . \model, z \models \phi_2$. Let $p$ be a $\mathfrak{N}$-path, with $p : x \pto{}{}$, and let $l$ be such that $\model, p(l) \models \lnot \phi_1$. Consider the set $K^- = \{ k \mid \forall h \in \{0,\ldots,k\} . p(h) \in A \}$.  Since $0 \in K^-$, we have $K^- \neq \emptyset$. Consider the complement of $K^-$, namely $K^+ = \nats \setminus K^-$.  Since all points in $A$ satisfy $\phi_1$, and $p(l) \models \lnot \phi_1$, we have $l \in K^+$, thus $K^+ \neq \emptyset$. By existence of $l$, $K^-$ is finite, thus, being non-empty, it has a greatest element. Being a non-empty subset of the natural numbers, $K^+$ has a least element. Let $k^- = \max K^-$ and $k^+ = \min K^+$. By definition of $K^-$, if $k \in K^-$ and $h \in [0,k)$, then $h \in K^-$. In particular, for all $h \leq k^-$, we have $h \in K^-$. By definition of $k^-$, we have $k^- + 1 \notin K^-$, that is, $k^- + 1 \in K^+$. Therefore, we have $k^- + 1 = k^+$, thus $(k^-,k^+) \in \succ$. 
 Let $S = \{ p(k) | k \in K^- \} \subseteq A$. By monotonicity of closure, we have $\closure(S) \subseteq \closure(A)$. By definition of $\closure_\succ$, we have $k^+ \in \closure_\succ(K^-)$, thus by closure-continuity $p(k^+) \in \closure(S)$ and therefore $p(k^+) \in \closure(A)$.  But it is also true that $p(k^+) \notin A$; if $p(k^+) \in A$, then we would have $k^+ \in K^-$, by definition of $K^-$. Thus, $p(k^+) \in \cboundary(A)$, therefore $p(k^+) \models \phi_2$. Note that in particular $k^+ \neq 0$ as $p(0) = x \in A$, and $k^+ \leq l$ as $l \in K^+$ and $k^+ = \min K^+$.

For the other direction, assume $\model = ((X,\closure_R),\eval)$ where $\closure_R$ is the closure operator derived by a relation $R$. Consider point $x$ with $\model, x \models \phi_1$, and assume that for each $p : x \pto{}{}$ and $l$ such that $\model, p(l) \models \lnot \phi_1$ there is $k \in \{1, \ldots, l\}$ such that $\model, p(k) \models \phi_2$. 
Define the following set: 
$$A_x = \{x\} \cup \{ y \in X \mid \exists p : x \pto{}{} . \exists l > 0. p(l) = y \land \forall k \in \{1,\ldots,l\} . \model, p(k) \models \phi_1 \land \lnot \phi_2\}$$

 We will use $A_x$ as a witness of the existence of a set $A$, in order to prove that $\model, x \models \phi_1 \mathcal{U} \phi_2$ according to \cite{Ci+14b}. Note that by definition of $A_x$, $x \in A_x$ and $\forall y \in A_x . \model, p(y) \models \phi_1$. We need to show that $\forall z \in \cboundary(A_x) . \model, z \models \phi_2$. Consider $z \in \cboundary(A_x)$. Since $\model$ is based on a quasi-discrete closure space, by \autoref{eqn:boundary-quasi-discrete-3} in \autoref{pro:interior-boundary-in-quasi-discrete}, we have $z \in \overline{A_x}$ and there is $y \in A_x$ such that $(y,z) \in R$. Suppose $y = x$. Let $p$ be the path defined by $p(0) = x$, $p(i \neq 0) = z$. If $\model, z \models \phi_1$, suppose $\model, z \nmodels \phi_2$; then $z \in A_x$, witnessed by the path $p$, with $l=1$; therefore, since $z \in \overline {A_x}$ we have $\model, z \models \phi_2$. If $\model, z \nmodels \phi_1$, then noting $p(1) = z$, by hypothesis, there is $k \in \{1,\ldots,1\}$ with $\model, p(k) \models \phi_2$, that is $\model, z \models \phi_2$. Suppose $y \neq x$. Then there are $p : x \pto{}{}$  and $l > 0$ such that $p(l) = y \land \forall k \in \{1,\ldots,l\} . \model, p(k) \models \phi_1 \land \lnot \phi_2$. Define $p'$ by $p'(l') = p(l')$ if $l' \leq l$, and $p'(l') = z$ otherwise. The rest of the proof mimics the case $y = x$. If $\model, z \models \phi_1$, then $\model, z \nmodels \phi_2$ implies $z \in A_x$, witnessed by $p'$ and $l'=l+1$, therefore $\model, z \models \phi_2$. If $\model, z \models \lnot \phi_1$, then by hypothesis there must be $k \in \{1, \ldots,l+1\}$ such that $\model, p'(k) \models \phi_2$. By definition of $p'$, it is not possible that $k \in \{1, \ldots,l\}$, thus $k = l+1$ and $\model, z \models \phi_2$. By this argument, we have $\model, x \models \phi_1 \lsurr \phi_2$ using the set $A_x$ to verify the definition of satisfaction.
\end{proof}

\begin{proof}(of \autoref{lem:collective-derived-operators})
The results in \autoref{lem:empty} and \autoref{lem:exists} easily follow from \autoref{lem:individually}. For \autoref{lem:individually}, we have:
$
\der \model, A \models_C \lForall \phi 
\jstp \iff {$Def. $ \lForall} \model, A \models_C \lnot \phi \lShare \lGroup \bot 
\jstp \iff {$Def. $ \lShare} \model, \{x \in A \mid \model, x \models \lnot \phi \} \models \lGroup \bot 
\jstp \iff {$Def. $ \lGroup} \exists B \subseteq X . \{x \in A \mid \model, x \models \lnot \phi \} \subseteq B \land B \text{ is path connected } \land \forall z \in B . \model, z \models \bot
\jstp \iff {\forall z \in B . \model, z \models \bot \iff B = \emptyset }
\{x \in A \mid \model, x \models \lnot \phi \} \subseteq \emptyset 
\stp \iff \forall x \in A . \model, x \models \phi 
$
\qed
\end{proof}

\begin{proof}(of \autoref{lemma:termination})
We prove by induction on the syntax of \slcs\ formulae that for any 
quasi-discrete closure model $\model=((X,\closure_{R}),\eval)$,
and for any formula $\phi$ function $\check$ terminates in at most $\mathcal{O}(size(\phi)\cdot(|X|+|R|))$ steps. 

\medskip
\noindent
\emph{Base of Induction.} If $\phi=\top$ or $\phi=p$ the statement follows directly from the definition 
of $\check$. Indeed, in both these cases function $\check$ computes the final result in just $1$ step.

\medskip
\noindent
\emph{Inductive Hypothesis.} Let $\phi_1$ and $\phi_2$ be such that  for any 
quasi-discrete closure model $\model=((X,\closure_{R}),\eval)$,  function $\check(\model,\phi_i)$, $i=1,2$, terminate
 in at most $\mathcal{O}(size(\phi_i)\cdot(|X|+|R|))$ steps.

\medskip
\noindent
\emph{Inductive Step.} 

\begin{description}
\item[$\phi=\neg\phi_1$] In this case function $\check$ first recursively computes the set $P=\check(\model, \phi_1)$,
then returns $X-P$. By inductive hypothesis, the calculation of $P$ terminates in at most $\mathcal{O}(size(\phi_1)\cdot(|X|+|R|))$ steps, while to compute $X-P$ we need $\mathcal{O}(|X|)$ steps. Hence, $\check(\model, \neg\phi_1)$
terminates in at most  $\mathcal{O}(size(\phi_1)\cdot(|X|+|R|))+\mathcal{O}(|X|)$. However:
\[
\begin{array}{cl}
 & \mathcal{O}(size(\phi_1)\cdot(|X|+|R|))+\mathcal{O}(|X|) \\
\leq & \mathcal{O}(size(\phi_1)\cdot(|X|+|R|))+\mathcal{O}(|X|+|R|) \\
= & \mathcal{O}((1+size(\phi_1))\cdot(|X|+|R|)) \\
= & \mathcal{O}(size(\neg\phi_1)\cdot(|X|+|R|))
\end{array}
\]

\item[$\phi=\phi_1\wedge\phi_2$] To compute  $P=\check(\model, \phi_1\wedge \phi_2)$ function $\check$
first computes $P=\check(\model, \phi_1)$ and $Q=\check(\model, \phi_2)$. Then the final result is 
obtained as $P\cap Q$. Like for the previous case, we have that the statement follows from inductive hypothesis and by using the fact that $P\cap Q$ can be computed in 
at most $\mathcal{O}(|X|)$. 

\item[$\phi=\lnear\phi_1$] In this case function $\check$ first computes, in at most $\mathcal{O}(size(\phi_1)\cdot(|X|+|R|))$ steps, the set $P=\check(\model,\phi_1)$. Then the final result is obtained as $\closure_{R}(P)$. Note that,
to compute  $\closure_{R}(P)$ one needs $\mathcal{O}(|X|+|R|)$ steps. According to 
\autoref{def:closure-operator-of-a-relation}, $\closure_{R}(P)$ is obtained as the union, 
computable in $\mathcal{O}(|X|)$ steps, of $P$ with $\{x \in X | \exists a\in P. (a,x)\in R\}$. The latter
can be computed in $\mathcal{O}(|R|)$ steps. Indeed, we need to consider all the \emph{edges} exiting from $P$. 
Hence, $\check(\model,\lnear\phi_1)$ terminates in a number of steps that is:
\[
\begin{array}{cl}
& \mathcal{O}(size(\phi_1)\cdot(|X|+|R|))+\mathcal{O}(|X|)
+\mathcal{O}(|R|)\\
= & \mathcal{O}(size(\phi_1)\cdot(|X|+|R|))+\mathcal{O}(|X|+|R|)\\
= & \mathcal{O}((1+size(\phi_1))\cdot(|X|+|R|))\\
= & \mathcal{O}(size(\lnear\phi_1)\cdot(|X|+|R|))\\
\end{array}
\]

\item[$\phi=\phi_1\lsurr\phi_2$] When $\phi=\phi_1\lsurr\phi_2$ function $\check$ recursively invokes function $\checkUntil$ that first computes the sets $V=\check(\model,\phi_1)$, $Q=\check(\model,\phi_2)$ and $T= \cboundary(V \cup Q)$.
By inductive hypothesis, the computations of $V$ and $Q$ terminate in at most $\mathcal{O}(size(\phi_1)\cdot(|X|+|R|))$
and $\mathcal{O}(size(\phi_2)\cdot(|X|+|R|))$ steps, respectively, while $T$ can be computed in $\mathcal{O}(|X|+|R|)$.
After that, the loop at the end of function $\checkUntil$ is executed. We can observe that:
\begin{itemize}
\item a point $x$ is added to $T$ only one time (i.e. if an element is removed from $T$, it is never reinserted in $T$);
\item all the points in $T$ are eventually removed from $T$;
\item each \emph{edge} in $\model$ is traversed at most one time.
\end{itemize}

The first two items, together with the fact that $\model$ is finite, guarantee that the loop terminates. The last item
guarantees that the loop terminates in at most $\mathcal{O}(|R|)$ steps\footnote{Note that this is the complexity for a DFS 
in a graph}. Summing up, the computation of $\check(\model,\phi_1\lsurr\phi_2)$ terminates in at most
\[
\begin{array}{cl}
& \mathcal{O}(size(\phi_1)\cdot(|X|+|R|))+\mathcal{O}(size(\phi_2)\cdot(|X|+|R|))\\
& +\mathcal{O}(|X|+|R|)+\mathcal{O}(|R|)\\
= & \mathcal{O}((size(\phi_1)+size(\phi_2))\cdot(|X|+|R|))+\mathcal{O}(|X|+|R|)\\
= & \mathcal{O}((1+size(\phi_1)+size(\phi_2))\cdot(|X|+|R|))\\
= & \mathcal{O}(size(\phi_1\lsurr\phi_2)\cdot(|X|+|R|))\\
\end{array}
\]

\item[$\phi=\phi_1\ldiff\phi_2$] Similarly to the previous case, when $\phi=\phi_1\ldiff\phi_2$ function $\check$ recursively invokes function $\checkDiff$ that first computes the sets $V=\check(\model,\phi_1)$, $Q=\check(\model,\phi_2)$ and $T= \cboundary(V)\cap Q$.
By inductive hypothesis, the computations of $V$ and $Q$ terminate in at most $\mathcal{O}(size(\phi_1)\cdot(|X|+|R|))$
and $\mathcal{O}(size(\phi_2)\cdot(|X|+|R|))$ steps, respectively, while $T$ can be computed in $\mathcal{O}(|X|+|R|)$.
After that, the loop at the end of function $\checkDiff$ is executed. We can observe that:
\begin{itemize}
\item a point $x$ is added to $T$ only one time (i.e. if an element is removed from $T$, it is never reinserted in $T$);
\item all the points in $T$ are eventually removed from $T$;
\item each \emph{edge} in $\model$ is traversed at most one time.
\end{itemize}

The first two items, together with the fact that $\model$ is finite, guarantee that the loop terminates. The last item
guarantees that the loop terminates in at most $\mathcal{O}(|X|+|R|)$ steps. Summing up, the computation of $\check(\model,\phi_1\ldiff\phi_2)$ terminates in at most
\[
\begin{array}{cl}
& \mathcal{O}(size(\phi_1)\cdot(|X|+|R|))+\mathcal{O}(size(\phi_2)\cdot(|X|+|R|))\\
& +\mathcal{O}(|X|+|R|)+\mathcal{O}(|X|+|R|)\\
= & \mathcal{O}((size(\phi_1)+size(\phi_2))\cdot(|X|+|R|))+\mathcal{O}(|X|+|R|)\\
= & \mathcal{O}((1+size(\phi_1)+size(\phi_2))\cdot(|X|+|R|))\\
= & \mathcal{O}(size(\phi_1\ldiff\phi_2)\cdot(|X|+|R|))\\
\end{array}
\]\vspace{-24.5 pt}
\end{description}
\end{proof}

\begin{proof}(of \autoref{thm:sound-compl})
The proof proceeds by induction on the syntax of \slcs\ formulae. 

\medskip
\noindent
\emph{Base of Induction.} If $\phi=\top$ or $\phi=p$ the statement follows directly from the definition 
of function $\check$ and from \autoref{def:closure-semantics}. 

\medskip
\noindent
\emph{Inductive Hypothesis.} Let $\phi_1$ and $\phi_2$ be such that  for any finite
quasi-discrete closure model $\model=((X,\closure_{R}),\eval)$,  function $x\in \check(\model,\phi_i)$ if and only if
$\model,x\models \phi_i$, for $i=1,2$.

\medskip
\noindent
\emph{Inductive Step.} 

\begin{description}
\item[$\phi=\neg\phi_1$] 

$x\in \check(\model, \neg\phi_1)
\jstp \iff {\mbox{Definition of $\check$}}
x\not\in \check(\model, \phi_1)
\jstp \iff {\mbox{Inductive Hypothesis}}
\model,x\not\models\phi_1
\jstp \iff {\mbox{\autoref{def:closure-semantics}}}
\model,x\models\neg\phi_1
$

\item[$\phi=\phi_1\wedge\phi_2$] 
$x\in \check(\model, \phi_1\wedge\phi_2)
\jstp \iff {\mbox{Definition of $\check$}}
x\in \check(\model, \phi_1)\cap \check(\model, \phi_2)
\stp \iff
x\in \check(\model, \phi_1)\mbox{ and } x\in \check(\model, \phi_2)
\jstp \iff {\mbox{Inductive Hypothesis}}
\model,x\models\phi_1  \mbox{ and } 
\model,x\models\phi_2 
\jstp \iff {\mbox{\autoref{def:closure-semantics}}}
\model,x\models\phi_1\wedge\phi_2
$

\item[$\phi=\lnear\phi_1$] 
$x\in \check(\lnear\phi_1)
\jstp \iff {\mbox{Definition of $\check$}}
x\in \closure_{R}(\check(\model, \phi_1))
\jstp \iff {\mbox{Definition of $\closure_{R}$}}
\exists A\subseteq \check(\model, \phi_1): x\in \closure_{R}(A)
\jstp \iff {\mbox{Inductive Hypothesis}}
\exists A\subseteq X. \forall y\in A. \model,y,\models \phi_i\mbox{ and } x\in \closure_{R}(A)
\jstp \iff {\mbox{\autoref{def:closure-semantics}}}
\model,x\models \lnear\phi_1
$

\item[$\phi=\phi_1\lsurr\phi_2$] We prove that $x\in \checkUntil(\model,\phi_1,\phi_2)$ if and 
only if $\model,x\models \phi_1\lsurr\phi_2$. 
Function $\checkUntil$ takes as parameters a model $\model$ and two \slcs\ 
formulas $\phi_1$ and $\phi_2$ and 
computes the set of points in $\model$ satisfying $\phi_1 \lsurr \phi_2$ by removing from $V=\check(\model,\phi_1)$
all the \emph{bad} points. 

A point is \emph{bad} if it can \emph{reach} a point satisfying $\neg\phi_1$ 
without passing through a point satisfying $\phi_2$.
Let $Q=\check(\model,\phi_2)$ be the set of points in $\model$ satisfying $\phi_2$. 
To identify the \emph{bad} points in $V$ the function \checkUntil performs a \emph{backward search} from 
$T=\cboundary(V\cup Q)$.  Note that any \emph{path exiting} from $V\cup Q$ has to pass through points
in $T$. Moreover, the latter only contains points that satisfy neither $\phi_1$ nor $\phi_2$, by definition.
Until $T$ is empty, function \checkUntil first picks all the elements $x$ in $T$ and then removes from $V$ 
the set of (bad) points $N$ that are in $V-Q$ and that can reach  $x$ in \emph{one step}.  
%
%
At the end of each iteration the set $T$ contains the set of \emph{bad} points discovered in the last
iteration.
The proof proceeds in two steps. The first step guarantees that if $x$ does not satisfy $\phi_1\lsurr \phi_2$,
then $x$ is eventually removed from $V$. The second step shows that if $x$ is removed from $V$
then $x$ does not satisfy $\phi_1\lsurr \phi_2$.

Note that, by Inductive Hypothesis, we have that: 

\begin{equation}
\label{eq:assumption1}
x\in V=\check(\model,\phi_1) \Leftrightarrow \model,x\models \phi_1
\end{equation}

\begin{equation}
\label{eq:assumption2}
x\in Q=\check(\model,\phi_2) \Leftrightarrow \model,x\models \phi_2
\end{equation}

For each $x\in X$ we let:
\[
\mathcal{I}_{x} = \{ i\in \mathbb{N} | \exists p:x\pto{}{}. \model,p(i) \models\neg\phi_1\wedge\forall j\in \{ 1,\ldots, i \}. \model,p(j)\models\neg\phi_2 \}
\]
Note that, by definition, we have that
$\model,x\models \phi_1\lsurr \phi_2$ if and only if $\model,x\models \phi_1$ and $\mathcal{I}_{x}=\emptyset$.

First we prove that 
if $\mathcal{I}_{x}\not=\emptyset$ and $\model,x\models \phi_1$, then $x$ is removed from 
$V$ at iteration $i=\min \mathcal{I}_{x}$. This guarantees that if $x$ does not satisfy $\phi_1\lsurr \phi_2$,
then $x$ is eventually removed from $V$. The proof of this result proceeds by induction on $i$:

\begin{description}
\item[Base of Induction] Let $x\in X$ such that $\model,x\models \phi_1$, $\mathcal{I}_{x}\not=\emptyset$ and 
$\min \mathcal{I}_{x}=1$. 
Since $\min \mathcal{I}_{x}=1$, we have that there exists $p:x\pto{}{}$ such 
that $\model,p(1)\models \neg\phi_1$ and $\model,p(1)\models \neg\phi_2$. 
By definition of paths, we also have that $x=p(0)$ and $(x,p(1))\in R$. This implies that $p(1)\in \cboundary(V\cup Q)$
and $x\in pre(p(1))$. By definition of function $\checkUntil$ we have that $p(1)$ is in $T$ and 
$x$ is removed from $V$ during the first iteration.
Note that $x$ will be added to $T$ only if it does not satisfy $\phi_2$ (i.e. if $x\not\in Q$).  

\item[Inductive Hypothesis] For each $x\in X$ be such that $\model,x\models \phi_1$, $\mathcal{I}_{x}\not=\emptyset$ and 
$\min \mathcal{I}_{x}=k$, $x$ is removed from $V$ at iteration $k$.

\item[Inductive Step] Let $x\in X$ be such that $\model,x\models \phi_1$, $\mathcal{I}_{x}\not=\emptyset$ and 
$\min \mathcal{I}_{x}=k+1$. 
If $\min \mathcal{I}_{x}=k+1$ then there exists $p:x\pto{}{}$ such that $\model,p(k+1)\models \neg\phi_1$ and 
for each $j\in \{1,\ldots,k+1\}$ $\model,p(j)\models \neg\phi_2$. 
We have also that $\model, p(1) \models \phi_1$ (otherwise $\min \mathcal{I}_{x}=1$) and 
$\min \mathcal{I}_{p(1)}=k$  (otherwise $\min \mathcal{I}_{x}\not=k+1$).
By inductive hypothesis we have that $p(1)$ is removed from $V$ at iteration $k$. However,
since $\model,p(1)\models\neg \phi_2$ we have that $p(1)\not \in Q$ and $p(1)$ is in the set $T$ 
at the beginning of iteration $k+1$. 
This implies that $x=p(0)$ is removed from $V$ at iteration $k+1$, since $x\in pre(p(1))$.
\end{description}\medskip

\noindent We now prove that if $x$ is removed from $V$ at iteration $i$, then $\mathcal{I}_{x}\not=\emptyset$ and $i=\min \mathcal{I}_{x}$. 
This ensures that if $x$ is removed from $V$ then $x$ does not satisfy $\phi_1\lsurr \phi_2$.
We proceed by induction on the number of iterations $i$:

\begin{description}
\item[Base of Induction] If $x\in V$ is removed in the first iteration we have that there exists a point 
$y\in \cboundary(V\cup Q)$ such that $(x,y)\in R$. From \autoref{eq:assumption1} 
and \autoref{eq:assumption2} we have that $\model, x \models \phi_1$ while $\model,y\models\neg\phi_1\wedge\neg\phi_2$. This implies that there exists a path $p:x\pto{}{}$ such that $p(1)=y$ and $1=\min \mathcal{I}_{x}$. 

\item[Inductive Hypothesis] For each point $x\in V$,  if $x$ is removed from $V$ at iteration $i\leq k$, then 
$\mathcal{I}_{x}\not=\emptyset$ and $i=\min \mathcal{I}_{x}$. 

\item[Inductive Step] Let $x\in V$ be removed at iteration $k+1$.  This implies that after $k$ iterations, 
there exists a point $y$ in $T$ such that $(x,y)\in R$. This implies that $y$ has been removed
from $V$ at iteration $k$ and, by inductive hypothesis, $\mathcal{I}_{y}\not=\emptyset$ and 
$k=\min \mathcal{I}_{y}$. Hence, there exists a path $p:y\pto{}{}$ such that $\model,p(k)\models \neg\phi_1$ and 
for each $j\in \{1,\ldots,k\}$ $\model,p(j)\models \neg\phi_2$.
Moreover, since $y\in T$, we have also that $y\not\in Q$ and, from \autoref{eq:assumption2}, 
$\model,y\models\neg\phi_2$. We can consider the path $p':x\pto{}{}$ such that, for each $j$, $p'(0)=x$ and
$p'(j+1)=p(j)$. We have that $\model,p'(k+1)\models\neg\phi_1$ and for each $j\in \{1,\ldots,k+1\}$, 
$\model,p'(j)\models\neg\phi_2$. Hence $\mathcal{I}_{x}\not=\emptyset$ and $k+1=\min \mathcal{I}_{x}$
(otherwise $x$ should be removed from $V$ in a previous iteration).
\end{description}

\item[$\phi=\phi_1\ldiff\phi_2$] We let $R_k$, $T_k$ and $Q_k$ denote the values of variables $R$, $T$ and $Q$ at iteration $k$ in  \checkDiff, respectively. 
Our proof proceeds in three steps. First (\textbf{Step 1}) we prove that:

\[
\forall k. x\in R_{k}\wedge post(x)\cap Q_{k}\not=\emptyset \Rightarrow x\in T_{k}
\]

\noindent
then (\textbf{Step 2}) we show that:

\[
\forall k. T_{k+1}=\cboundary(R_{k})\cap Q_{k}
\]

\noindent
finally (\textbf{Step 3}) we prove that 

\[
\begin{array}{rcr}
\forall k. x\in R_k &\Leftrightarrow &\model,x\models \Phi_2\land \exists y . \exists l\leq k+1.\exists p : y \pto {l}{x} .
\model, y \models \phi_1\\
& &  \land\forall i . 0 < i < l \implies \model, p(i) \models \phi_2
\end{array}
\]

\noindent 
After that the statement directly follows from Def.~\ref{def:closure-semantics}.
However, before proceeding further, we can notice that, by Inductive Hypothesis, the following hold: 

\begin{equation}
\label{eq:prop_assumption1}
x\in V=\check(\model,\phi_1) \Leftrightarrow \model,x\models \phi_1
\end{equation}

\begin{equation}
\label{eq:prop_assumption2}
x\in Q=\check(\model,\phi_2) \Leftrightarrow \model,x\models \phi_2
\end{equation}

Moreover, we can also notice that:

\begin{equation}
\label{eq:prop_fact1}
\forall k. R_{k}\cap Q_{k}=\emptyset
\end{equation}

\begin{equation}
\label{eq:prop_fact2}
\forall k. R_{k}\cap Q_{k}=Sat(\phi_2)
\end{equation}

\noindent 
both the fact above can be derived directly from the definition of \checkDiff in Fig.~\ref{alg:check-diff-quasi-discrete}.
Indeed, at the beginning $R_0=T_0$ while $Q_0=Sat(\phi_2)\setminus T_0$. Moreover,
at every iteration $R_{k+1}=R_{k}\cup T'$ while $Q_{k+1}=Q_{k}\setminus T'$.

\paragraph{Step 1:} We prove by induction on $k$ that:

\[
\forall k. x\in R_{k}\wedge post(x)\cap Q_{k}\not=\emptyset \Rightarrow x\in T_{k}
\]

\begin{description}
\item[Base of Induction] Let $k=0$. The statement follows directly from the fact that $R_{0}=T_{0}$. Hence:

\[
x\in R_{0}\wedge post(x)\cap Q_{0}\not=\emptyset \Rightarrow x\in T_{0}
\]

\item[Inductive Hypothesis] For any $k\leq n$:

\[
x\in R_{k}\wedge post(x)\cap Q_{k}\not=\emptyset \Rightarrow x\in T_{k}
\]

\item[Inductive Step] Let $k=n+1$:

$x\in R_{n+1}\wedge post(x)\cap Q_{n+1}=\emptyset \wedge x\not\in T_{n+1}
\jstp \iff {x\not\in T_{n+1}\wedge R_{n+1}=R_{n}\cup T_{n+1}}
x\in R_{n}\wedge post(x)\cap Q_{n+1}=\emptyset
\jstp \iff {Q_{n+1}= Q_{n}\setminus T_{n+1}}
x\in R_{n}\wedge post(x)\cap Q_{n}=\emptyset
\jstp \iff {\mbox{Inductive Hypothesis}}
x\in T_{n}
\jstp \iff {\mbox{Def. of \checkDiff in Fig.~\ref{alg:check-diff-quasi-discrete}}}
post(x)\cap Q_{n}\subseteq T_{n+1}
\jstp \iff {Q_{n+1}= Q_{n}\setminus T_{n+1}}
post(x)\cap Q_{n+1}=\emptyset \mbox{\qquad (\emph{RAA})}
$
\end{description}

\paragraph{Step 2:} We prove that:

\[
\forall k. T_{k+1}=\cboundary(R_{k})\cap Q_{k}
\]

We first show that $T_{k+1}\subseteq \cboundary(R_{k})\cap Q_{k}$. Let $x\in T_{k+1}
\jstp \iff {\mbox{Def. of \checkDiff in Fig.~\ref{alg:check-diff-quasi-discrete}}}
\exists y\in T_{k}\wedge x\in post(y)\cap Q_{k}
\jstp \Longrightarrow {T_{k}\subseteq R_{k}\mbox{ and Def. of $\closure$} }
x\in \closure(R_{k}) \wedge x\in Q_{k}
\jstp \Longrightarrow {Eq.~\ref{eq:prop_fact1}}
x\in \closure(R_{k}) \wedge x\not\in R_{k}\wedge x\in Q_{k}
\jstp \Longrightarrow {Def. of \cboundary}
x\in \cboundary(R_{k}) \wedge x\in Q_{k}
\stp \Longrightarrow 
x\in \cboundary(R_{k}) \cap Q_{k}
$

Now we show that $\cboundary(R_{k})\cap Q_{k}\subseteq T_{k+1}$. Let $x\in \cboundary(R_{k})\cap Q_{k}
\jstp \Longrightarrow {\mbox{Def. of $\cboundary$ and Def. of $\closure$}}
\exists y\in R_{k}: x\in post(y)\wedge x\not \in R_{k}\wedge x\in Q_{k}
\jstp \Longrightarrow {\mbox{\textbf{Step 1}}}
\exists y\in T_{k}: x\in post(y)\wedge x\in Q_{k}
\jstp \Longrightarrow {\mbox{Def. of \checkDiff in Fig.~\ref{alg:check-diff-quasi-discrete}}}
x\in T_{k+1}
$

\paragraph{Step 3:} We can now prove by induction on $k$ that:

\[
\begin{array}{rcr}
\forall k. x\in R_k &\Leftrightarrow &\model,x\models \phi_2\land \exists y . \exists l\leq k+1.\exists p : y \pto {l}{x} .
\model, y \models \phi_1\\
& &  \land\forall i . 0 < i < l \implies \model, p(i) \models \phi_2
\end{array}
\]

\begin{description}
\item[Base of Induction] Let $k=0$ and $x\in R_0
\jstp \iff {\mbox{Definition of $\checkDiff$}}
x\in \closure_{R}(V) \cap Sat(\phi_2)
\jstp \iff {\mbox{Definition of $\closure_{R}$}}
x\in (V\cap \cboundary(V)) \cap Sat(\phi_2)
\stp 
\iff x\in (Q\cap V)\cup (Q\cap \cboundary(V))
\jstp \iff {\mbox{Definition of $\cboundary(V)$}}
 x\in (Sat(\phi_2)\cap V)  
\stp {\mbox{ or }} x\in Sat(\phi_2) \wedge \exists y\in V: (y,x)\in R
\jstp \iff {\mbox{From \ref{eq:prop_assumption1} and \ref{eq:prop_assumption2}}}
 \model,x \models \phi_2 \wedge \model,x \models \phi_2
\stp {\mbox{ or }} \model,x \models \phi_2 \wedge \exists y. \model,y\models \phi_1: (y,x)\in R
\jstp \iff {\mbox{From Def.~\ref{def:path} and Lemma~\ref{lem:paths-are-paths}}}
\model,x\models \phi_2\land
\stp {} \exists y . \exists l\leq 1.\exists p : y \pto {l}{x} .
\model, y \models \phi_1\land\forall i . 0 < i < l+1 \implies \model, p(i) \models \phi_2
$

\item[Inductive Hypothesis] For any $k\leq n$: 
\[
\begin{array}{rcl}
x\in R_k &\Leftrightarrow &\model,x\models \phi_2\land \exists y . \exists l\leq k+1.\exists p : y \pto {l}{x} .
\model, y \models \phi_1\\
& & \qquad \land\forall i . 0 < i < l \implies \model, p(i) \models \phi_2
\end{array}
\]

\item[Inductive Step] Let $k=n+1$ and $x\in R_{n+1}
\jstp \iff {\mbox{Def. of \checkDiff in Fig.~\ref{alg:check-diff-quasi-discrete}}}
x\in R_{n} \cup T_{n+1}
\stp \iff
x\in R_{n}
\stp {\mbox{or}}
x\in T_{n+1}
\jstp \iff {\mbox{\textbf{Step 2}}}
x\in R_{n}
\stp {\mbox{or}}
x\in \cboundary(R_{n})\cap Q_{n}
\jstp \iff {\mbox{Def. of $\cboundary$}}
x\in R_{n}
\stp {\mbox{or}}
\exists x' \in R_{n}. x\in post(x')\cap Q_{n}
\jstp \iff {\mbox{\ref{eq:prop_fact2} and \ref{eq:prop_assumption2}}}
x\in R_{n} 
\stp {\mbox{or}} 
\exists x' \in R_{n}. x\in post(x') \wedge \model,x\models \phi_{2}$ 

$
\jstp \iff {\mbox{Inductive Hypothesis}}
\model,x\models \phi_2\land \exists y . \exists l\leq n+1.\exists p : y \pto {l}{x} .
\model, y \models \phi_1 
\stp {} \qquad\qquad\qquad\qquad \land\forall i . 0 < i < l \implies \model, p(i) \models \phi_2
\stp {\mbox{or}}
\exists x'. \in R_{n}.  \model,x'\models \phi_2\land \exists y . \exists l\leq n+1.\exists p : y \pto {l}{x'} .
\model, y \models \phi_1 
\stp {} \qquad\qquad\qquad\qquad \land\forall i . 0 < i < l \implies \model, p(i) \models \phi_2
\stp {} x\in post(x')\wedge \model,x\models \phi_{2}
\stp \iff {}
\model,x\models \phi_2\land \exists y . \exists l\leq n+1.\exists p : y \pto {l}{x} .
\model, y \models \phi_1 
\stp {} \qquad\qquad\qquad\qquad \land\forall i . 0 < i < l \implies \model, p(i) \models \phi_2
\stp {\mbox{or}}
\model,x\models \phi_2\land \exists y . \exists 0<l\leq n+2.\exists p : y \pto {l}{x} .
\model, y \models \phi_1 
\stp {} \qquad\qquad\qquad\qquad \land\forall i . 0 < i < l \implies \model, p(i) \models \phi_2
\stp \iff {}
\model,x\models \phi_2\land \exists y . \exists l\leq n+2.\exists p : y \pto {l}{x} .
\model, y \models \phi_1 
\stp {} \qquad\qquad\qquad\qquad \land\forall i . 0 < i < l \implies \model, p(i) \models \phi_2
$\qedhere
\end{description}

\end{description}

\end{proof}

\begin{proof}(of \autoref{thm:check-collective})
 We provide a sketch, as the core of the proof is that of Tarjan's algorithm, which we assume given. The proof is by induction on the structure of formulas. The only case where the algorithm is not a direct implementation of its mathematical definition is the one for $\psi = \lGroup \phi$. If $A = \emptyset$ the algorithm returns $True$. This is correct by definition of $\models$, as the empty set is strongly connected. Otherwise, the set of points $B$ satisfying $\phi$ is computed using function $\Sat$, and the algorithm returns $\mathit{False}$ if $A \nsubseteq B$. This is correct since all elements of $A$ must satisfy $\phi$. Under the hypothesis that $0 \neq A \subseteq B$,  an element $x$ is chosen from $A$, and the algorithm executes a depth-first search according to \cite{Tar72}, modified to only follow successors of $x$ that are in $B$.  Note that the start node $x$ is in $B$, therefore the algorithm only visits nodes in $B$. For each strongly connected component $C$ reachable from $x$ in the subgraph defined by $B$, the algorithm checks whether $A \subseteq C$. If this is the case, then $\model, A \models \lGroup \phi$ and the algorithm returns $\mathit{True}$. Conversely, if there is at least one point in $A \cap C$, but not all points of $A$ are in $C$, then $\model, A \nmodels \lGroup \phi$. To see this, consider $y \in A \cap C$ and $z \in A \cap (X \setminus C)$. It cannot be the case that there are a path from $y$ to $z$ and a path from $z$ to $y$ both only crossing nodes in $B$, otherwise we would have $z \in C$. Therefore, the algorithm returns $\mathit{False}$. If a strongly connected component is found, but no node of $A$ belongs to it, the algorithm returns $\mathit{undefined}$ and the depth-first search continues.  One of the first two conditions necessarily happens along the execution of \autoref{alg:check-group}, when invoked from \autoref{alg:model-checking}, since there is at least one strongly connected component reachable from $x$ and containing $x$ itself, with $x \in A$. Therefore, \autoref{alg:check-group} never returns $\mathit{undefined}$ when $x \in A$. Termination, and the fact that the algorithm effectively finds strongly connected components, is a consequence of correctness of Tarjan's procedure. 
 The worst case time complexity of Tarjan's algorithm is $\mathcal{O}(|X|+|R|)$ steps. This, the fact that the definition of \autoref{alg:model-checking} is by induction on the structure of formulas, and \autoref{thm:sound-compl}, cause the algorithm to have time complexity $\mathcal{O}(k\cdot(|X|+|R|))$.
\end{proof}

\end{document}